\documentclass[12pt,oneside,onecolumn]{report}
\usepackage{graphicx,bm,floatflt,amssymb,epsf,psfig,rotate} 
\newsavebox{\fmbox}                       

\def\MZ{M_Z}

\def\MR{M_R}
\def\MU{M_U}

\def\ep{\epsilon}

\def\gsim{\mathop{\smash{>}}\limits_\sim}

\def\lr{{\cal G}_{LR}}
\def\std{{\cal G}_{std}}

\def\l{\lambda}
\def\L{\Lambda}
\def\m{\mu}
\def\n{\nu}
\def\r{\rho}
\def\g{\gamma}
\def\G{\Gamma}
\def\d{\delta}
\def\D{\Delta}

\def\P{\partial}
\def\a{\alpha}

\def\b{\beta}

\def\e{\epsilon}
\def\s{\psi}
\def\S{\Psi}
\def\t{\theta}
\def\T{\Theta}
\def\r{\rho}

\def\ep{\epsilon}

\def\op{\oplus}
\def\ov{\overline}

\def\ra{\rightarrow}

\def\bc{\begin{center}}
\def\ec{\end{center}}
\def\be{\begin{equation}}
\def\ee{\end{equation}}
\def\bi{\begin{itemize}}
\def\ei{\end{itemize}}

\newcommand{\br}{\begin{array}}
\newcommand{\er}{\end{array}}
\newcommand{\ba}{\begin{eqnarray}}
\newcommand{\ea}{\end{eqnarray}}

\newcommand{\CL}{{\cal L}} 
 
\newcommand{\CO}{{\cal O}} 

\def\bpmat{\begin{pmatrix}}
\def\epmat{\end{pmatrix}}

\def\ra{\rightarrow}
\def\stilde{\widetilde}

\def\lim{\raisebox{-.9ex}{\rlap{\tiny $\e \rightarrow {\tiny 0}$}} \raisebox{.9ex}{lim}}

\def\ltap{\raisebox{-.4ex}{\rlap{$\sim$}} \raisebox{.4ex}{$<$}} 
\def\gtap{\raisebox{-.4ex}{\rlap{$\sim$}} \raisebox{.4ex}{$>$}}
             
\begin{document}
\thispagestyle{empty}


\vskip4cm
\begin{center}
{\huge \bf{Some Explorations of New Physics \\ Beyond The Standard Model}}

\end{center}

\vspace{4cm}

\begin{center}

{\bf{       
 Thesis Submitted to \\
  The University of Calcutta \\
       for The Degree of \\
Doctor of Philosophy (Science) \\}}

\vspace{7cm}
            
{\large By}\\

{\bf {\Large Swarup Kumar Majee}}\\
{\large {Department of Physics,\\ University of Calcutta, \\ 92 Acharya Prafulla Chandra Road,  \\ Kolkata 700 009, India}}

\vspace{1cm}      

{\large March, 2008     }
\end{center}


\pagenumbering{roman} 
\newpage
\mbox{}
\newpage
\begin{center}
{\large {\bf ACKNOWLEDGMENTS}}
\end{center}
\vskip 2\baselineskip
 
\begin{normalsize}
I gratefully acknowledge the constant and invaluable academic
and personal support received from my supervisor Professor 
Amitava Raychaudhuri. I am really thankful and indebted to him 
for having been the advisor everyone would like to have, for his  
clear and enthusiastic discussion, for helping me to find my elusive
grounding in the research problem and without whose dedicated 
supervision it would have been never possible to complete this thesis.

I have no words to thank my senior-cum-collaborators Professor 
Gautam Bhattacharyya (SINP) and Dr. Anindya Datta (CU) for 
their encouragement, their support, their collaboration  and for
constantly motivating me to be in Physics. I have really enjoyed 
by working with them. They have been substantial sources of inspirations 
and profound insight during these years.

I am very much indebted to Professor Mina Ketan Parida (NISER) not
only for his collaboration but his constant teaching on various subjects
during his stay at Harish-Chandra Research Institute. I would also like 
to thank Professor Utpal Sarkar (PRL) for his collaboration and beautiful
discussion. Without their help it would not be possible to reach at the 
present status of the thesis. 

It is my great pleasure to thank Professor Biswarup Mukhopadhyaya, Dr.  
Srubabati Goswami, Dr. Asesh Krishna Datta, Dr. V. Ravindran, Professor Raj 
Gandhi, Dr. Sandhya Choubey and Dr. Andreas Nyffeler for 
their encouragement and support at different stages during my stay at the 
Harish-Chandra Research Institute. I have really enjoyed the discussion 
with them on various physics issues. In addition, I would like to give 
my sincere thanks to Professor  Ashoke Sen, Professor Sumathi Rao, Professor 
Debashis Ghoshal, Professor Rajesh Gopakumar, Professor Dileep Jatkar, 
Professor S. Naik for their helps and suggestions at various stages. 

I am really grateful to Professor Tapan Kumar Das, Professor Subinay Dasgupta, 
Dr. Anirban Kundu, Dr. Parongama Sen and Dr. Gautam Gangopadhyay from Calcutta 
University who have helped me in various ways during last four years. I 
would also like to thank Professor Amitava Datta (JU), Professor Sreerup 
Raychaudhuri (TIFR), Dr. Subhendu Rakshit (DU, Germany), 
Professor Debajyoti Choudhury (DU), Professor Soumitra Sengupta (IACS) 
and Dr. Rathin Adhikari (CTP, JMI) for physics discussion at various 
stages during my research work. 

I would like to thank Dr. Abhijit Samanta (SINP) and Dr. Arunansu Sil (Saclay)
 for their help not only in learning the computational packages and 
discussing physics but also for their clear positive ideas in both physics 
and non-physics issues which made my life smooth. At this juncture I would 
like  to extend my thanks to Sanjib Kumar Agarwalla (CU) and 
Shashank Shalgar for their help in learning different computational packages,
great friendship and sharing different thoughts for all these years.

 My special thanks to Soumitra Nandi, Biplob Bhattacharjee, Kamalika 
Basu Hazra, Anasuya Kundu, Pratap K. Das, Anjan Kumar Chandra,  
Kirtiman Ghosh from Calcutta University and Tirtha Shankar Ray from
SINP who have helped in various ways and shared their thoughts and provided
 me the moral supports. 

 I also thank to Dr. Abhijit Bandyopadhyay and Dr. Paramita Dey and all 
the Ph.D. students, in particular Arijit Saha, Kalpataru Pradhan, 
R. Srikanth H., Sudhir Gupta,
 Subhaditya Bhattacharyya and Priyotosh Bandyopadhyay from Harish-Chandra 
Research Institute for interesting on-train discussions and wonderful 
explanations, for their enjoyable company and support.

 I would like to acknowledge the Council of Scientific and Industrial 
Research, India for providing the financial support in the initial stage
of this work.

  Last but not least I would like to thank my parents and my brothers and
sister for believing in me, for their patience,  
for their constant support and for inspiring me not only to pursue my 
research work during all these years but to continue my work in physics
for the rest of my life.

\vskip3cm
\begin{flushright} 
Swarup Kumar Majee\\ 
Kolkata, India ~~~~
\end{flushright}
\end{normalsize}

\newpage
\begin{center}
{\Large \bf \underline{Abstract of The Thesis}}\\
\vskip2mm
\end{center}

We study the power law running of gauge, Yukawa and quartic scalar couplings
in the universal extra dimension scenario where the extra dimension is
accessed by all the standard model fields. Assuming compactification on an $S^1
/Z_2$ orbifold, we compute one-loop contributions of the relevant Kaluza-Klein
 towers to the above couplings up to a cutoff scale $\Lambda$. We get a 
low unification scale around 30 TeV for a radius $R \sim 1\; \rm
TeV^{-1}$. We also examine the consequences of power law running on the 
triviality and 
vacuum stability bounds on the Higgs mass.  Supersymmetric extension of the 
scenario requires $R^{-1}$ to be larger than $\sim 10^{10}$ GeV in order that 
the gauge couplings remain perturbative up to the scale where they tend to 
unify. 

Restricting the 
first two fermion generations in the brane, we derive, using the effective
potential approximation technique, an upper 
limit on the mass of the lightest CP-even neutral Higgs in
the minimal supersymmetric standard model in the presence of extra dimensions.
We observe that the lightest Higgs, whose upper bound in four dimensions is 
$\sim 135$ GeV, may comfortably weigh around 200 GeV (300 GeV) with one (two) 
extra dimension(s).

The $SO(10)$ Grand Unified Theory (GUT) is a preferred choice for the 
unification of
different standard model gauge groups. A low intermediate scale within minimal
supersymmetric $SO(10)$ GUT is a desirable feature to accommodate 
leptogenesis. We point out that any one of
three options --  threshold corrections due to the mass spectrum
near the unification scale, gravity induced non-renormalizable
operators near the Planck scale, or presence of additional light
Higgs multiplets -- can permit unification along with
much lower values of $M_R$ in both the doublet and triplet higgs
scalar models. In the
triplet model, independent and irrespective of these corrections, we 
find a lower bound
on the intermediate scale,
 $M_R > 10^9$ GeV, arising from the requirement that the theory
must remain perturbative at least upto the GUT scale. We show that
in the doublet model $M_R$ can even be in the TeV region which,
apart from permitting resonant leptogenesis, can be tested at LHC and ILC.

We have also explored the quark model interpretation of the pentaquark state. 
We estimate the pentaquark $(qqqq\bar{q})$ mass 
after calculating the $SU(6)$ unitary scalar factors and Racah
coefficients to incorporate proper colour-spin symmetry properties for 
the triquark $(q q \bar{q})$ state. When hyperfine interactions are 
assumed to be quark flavour independent and of the same strength for diquarks
and triquarks, extracting it from the baryon sector yields a
$\theta^+$ mass prediction of 1534 MeV. In this framework, other
pentaquark states $\Xi$ with S=--2 and $\theta^c $ with C=-1 are
expected at 1558 MeV and 2895 MeV respectively.


\newpage
\mbox{}
\tableofcontents
\newpage
\mbox{}
\listoffigures
\addcontentsline{toc}{chapter}{List of Figures}
\newpage
\listoftables
\addcontentsline{toc}{chapter}{List of Tables}
\newpage
\thispagestyle{empty}
\mbox{}
\pagenumbering{arabic}
\chapter{ Introduction}
\section{Introduction} \label{introintro}
The Standard Model (SM) of particle physics describes the dynamics of the elementary particles. 
It is a gauge field theory based on the group $SU(2)_L\times U(1)_Y\times SU(3)_C$. 
The electroweak theory ($SU(2)_L\times U(1)_Y$), proposed by Glashow-Salam-Weinberg \cite{sm},
describes the weak and electromagnetic interactions between the fundamental particles (quarks and
leptons). The colour gauge group $SU(3)_C$ acts only on the quark sector. Under the 
$SU(2)_L$ gauge group the left-handed particles are charged ones but the right-handed particles 
transform trivially. The electromagnetic interaction, as like the 
gravitational interaction, is of infinite range but the ranges of the weak and strong forces are 
finite. 
The masslessness of the photon field explains the long range 
behaviour of the electromagnetic field. Experiments revealed the weak gauge bosons as 
massive as required by  
the short-range characteristic of the weak interaction. To explain the same for the strong 
interaction 
we need the principle of colour confinement which states that the only observable states 
 are the colour singlet hadrons. Thus, in spite of the fact that the gluons, the carrier of strong
interactions, are massless the strong interaction is of finite range. Although, we need massive 
weak gauge bosons to explain the short range behaviour of the weak interaction,
 the $SU(2)_L$ gauge symmetry does not permit them, and fermions as well, to have a mass term 
in the Lagrangian. 

	The spontaneous symmetry breaking mechanism is a way out to generate the weak gauge boson
and fermion masses in the standard model by introducing an additional weak isodoublet complex 
scalar field. Weak gauge bosons get masses by absorbing three Goldstone bosons, three 
components of the scalar field, the remaining degree of freedom corresponds to a physical 
particle, the Higgs boson, the most wanted member for the present particle physics collider 
search. Once we choose a ground state, out of infinite possibilities, as the physical one, 
the electro-weak $SU(2)_L\times U(1)_Y$ symmetry
  breaks to $U(1)_Q$ symmetry. As a result, via the spontaneous symmetry breaking, the 
weak gauge bosons and the fermions acquire non-zero masses. 
In most versions of new physics
beyond the standard model nowadays the Higgs sector plays a key role.

	The standard model, till now, is in very good agreement with different experiments,
 like LEP, Tevatron run-I \& -II, HERA etc. It has predicted different weak gauge
boson masses very precisely, made several predictions for testing quantum electroweak corrections,
 etc. which 
 have all been verified.
	Despite the tremendous success of the standard model it has a few shortcomings. 
First of all the Higgs boson is not found in any of the
present or past experiments. There is no satisfactory explanation of why should there 
be any gauge symmetry $i.e.$ why should the Lagrangian be invariant under the local gauge 
transformations?  Why only three generations of fermions are there? All the fermions and 
Higgs boson masses and the gauge coupling constants 
are only parameters in the standard model. The clear evidence for physics beyond 
the standard model is the small nonzero neutrino  mass. Introducing a heavy right-handed 
neutrino in the see-saw mechanism one can explain the light neutrino mass. To reduce the large 
number of parameters of the standard model the Theory of Grand Unification has been introduced. 
According to this theory the difference in gauge coupling strengths is a low energy behaviour,
the coupling $`$constants' are functions of the energy scale and all gauge couplings will 
unify to a single one at an 
energy scale, the GUT scale ($\sim 10^{15}$ GeV), much higher than the electro-weak scale. In 
that high scale not only the gauge couplings but all the fermions from a generation can be 
put in a single (or a finite) multiplet(s) which leads to a few mass parameters and hence only 
a few 
Yukawa couplings. 
The problem which causes an itch to the high energy particle physicists is the huge 
difference between the Planck scale ($10^{19}$ GeV) and the electroweak symmetry breaking 
scale ($\sim$ 100 GeV). The Higgs boson mass receives a quadratically divergent 
quantum loop correction of the order of the Planck scale. This huge correction one can
remove by introducing TeV scale new physics like Supersymmetry, Extra Dimensions etc.
The standard model does not include the gravitational interaction. 

	Beyond the standard model, thus, is an obvious area we have to look into in order 
to explain the present and future experimental data as well as to have a clear picture about
the physics. So, a detailed discussion of some of the new physics is first presented. 
Based on the current experimental data we put some constraints on different parameters of the new 
physics and have also discussed how different standard model phenomena 
change their characteristic in the presence of such new physics. 
So let us begin with a short discussion of the 
standard model and new physics beyond it for a better understanding of 
 the work reported in this thesis.

\section{The standard model}\label{theSM}
The standard model, as we stated in the previous section, is a gauge field theory based on the group $SU(2)_L\times U(1)_Y\times SU(3)_C$. The particle content of the SM is enlisted in 
Table \ref{particle} with their corresponding gauge group representations. The left-handed 
particles are doublet under the $SU(2)_L$ gauge
transformation while the right-handed ones are singlet. As $SU(3)_C$ group does not distinguish
 left or right chirality, so both type of quarks are triplet while letpons are singlet. The 
hypercharge quantum number $Y$ is normalised to 
\be
Q = I_3 + Y,
\label{hypercharge}
\ee

\bc
\begin{table}
\begin{tabular}{|c|c|c|}  \hline 
Nature & particles&$SU(3)_C\times SU(2)_L\times U(1)_Y$ \\ \hline
Quarks (left-handed) : $Q_L$& $\left(\br{c}u\\d\er \right)_L$, $\left(\br{c}c\\s\er \right)_L$, 
$\left(\br{c}t\\b\er \right)_L$ & $(3,2, {1\over 6})$ \\ \hline 
Quarks (right-handed) : $\br{c}q^u_R\\q^d_R\er $& $\br{ccc}u_R,&c_R,&t_R \\ d_R, & s_R, & b_R\er $ 
& $\left(3,1,\br{c}{2/3}\\{-1/3}\er \right)$ \\ \hline
Leptons (left-handed) : $L_L$ & $\left(\br{c}\nu_e\\e^-\er \right)_L$, $\left(\br{c}\nu_{\mu}\\ \mu^-\er \right)_L$, 
$\left(\br{c}\nu_{\tau}\\ \tau^-\er \right)_L$ & $(1,2, -{1\over 2})$\\ \hline
Leptons (right-handed) : $l_R$ & $\br{ccc} e^-_R,& {\mu}^-_R,&{\tau}^-_R \er $ & $ (1,1,-1)$ \\ \hline
$SU(3)_C$ gauge boson & $G^a_\m$ (a=1,..,8) & (8,1,0) \\ \hline
$SU(2)_L$ gauge boson & $W^i_\m$ (i=1,2,3) & (1,3,0) \\ \hline
$U(1)_Y$ gauge boson & $B_\m$  & (1,1,0) \\ \hline
\end{tabular}
\caption{\sf \small{The fundamental matter and mediator members of particle physics.}}
\label{particle}
\end{table}
\ec

where $Q$ is the electric charge and $I_3$ is the third component of the 
isospin vector. The standard model also contains a yet to be observed 
$SU(2)_L$ doublet scalar, the essential ingredient for the Higgs 
mechanism,
\be
\Phi = \left(\br{c} \phi^+ \\ \phi^0\er \right) \equiv (1,2,{1\over 2}).
\label{hig1}
\ee 

Both the $\phi^+$ and $\phi^0$ are complex fields which can be expressed in terms of real
scalar fields $\phi _i$ 
\be
\phi^+ = {1\over{\sqrt 2}}(\phi _1 + i\phi _2) ~~~~~ {\rm and}~~~~~ \phi^0 = {1\over{\sqrt 2}}(\phi _3 + i\phi _4).
\label{hig2}
\ee

\section{Gauge invariance of the SM}\label{ginv}

Let us consider a Dirac field $\S$ and assume that the theory is invariant 
under the transformation 
\be
\S(x) \ra U\S(x)
\label{gtrnasform}
\ee 

where, $U=e^{i\a (x)}$. This is a phase rotation through an angle $\a(x)$ that 
itself depends on the space-time point. 

Let us start with the Lagrangian of a free Dirac field which can be written as 
\be
\CL = \bar{\S}(i\g ^\m\P _\m -{\rm m})\S
\label{freedirac}
\ee

Independence on space-time of $\a$ leaves eqn.(\ref{freedirac}) invariant under the 
transformation eqn.(\ref{gtrnasform}). Will the Lagrangian be invariant when $\a$ depends
on the space-time point as well? The mass term ${\rm m}\bar{\S}\S$ is invariant under both the global
as well as the local phase rotations. The problem will arise with the kinetic term; it is not 
invariant under the local phase transformation. But, one can make a co-variant kinetic term
 \cite{QFT} as follows:

	The derivative of the Dirac field along $n^\m$ direction is given by 
\be
n^\m\P _\m\S(x) = \lim ~~ {1\over \e} [\S(x+\e n) - \S(x)]
\label{gtrnasform2}
\ee

As $\S(x+\e n)$ and $\S(x)$ transform differently under eqn.(\ref{gtrnasform}), so eqn.(\ref{gtrnasform2}) is not a meaningful one. To make the difference sensible we should define a scalar
quantity $U(y,x)$ which will compensate the phase transformation from one point to another one by it's
transformation between two points as
\be
U(y,x) \ra e^{i\a(y)}U(y,x)e^{-i\a(x)}
\label{gtrnasform2P}
\ee

whenever the Dirac field $\S$ will transform as eqn.(\ref{gtrnasform}). An obvious requirement is
$U(x,x)=1$, as a generalization it implies $U(y,x)$ to be a pure phase only. Now both the fields
$\S(x+\e n)$ and $\S(x)$ transform the same way and the $covariant ~ derivative$ can be defined as
follows:
\be
n^\m D _\m\S(x) = \lim ~~ {1\over \e} [\S(x+\e n) - U((x+\e n),x)\S(x)]
\label{gtrnasform3}
\ee

For a continuous local phase transformation, as the gauge transformations are continuous, one can expand the function $U(y,x)$ between two points as
\be
U((x+\e n),x) = 1 - ie\e n^\m A_\m(x) + \CO(\e ^2).
\label{gtrnasform4}
\ee

Here $e$ is an arbitrary constant. It will appear as the gauge coupling constant in the context of the standard model. The coefficient of the term $(\e n^\m)$
is a new field $A_\m(x)$, the gauge field, introduced to keep the kinetic term of the Lagrangian
invariant under the gauge transformation. Thus the covariant derivative, now, can be written
as 
\be
D _\m\S(x) = \P _\m\S(x) + ieA_\m(x)\S(x).
\label{covdev}
\ee

Using eqn.(\ref{gtrnasform4}) on eqn.(\ref{gtrnasform2P}) we see that the gauge field transforms
as 
\be
A_\m(x) \ra A_\m(x) - {1 \over e} \P _\m\a(x).
\label{gaugeT}
\ee

Now, it is easy to check the invariance of covariant derivative $D _\m\S(x)$, using 
eqn.(\ref{gtrnasform}) and eqn.(\ref{gaugeT}), under the gauge transformation. So, in summary
we need a gauge field $A_\m(x)$, transforming as eqn.(\ref{gaugeT}), to keep the Lagrangian
invariant under the local gauge transformation. Immediately, we see that although the term 
$F_{\m\n} = \P _\m A_\n - \P _\n A_\m $, which will generate the kinetic term for the gauge field, 
 is invariant under the gauge transformation the mass term for the gauge field 
${\rm m}^2_AA_\m A^\m$ is not. Finally, the gauge invariant Lagrangian can be written as 
\be
\CL = -{1\over 4}F_{\m\n}F^{\m\n} + \bar{\S}[i\g ^\m(\P _\m +ieA_\m)-{\rm m}]\S
\label{freedirac2}
\ee

\section{Spontaneous symmetry breaking}\label{ssb}
To address the problem of the vanishing mass term for a gauge field in a gauge invariant theory we 
have to incorporate the mechanism of {\it Spontaneous Symmetry Breaking} (SSB).
It means the Lagrangian or the equation of motion has some symmetry but it's solution, 
the ground state, does not. Introduced by Heisenberg, in 1928, to explain the property of 
ferromagnetism whose spin states below a certain critical temperature choose a specific direction out 
of infinite possibilities for the ground state, a similar situation also arises in case of quantum field theory. 

\subsection{SSB for a global $U(1)$ symmetry}\label{gssb}
Let us consider, to start with, the potential of a complex scalar field 
$\Phi \equiv {{\left(\phi_1 + i\phi_2\right)}\over{\sqrt{2}}}$, as
\be
V(\phi) = {\rm m}^2\Phi^*\Phi + \lambda(\Phi^*\Phi)^2,
\label{ssbP}
\ee

where $\lambda>0$, so as not to make the potential unbounded from below and, hence, the
Lagrangian for this field can be written as 
\be
\CL = \partial_{\mu}\Phi^*\partial^{\mu}\Phi - {\rm m}^2\Phi^*\Phi - \lambda(\Phi^*\Phi)^2.
\label{ssbL}
\ee

\begin{center}
\begin{figure}[thb]
\hskip 0.8cm
\psfig{figure=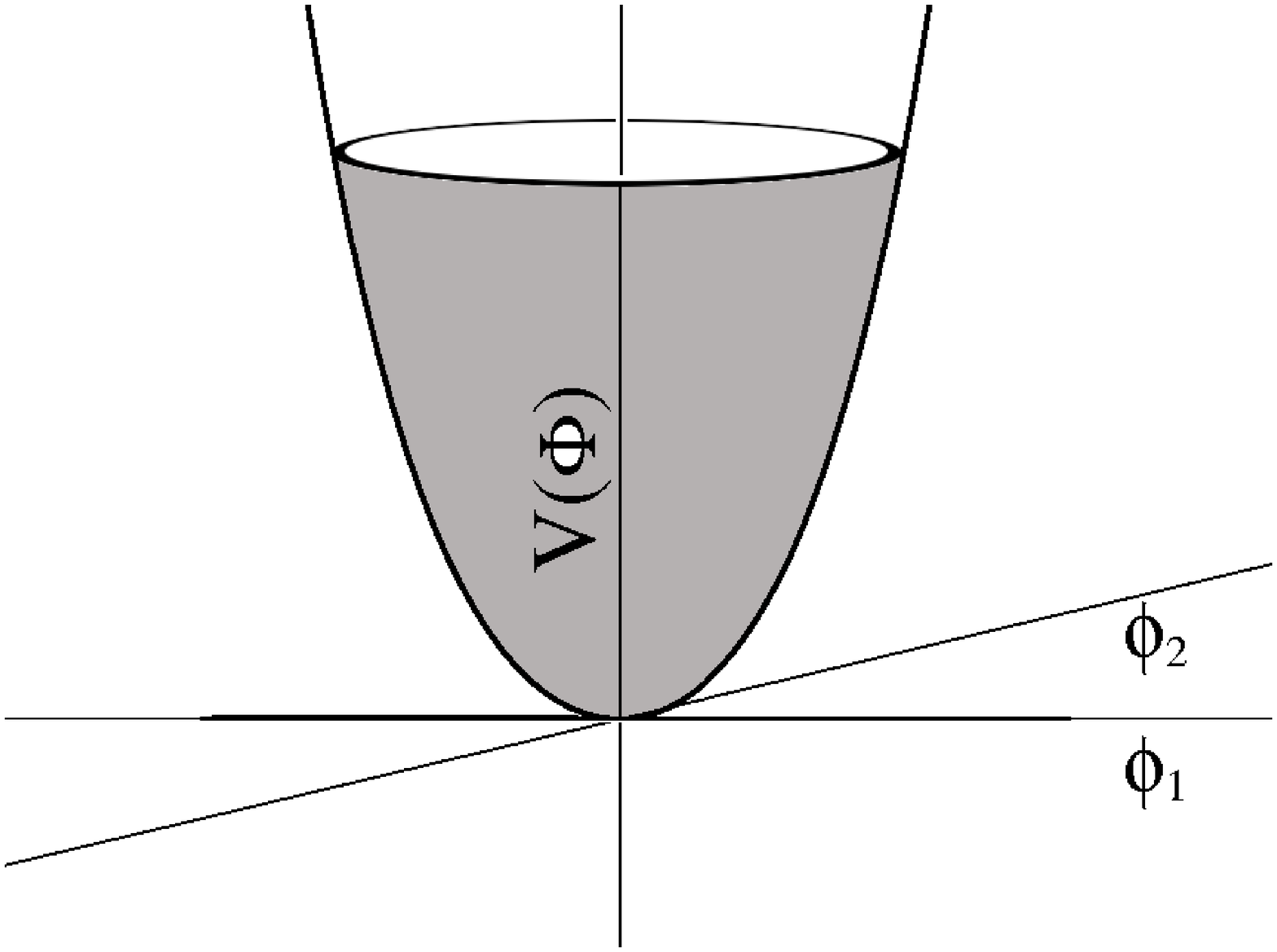,width=7.2cm,height=5.2cm,angle=0}
\vskip -5.23cm
\hskip 7.5cm
\hskip 1.50cm
\psfig{figure=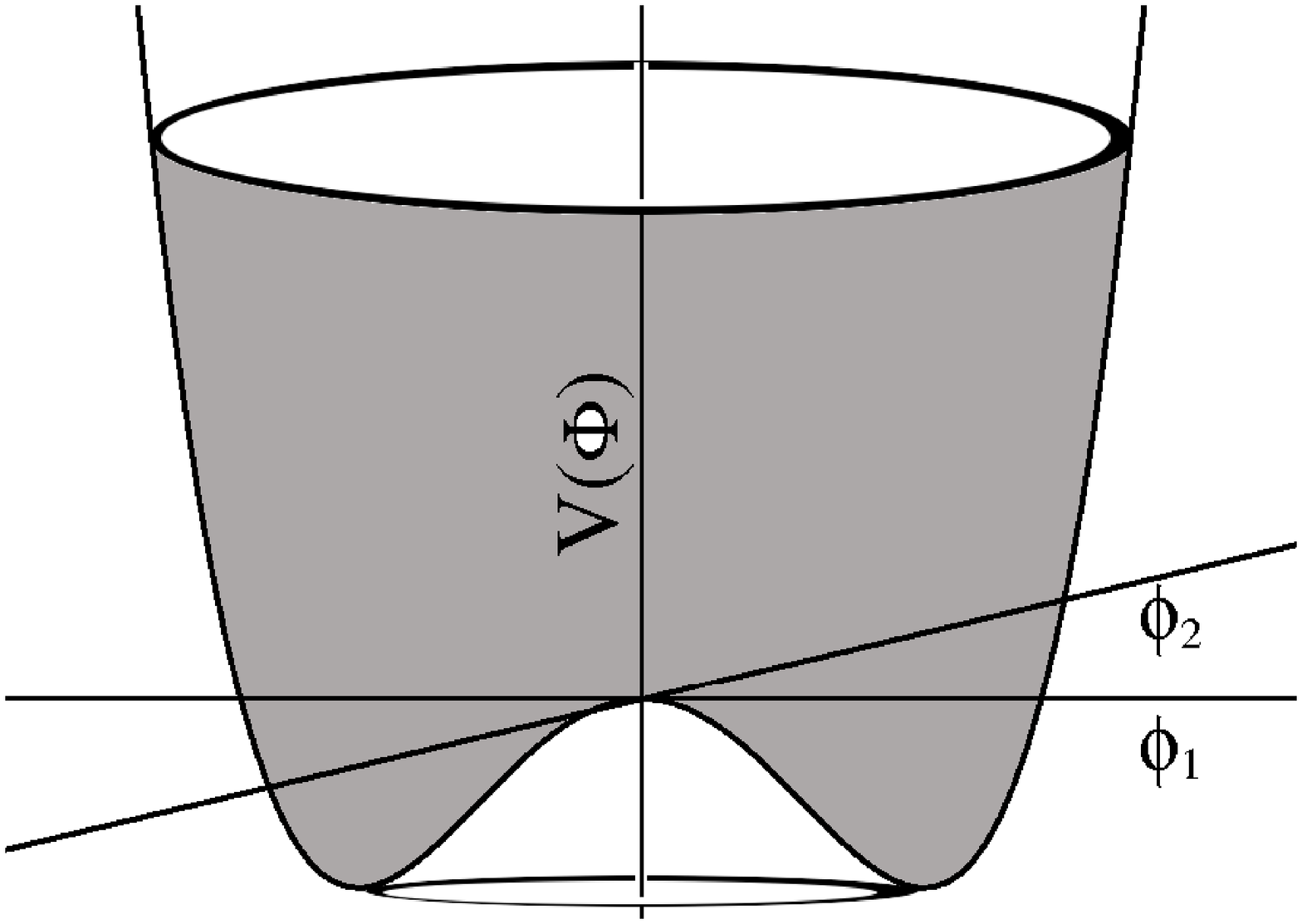,width=7.2cm,height=5.2cm,angle=-90}
\caption{\sf \small {The Higgs Potential}}
\label{f:poten}
\end{figure}
\end{center}

For the case ${\rm m}^2>0$, left figure of Fig. \ref{f:poten}, $`$m' will represent the mass of the scalar field $\Phi$, and the ground state of the potential 
will obviously be at $\Phi=0$. 

To discuss the case with ${\rm m}^2<0$, right figure of Fig. \ref{f:poten}, let us rewrite the 
Lagrangian as
\be
\CL = {1\over 2}(\partial_{\mu}\phi_1)^2+ {1\over 2}(\partial_{\mu}\phi_2)^2 
- {1\over 2}{\rm m}^2(\phi_1^2 + \phi_2^2) 
-{1\over 4} \lambda(\phi_1^2+\phi_2^2)^2.
\ee

In this case the minima will be situated at all 
$\phi_1$ and $\phi_2$'s satisfying the condition 
\be
(\phi_1^2+\phi_2^2)= v^2 = {-{\rm m}^2/\l},
\ee

with the vacuum expectation value (\emph{vev}) $v=\sqrt{-{\rm m}^2/\lambda}$.

The minimum of the potential, now, is not unique and also not at  
$\Phi=0$, as a result perturbation theory will not be applicable around that point. 
To pursue it one should shift the field $\Phi (x)$ to 
\be
\Phi(x)={1\over\sqrt 2}[v + \eta(x) + i\xi(x)]
\label{newphi}
\ee

at one of the minima of the potential.
Replacing the $\Phi(x)$ field by eqn.(\ref{newphi}) in the above Lagrangian of eqn.(\ref{ssbL}) 
we get 
\be
\CL =  {1\over 2}\partial_{\mu}\xi\partial^{\mu}\xi + {1\over 2}\partial_{\mu}\eta\partial^{\mu}\eta {- \lambda v^2\eta^2} - \lambda v \eta^3 + {\rm other~terms}.
\ee

Due to the above shift of the field $\Phi$, the third term of the modified Lagrangian is, now,
 the mass term of the field $\eta$ with the mass ${\rm m}_{\eta}=\sqrt{2\lambda v^2}=\sqrt{-2{\rm m}^2}$. 
The first term of the modified Lagrangian is the kinetic energy of the $\xi$ 
field. Note, there is no corresponding mass term for this field which implies
that the
theory contains a new massless scalar field. If physically there were any
such particle we should have detected it. Experimentally, we did
not observe any such particle; so does it mean that the spontaneous symmetry
breaking mechanism is incorrect? We can easily see from the right figure of Fig. \ref{f:poten} that the potential has a flat (circular) direction at the minimum implying the presence of a massless mode.  
It is a simple example of the Goldstone theorem \cite{halzen}, which states 
that massless scalars occur whenever a continuous symmetry of a physical system
is $``$spontaneously broken" (or, more accurately, is $``$not apparent in the ground
state"). 

\subsection{SSB for a local $U(1)$ gauge symmetry}
Let us now discuss the spontaneous symmetry breaking mechanism for a local $U(1)$ gauge 
symmetry which, basically, is known as the \emph{Higgs mechanism} \cite{higgsmech}. 
What we did in the previous 
subsection, sec.~\ref{gssb}, is that we had a Lagrangian with a negative (mass)$^2$ term and 
invariant under the global $U(1)$  transformation. Later, we shifted the scalar field to
 accommodate proper \emph{vev}. We need a Lagrangian, here, which will be invariant under the 
\emph{local U(1) gauge transformation}, eqn.(\ref{gtrnasform}). The Lagrangian of eqn.(\ref{ssbL}) 
is not invariant under this local gauge transformation. To make it invariant, as discussed
 in sec.~\ref{ginv}, we need a covariant derivative $D_\m$, defined in eqn.(\ref{covdev}), as
\bc
$\P _{\m} \ra D_{\m} \equiv \P _{\m} +ieA_\m $
\ec
instead of $\P _{\m}$ and a gauge field $A_\m$ which simultaneously has to be transformed as
\be
A_\m(x) \ra A_\m(x) - {1 \over e} \P _\m\a(x).
\label{gaugeT1}
\ee

For the ${\rm m}^2>0$ scenario, $`$m' will represent the mass of the scalar field $\Phi$, but we are
interested in SSB for which ${\rm m}^2<0$. The local $U(1)$ gauge invariant Lagrangian using eqn.(\ref{freedirac2}) and eqn.(\ref{ssbL}), thus, is given by 
\be
\CL = -{1\over 4}F_{\m\n}F^{\m\n} + (\partial_{\mu}-ieA_\m)\Phi^*(\partial^{\mu}+ieA^\m)\Phi - {\rm m}^2\Phi^*\Phi - \lambda(\Phi^*\Phi)^2.
\label{locUinvL}
\ee

For SSB, we need to transform the scalar field by eqn.(\ref{newphi}). The Lagrangian will, 
thus, be given by
\be
\CL = - {1\over 4}F_{\m\n}F^{\m\n} + {1\over 2}\partial_{\mu}\xi\partial^{\mu}\xi + {1\over 2}\partial _{\mu}\eta\partial ^{\mu}\eta - \lambda v^2\eta^2 + {1\over 2}e^2v^2A_ \m A^ \m - evA_\m \P ^\m\xi + {\rm other~terms}.
\label{locUinvL1}
\ee

Eqn.(\ref{locUinvL1}) describes the interaction of a masless boson $\xi$, a massive scalar 
field $\eta$ of mass $m_{\eta}=\sqrt{2\lambda v^2}$ and a massive gauge boson of mass $m_A=ev$.
So, with the help of the SSB mechanism for a local gauge symmetry we succeeded to generate a
mass for the gauge boson. The problem is with the unwanted massless scalar field $\xi$ as we 
have seen in sec.~\ref{gssb}. Actually, presence of the off-diagonal term $ evA_\m\P ^\m\xi$ 
in the transformed Lagrangian implies that the fields are not in the physical mass basis. We
have to reinterpret the particles described by the Lagrangian eqn.(\ref{locUinvL1}).

For this purpose, it is convenient to use an alternative but equivalent 
parameterisation of the shifted field $\Phi(x)$. Instead of eqn.(\ref{newphi})
 we write 
\be
\Phi(x)={1\over\sqrt 2}[v + h(x)]e^{-i\t (x)/v}
\label{newphi1}
\ee

Since the theory is invariant under local $U(1)$ gauge transformations, 
consider the following transformations, for the set of real fields $h,~A_\m$ and $\t$ as
\be
\Phi(x) \ra e^{i\t (x)/v} \Phi(x) ~~~~~~~{\rm and}~~~~~~~~~~
A_\m(x) \ra A_\m(x) - {1\over {ev}}\P _{\m}\t (x).
\label{bgt}
\ee

This gauge transformation with the condition that the theory will be independent of the 
field $\t (x)$, will help to keep this extra unwanted massless scalar 
field away. After this transformation, the unwanted field $\t (x)$ is removed
from the theory and using the transformation eqn.(\ref{bgt}) on eqn.(\ref{locUinvL}) we obtain the new transformed Lagrangian as 
\be
\CL = - {1\over 4}F_{\m\n}F^{\m\n} + {1\over 2}\partial_{\mu}h\partial^{\mu}h - \lambda v^2h^2 +{1\over 2}e^2v^2A_\m A^\m  +{1\over 2}e^2vA_\m A^\m h + {\rm other~terms}.
\ee

So as a result of this gauge transformation the unwanted massless scalar field has
been absorbed as the longitudinal component of the gauge field $A_\m$. Finally our theory
is, now, described by a massive Higgs scalar field $h$ with mass ${\rm m}_h = \sqrt{2\l v^2}$ and 
the massive gauge boson $A_\m$ with mass ${\rm m}_A = ev$.

\subsection{Higgs mechanism in SM}{\label{higgsSM}}

Let us now come to a more realistic case where, in the SM, the local $SU(2)_L\times U(1)_Y$
 gauge symmetry will spontaneously be broken by   
the Higgs field, as

\bc
$SU(2)_L\times U(1)_Y \longrightarrow U(1)_Q$.
\ec

Let us consider the Higgs field, as introduced by Weinberg and defined in eqn.(\ref{hig1}) 
and eqn.(\ref{hig2}), as 
\bc
$\Phi \equiv \left(\br{c} \phi^+ \\ \phi^0 \er\right) \equiv (2,1) \subset (SU(2)_L\times U(1)_Y)$,
\ec
where, 
\bc
$\phi^+ = {1\over{\sqrt 2}}(\phi _1 + i\phi _2) ~~~~~ {\rm and}~~~~~ \phi^0 = {1\over{\sqrt 2}}(\phi _3 + i\phi _4).$
\ec
Let us consider the Lagrangian for the field $\Phi$ as
\be
\CL = \partial_{\mu}\Phi^{\dagger}\partial^{\mu}\Phi - {\rm m}^2\Phi^{\dagger}\Phi - \lambda(\Phi^{\dagger}\Phi)^2.
\label{higspot}
\ee

The Lagrangian, eqn.(\ref{higspot}), is invariant under the global $SU(2)_L\times U(1)_Y$ gauge transformation 
but not under the local $SU(2)_L\times U(1)_Y$ gauge transformation 
\be
\Phi(x) \ra e^{i({\vec{\a}(x).\vec{\tau}} + \b (x){Y\over 2})}\Phi(x),
\ee

where, ${\bf \tau _a}$ (a=1,2,3), are the $SU(2)_L$ generators and $Y$ the hypercharge as defined in 
eqn.(\ref{hypercharge}). 

To make it invariant under the local $SU(2)_L\times U(1)_Y$ gauge transformation we require, as before, to replace 
\be
\P _\m \ra \partial_{\mu} + i g_2 {\vec{\tau}\over 2} .\vec{W}_{\mu} + i g_1{Y\over 2}B_{\mu}.
\label{covdevsm}
\ee

 In eqn.(\ref{covdevsm}) $g_1$ and $g_2$
are the $U(1)_Y$ and $SU(2)_L$ gauge coupling constants respectively. In this case, the gauge field $B_\m$ 
transforms, as in eqn.(\ref{gaugeT}) 
\be
B_\m \ra B_ \m - {1\over {g_1}}\P _\m \b (x).
\ee

The $SU(2)_L$ gauge boson ${\vec W_\m}$ transforms, due to the non-abelian character, as
\be
\vec{W}_\m \ra \vec{W}_\m - {1\over {g_2}}\P _\m \vec{\alpha}(x) - \vec{\alpha}\times \vec{W}_\m.
\ee

This indicates that the rotation of the weak gauge boson, $\vec{W}_\m$, will be affected due to two factors,
one, due to the vector nature of the field and another due the variation of the space-time point.

The local $SU(2)_L\times U(1)_Y$ gauge invariant Lagrangian, thus, can be written as 
\ba
\CL = \left[\left(i\partial_{\mu} - g_2{\vec{\tau}\over 2}.\vec{W}_\m - {g_1\over 2}B_{\mu} \right)\Phi\right]^{\dagger}\left[\left(i\partial^{\mu} - g_2{\vec{\tau} \over 2}.{\vec{W}^{\mu}} - {g_1\over 2}B^{\mu} \right)\Phi\right]\nonumber \\
-V(\Phi^{\dagger}\Phi)
- {1\over 4}{B_{\mu\nu}B^{\mu\nu}}-{1\over 4}{\vec{W}_{\mu\nu} \vec{W}^{\mu\nu}},
\ea

where $Y=1/2$ is used for the Higgs scalar field.

The scalar potential, $V(\Phi^{\dagger}\Phi)$, is given by
\be
V(\Phi^{\dagger}\Phi)= {\rm m}^2(\Phi^{\dagger}\Phi) +\lambda(\Phi^{\dagger}\Phi)^2. 
\ee 

The tensors $B_{\mu\nu}$ and $\vec{W}_{\mu\nu}$ are defined as
\be
B_{\mu\nu} =  \partial_{\m} B _\n - \partial_{\n} B _\m
\ee

and 
\be
\vec{W}_{\mu\nu} =  \partial_{\m}\vec{W}_\n - \partial_{\n}\vec{W}_\m - ig_2 \vec{W}_\m \times 
\vec{W}_\n.
\ee

The condition for the spontaneous symmetry breaking is ${\rm m}^2<0$ and $\lambda>0$. 
The minima of the potential are at all those points of $\phi_i$s which satisfy the following 
condition 
\be
\Phi^{\dagger}\Phi={1\over 2}(\phi_1^2+\phi_2^2+\phi_3^2+\phi_4^2) = {v^2\over 2}={-{\rm m}^2\over{2\lambda}},
\ee

which implies an infinite number of ground states. The symmetry will spontaneously break once 
one of it is arbitrarily chosen. 
Keeping in mind that any unphysical term in the Lagrangian should not be allowed,
let us write the scalar field $\Phi$ in terms of four fields $\theta_1(x)$, $\theta_2(x)$, $\theta_3(x)$  and $h(x)$ as: 

\ba
\Phi(x) ={1\over{\sqrt{2}}} \left( \br{c} \theta_2 - i \theta_1 \\ 
( v + h)  - i \theta_3 \er \right) \simeq 
e^{i {\bf\theta_a (x) \tau^a/}{\it v}} \,  \left( \br{c} 0 \\ 
\frac{1}{\sqrt{2}} (v + h(x) \, )  \er \right) 
\ea

Once we put this transformed field $\Phi$ in the Lagrangian, we will get a massive Higgs 
field $h$ while the three massless unwanted 
bosons will disappear from the potential. By an appropriate local gauge transformation - a 
generalisation of eqn.(\ref{bgt}) - they may be removed from the theory -- effectively 
absorbed by the $\vec{W}$ bosons as longitudinal components. 

\subsection{Gauge boson and fermion masses}\label{gfmass}
 In order to see how the spontaneous breaking of the $SU(2)_L\times U(1)_Y$ gauge symmetry 
produces massive $W^\pm$ and $Z$ boson while leaving the photon field, 
$\gamma$, massless let us expand the relevant part of the Lagrangian 
explicitly: 

\ba
\bigg| \bigg( -i g_2 {\vec{\tau} \over 2}.\vec{W}_\mu 
- i g_1 {1\over 2} B_\mu \bigg)\Phi  \bigg|^2  
&=& {1\over 8} \left| \left( \br{cc} g_2 W_\mu^3 + g_1 B_\mu & g_2(W_\mu^1 -iW^2_\mu) \\ g_2(W_\mu^1 + iW^2_\mu) & -g_2 W_\mu^3  
+ g_1 B_\mu \er \right) \left( \br{c} 
0 \\ v  \er \right) \right|^2 \nonumber \\
&=& {1\over 8}g_2^2v^2[(W_\mu^1)^2+(W_\mu^2)^2]
+ {1\over 8}v^2 [g_2 W_\mu^3- g_1 B_\mu]^2. 
\ea

Let us define the new fields $W^\pm_\mu$ and $Z_\mu$ and it's orthogonal 
partner $A_\mu$ as
\ba
W^\pm = {1\over \sqrt{2}} (W^1_\mu \mp i W^2_\mu) \  , \ 
Z_\mu = \frac{g_2 W^3_\mu- g_1 B_\mu}{\sqrt{g_2^2+g_1^2}} \ , \ 
A_\mu = \frac{g_1 W^3_\mu+ g_2 B_\mu}{\sqrt{g_2^2+g_1^2}} \  
\label{gauge-fields}
\ea

to arrive to the form
\ba
{\rm m}_W^2 W^+_\mu W^{-\mu} + \frac{1}{2} {\rm m}_Z^2 Z_\mu Z^\mu + \frac{1}{2} {\rm m}_A^2 
A_\mu A^\mu. 
\ea

Finally we have three massive gauge fields $W^\pm$ and $Z$ and one
massless, the photon field, as needed: 
\ba
{\rm m}_W =\frac{1}{2}vg_2 \  , \ {\rm m}_Z= \frac{1}{2} v  \sqrt{g^2_2+g_1^2} \ , 
 \ {\rm m}_A=0 . 
\ea

It is useful to introduce the electroweak mixing angle $\t_W$ defined in terms of the 
gauge coupling constants $g_1$ and $g_2$ as
\be
{\rm sin}\t_W = {g_1\over{\sqrt{g_1^2+g_2^2}}}.
\label{Wangle}
\ee

It is worthwhile to define a few quantities at this point in terms of the mixing angle $\t_W$.
The charged current interactions are 
\be
{1\over 4}\CL _{CC} = {G_F\over \sqrt 2}J^+_\m{J^-}^\m,~~~~~ {\rm with}~ {G_F\over \sqrt2}= {g_2^2 \over {8{\rm m}_W^2}}
\ee

and the neutral current interactions are given by 

\be
{1\over 4} \CL _{NC} = {G_F^{NC}\over \sqrt 2}J^0_\m{J^\m}^0,~~~~~{\rm with}~ G_F^{NC} = {{g_1^2+ g_2^2}\over {8{\rm m}_Z^2}}.
\ee

An important parameter is the ratio of neutral and charged current interaction strengths, 
which equals to 1 in the standard model, expressed as 

\be
\r = {G_F^{NC}\over{G_F}} = {{{\rm m}^2_W}\over{{\rm m}^2_Z {\rm cos}^2\t_W}}.
\ee

Let us go back to the problem of gauge invariance for the fermion mass. As the $SU(3)_C$ 
gauge group is chirally blind, without indicating left or right subscripts let us denote
the quark as $Q$ and lepton as $L$. Quark is in the triplet $({\bf 3})$ and lepton is singlet
$({\bf 1})$ in the $SU(3)_C$ group representation. Then the antiquark, $\bar Q$, will be in the 
anti-triplet $({\bf \bar 3})$ representation. In $SU(3)$ we have,
\bc
${\bf 3 \otimes \bar 3 \equiv 1 \oplus 8}$.
\ec 
Hence we see that the mass term $\bar Q Q$ is invariant under the $SU(3)_C$ gauge transformation.
For the \emph{colour singlet} leptons it is an obvious one. 

It is quite different for the case of the $SU(2)_L$ gauge group. In this case the left and right
chiral fermions transform differently as pointed in Table \ref{particle}. A Dirac fermion field
can be decomposed as 
\ba
\S & = & \left({{1-\g _5}\over 2}\right) \S + \left({{1+\g _5}\over 2}\right)\S \nonumber \\
   & = & P_L\S + P_R\S ~~=~~ \psi _L + \psi _R,	 
\label{chiralf}
\ea

where $\psi _ L$ and $\psi _ R$ are respectively known as left-chiral and 
right-chiral fermions.

In the Weyl representation a Dirac fermion field can be written as
\ba
\S = \left(\br{c}\psi_L \\ \psi_R \er \right).
\ea

Using eqn.(\ref{chiralf}) the fermion mass term, thus, can be written as
\be 
{\rm m}\bar{\S}\S \equiv {\rm m}(\bar{\psi}_L\psi_R+ \bar{\psi}_R\psi_L).
\ee

Now, from Table \ref{particle} we see that left-handed fermions are doublet while the 
right-handed are singlet under the $SU(2)_L$ gauge transformation. So neither the term 
$\bar{\psi}_L\psi_R$ nor $\bar{\psi}_R\psi_L$ is invariant, and hence neither is $\bar{\S}\S$.
Thus we see that the fermion mass term ${\rm m}\bar{\S}\S$ is invariant under the 
$SU(3)_C$ gauge transformation but not under $SU(2)_L$. 

This problem can be cured with the help of the Higgs scalar multiplet. Using the Higgs doublet we can 
write an Yukawa interaction term 
\be
y\bar{\S}\S\Phi + h.c.
\ee 
where $y$ is the Yukawa coupling. To be more precise, for the first generation lepton sector,
 we can write,
\ba
L_{\rm electron} = -y_e (\bar{\nu}_e~~\bar{e})_L\left(\br{c} \phi^+ \\ \phi^0 \er\right) e_R + h.c.
\ea
Once we replace the Higgs field, due to spontaneous symmetry breaking, by
\ba
\Phi(x) =\left( \br{c} 0 \\ 
\frac{1}{\sqrt{2}} (v + h(x) \, )  \er \right) 
\ea
we will have a mass term in the Lagrangian as
\ba
L_{\rm electron} = -{\rm m}_e(\bar{e}_Le_R+\bar{e}_Re_L) ~~~{\rm with,}~~~ {\rm m}_e={y_ev\over {\sqrt 2}}.
\label{electronmass}
\ea

To generalize for all the matter fields we can write the Yukawa interaction terms, using
the notation used in Table \ref{particle}, as 
\be
\CL = - Y^u_{ij}\bar{Q}_{L_i}{q^u_{R_j}}\tilde{\Phi} - Y^d_{ij}\bar{Q}_{L_i}{q^d_{R_j}}{\Phi} - Y^l_{ij}\bar{L}_{L_i}{l_{R_j}}\Phi + h.c 
\ee

where, $\tilde{\Phi}=-i\sigma_2\Phi^*$, $Y^u$,  $Y^d$, $Y^l$ are the up-quark, down-quark and
charged lepton Yuakwa coupling constant matrices respectively. 
One point to be noted is that the particle content, listed in the Table \ref{particle},
in the SM does not contain any right-handed neutrino. It was conspired just to explain 
the then accepted zero mass of the neutrino. Once, the Higgs field gets a $vev$, $v$, then the
Lagrangian takes the form $\overline{f_L}m_f f_R$ with the mass matrices 
\be
{({\rm m}_u)}_{ij} =  Y^u_{ij}v, {({\rm m}_d)}_{ij} =  Y^d_{ij}v, {({\rm m}_l)}_{ij} =  Y^l_{ij}v,
\ee
where, $f = (u,c,t)^T~{\rm or}~ (d,s,b)^T ~{\rm or}~ (e,\m,\tau)^T$ represent the three 
generations of
fermion fields. These mass matrices are in the flavour basis, not the mass basis. 

\section{Shortcomings of the standard model}
Some unattractive features of the standard model have been noted in sec.~\ref{introintro}. 
Besides the non-zero neutrino mass, there are several conceptual shortcomings 
of the standard model. The standard model contains 19 parameters - three gauge couplings $g_1$, $g_2$, $g_3$,
six quarks and three charged-leptons masses, one CP-violating phase, three 
CKM mixing angles, the quadratic and quartic coupling constants for the Higgs scalar potential and strong CP parameter.
The standard model does not say anything about the fourth force, namely the gravitational interaction. 
At the scale of Planck mass, $M_P$, we need a theory of quantum gravitation which will also 
describe the dynamics of particles governed by the gravitational interaction in addition with other
forces. Although, to reduce the large 
number of parameters one can extend the theory of unification, but the point one may ask is can 
it predict different observables like mixing angles, fermion masses etc properly? What is
the origin of three generations of fermions? 

In order to remove these shortcomings, physicists have come up with 
different new options like Grand Unified Theory (GUT), Supersymmetry, 
Extra Dimensions etc. to extend the standard model. A brief introduction to 
these are given below.

\section{Grand unified theory}
A general aesthetic of physics is that the more 
symmetrical a theory is, the more $``$beautiful" and $``$elegant" it is. 
In this view, the Standard Model gauge group, which is the direct 
product of three groups, is not a truly satisfactory one. 
In analogy with the 19th-century unification of electricity 
with magnetism into electromagnetism, and especially the success of the 
electroweak theory, which utilizes the idea of spontaneous symmetry breaking 
as discussed in previous sections, to unify electromagnetism with the weak 
interaction, it is natural to attempt to unify all three 
groups in a similar manner. Three independent gauge 
coupling constants and a huge number of Yukawa coupling coefficients require 
far too many parameters, and it would be elegant if these coupling 
constants could be 
explained by a theory with fewer parameters. A gauge theory based on  
a simple group has only one gauge coupling constant, and since 
the fermions are now grouped together in larger representations, there are 
fewer Yukawa coupling coefficients as well. In order to get unification \cite{rabiM} 
of all three interactions we need a bigger group which will contain the standard
 model gauge group $SU(3)_C\times SU(2)_L\times U(1)_Y$ and, in addition, this 
gauge group has to break down to the standard model gauge group at some higher scale
in such a way that it will predict different mass and mixing angles at the low energy
scale. 





\subsection{$SU(5)$ GUT model} \label{su5gut}
The first attempts of grand unification were made by Pati 
and Salam \cite{Pati} to unify quarks and leptons within  
$SU(2)_L\times SU(2)_R \times SU(4)_C$, known as the Pati-Salam gauge group $({\cal
G}_{PS})$.  
In this scenario lepton is treated as the fourth component of 
the colour quantum number of the $SU(4)_C$ gauge group.
Another approach independently proposed by Georgi and Glashow in 1973 \cite{Georgi} considers 
$SU(5)$, which is of rank 4 (same as the SM gauge group), as the unified group. It has a few advantages; like, it gives a beautiful way of unifying all the three standard model gauge couplings.
In this $SU(5)$ GUT model there is a unique way to accommodate all the fifteen quarks 
and leptons in the $\bar 5$ and $10$ representations. The break up of these two multiplets
of the $SU(5)$ group in terms of the SM gauge group $SU(3)_C\times SU(2)_L \times U(1)_Y$ are:
\be
5 \equiv (3,1,-{1\over 3}) \oplus (1,2,{1\over 2}) ~~~{\rm and}~~~10 \equiv (1,1,1)\oplus 
(\bar 3,1,-{2\over 3}) \oplus (3,2,{1\over 6}) .
\label{5plet}
\ee

The right-handed down quark $d\equiv(d^r,d^g,d^b)$ and right-handed $(e^+,\tilde{\n_e})$ doublet
 can preferably be put into the $\bar{5}$ 
representation respectively. On the other hand the singlet charged left-handed anti-lepton $e^+$, the 
left-handed $u,d$ quark doublet and left-handed anti-u quark singlet $u^c$ will be in $10$, the 
antisymmetric part of the product of two 5 plets. 

Similarly, $24 (= 5^2 - 1)$ gauge bosons associated with the $SU(5)$ gauge group can be decomposed
as follows:
\be
24 \equiv (8,1,0) + (1,3,0) + (1,1,0) + (3,2,-{5\over 6}) + (\bar 3,2,{5\over 6})
\label{24plet}
\ee

which are the gluons, electro-weak gauge bosons and the new heavy X,Y gauge bosons. These  
new gauge bosons, X and Y, mediate the proton decay. One can have, for example, for the decay
mode,
\be
M(p\ra e^+\pi^0) \sim {g^2\over {{\rm m}^2_{X}}}, 
\ee
where g is the GUT gauge coupling constant. Hence, the proton lifetime is
\be
\tau _p \sim {{{\rm m}^4_{X}}\over{g^4{{\rm m}_p^5}}}.
\ee

Non-observation of proton decay puts a lower limit on these 
heavy gauge boson masses
\be
{\rm m}_{X,Y}~>~ 10^{15}~{\rm GeV} 
\ee

We have the normalisation of the generators of the GUT gauge group as 
\be
{\rm Tr}(t^a t^b) = N\d^{ab}
\label{normgen}
\ee
where, N is the normalisation constant. Invariance under the gauge group G 
means all 
observed gauge couplings are the same as that of the unified gauge group.
Unlike the $SU(2)_L$ and $SU(3)_C$ couplings, eqn. (\ref{normgen}) does not
fix the scale of the $U(1)_Y$ coupling constant. It would not change the physics if we divide it by a 
constant factor $c$ and simultaneously multiply the hypercharge $Y$ by the same factor. 

As stated above, considering ${{\tau ^{a}}\over{2}}$ as the $SU(2)$ generators we choose, 
\ba
{\rm Tr}\big({\tau^a\over 2}{\tau^b\over 2}\big) = {1\over 2}\d ^{ab}.
\label{su2gen}
\ea

So, let us assume that the $Y=c({y\over 2})$ is the generator of the unified gauge group, 
in addition with the unchanged $SU(2)_L$ and $SU(3)_C$ generators. Now, in the unified scenario
all the generators have common normalisation factor as a result we have,
\be 
{\rm Tr} (c^2 (y/2)^2) = {\rm Tr} (T_3)^2 
\ee
with $T_3$, the third $SU(2)_L$ generator. The trace is over all particle states in the 
representation. In the SM framework for one generation these are u, d, $\n_e$ and $e^-$. Thus for the above relation we have,
\be 
3(\frac{1}{4} + \frac{1}{4}) + \frac{1}{4} + \frac{1}{4} =
c^2 \big(3({1\over 6})^2 + 3({1\over 6})^2 + 3({2\over 3})^2 + 3(-{1\over 3})^2 + (-{1\over 2})^2 + (-1)^2 +  (-{1\over 2})^2 \big) . 
\ee
This implies, 
\be 
c=\sqrt{3\over 5},
\ee
and, hence the properly normalized generator is 
\be 
Y=\sqrt{3\over5} {y\over 2}.
\ee 

Thus, for the standard model scenario, we have 
\be
 g_{SU(5)} = g_1\sqrt{5\over3} ~~~~ {\rm and}~~~~~ g_{SU(5)} = g_2,
\ee
and (eqn.(\ref{Wangle})) leads to ${\rm tan \t_W} = \sqrt{3\over 5}$, hence,  ${\rm sin^2 \t_W} = {3\over 8}$.

Generally, the $SU(5)$ symmetry is broken down to the low energy 
$SU(3)_C\times U(1)_Q$ by two Higgs scalars $\Phi_{24}$ and $H_{5}$ which are in 
the adjoint 24 and 5 of $SU(5)$.
The breakdown of these two Higgs multiplets in the $SU(3)_C\times SU(2)_L \times U(1)_Y$ representation are given in eqns. (\ref{24plet}) and (\ref{5plet}) respectively.

When the neutral component $(1,1,0)$ of the the $\Phi _{24}$ gets a vev at the GUT scale, 
$SU(5)$ breaks to the SM gauge group while getting a nonzero vev for $H_5$ at the electro-weak 
scale breaks the SM down to $SU(3)_C\times U(1)_Q$.

 The stepwise breakdown of the gauge symmetry in this case, thus, is
\be
{SU(5)}~{\Phi_{24}\over (M_U)}{\hspace*{-3mm}\ra}~ {SU(3)_C \times SU(2)_L \times U(1)_Y}~{H_{5}\over (M_Z)}{\hspace*{-3mm}\ra}~ { SU(3)_C \times U(1)_Q}. 
\ee

\subsubsection{Gauge hierarchy problem}\label{gaugehier}
A major difficulty of the standard model is the gauge hierarchy problem \cite{hierarchy}.
In order to realise this hierarchy between $M_U$ and $M_Z$ and hence the problem of naturalness 
let us calculate
the quadratic divergence for the Higgs mass due to standard model fermions.  

\begin{center}
\begin{figure}[thb]
\hspace*{5cm}\psfig{figure=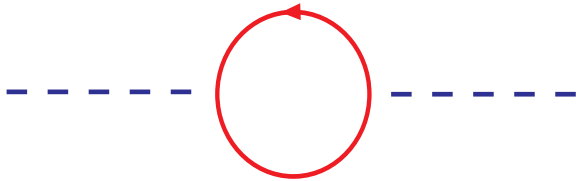,width=8.0cm,height=2.5cm,angle=0}
\caption{\sf \small {One loop fermionic correction to the Higgs mass}}
\label{f:unnatural}
\end{figure}
\end{center}

The one loop correction to the Higgs mass ${\rm m}_H$ is obtained by calculating the
two point function:
\ba
\Pi^f_{hh} & = &(-1)\int {d^4k\over(2\pi)^4}{\rm Tr}\left\{({-i\l_f\over\sqrt 2}){i\over{\not k -{\rm m}_f}}({-i\l_f\over\sqrt 2}){i\over{\not k -{\rm m}_f}}\right\}, 
\ea

where $\l _f$ is the fermion-scalar-fermion coupling constant. The loop 
momentum $k$ can take any value from zero to infinity. This leads to a 
correction which is infinite and makes the theory ill-defined. So, we assume that
our theory is valid upto a cut-off scale $\L$. The above integration,
thus, becomes
\ba
\Pi^f_{hh} & = & -2\l_f^2\int_0^{\L} {d^4k\over{(2\pi)^4}}\left[{1\over{k^2-{\rm m}_f^2}}+{2{\rm m}_f^2\over{(k^2-{\rm m}_f^2)^2}}\right] \nonumber \\
& = & { -{\lambda_f^2\over{8\pi ^2}}\Lambda ^2} + ... 
\ea

Thus the corrected Higgs $($mass${\rm )}^2$ is 
\be
{\rm m}_H^2 = {\rm m}_{H_0}^2 + \d {\rm m}_H^2
\ee
where the correction $\d {\rm m}_H^2$ is proportional to the $\Pi^f_{hh}$. In GUT we 
have a new scale at $10^{16}~~{\rm GeV}$. If there is no new physics before this scale 
then $\L \sim 10^{16}~~{\rm GeV}$ and to have a Higgs mass 
of $\CO(100~{\rm GeV})$ a fine-tuning of the co-efficient $\l_f$ to 1 part in $10^{26}$
is needed.




\subsection{$SO(10)$ GUT model}\label{so10gutM}

$SO(10)$ is a possible useful GUT gauge group for the unification 
of the SM \cite{so10}. It is a group of rank 5, unlike the SM gauge group which is
of rank 4. As the rank is the maximum number of diagonal generators of the group, 
so the extra diagonal generator of this unified gauge group will define another quantum 
number, which can be identified as $(B-L)$ for the left-right symmetric version of this 
theory. Due to the presence of an extra 
diagonal generator, $SO(10)$ can be broken to the standard model in various ways.

One good feature for the group is that in a single multiplet $16$ it can accommodate all the 
standard model fermions in addition with a standard model gauge singlet right-handed neutrino, 
needed to explain the tiny neutrino mass. This group can support 
 left-right symmetry, represented by the gauge group $SU(3)_C \times
SU(2)_L \times SU(2)_R \times U(1)_{(B-L)} \equiv {\cal G}_{LR}$. We consider this class of 
$SO(10)$ models below.

There are two broad classes of minimal $SO(10)$ models: those
with only doublet Higgs scalars (Model I) and the conventional
left-right symmetric model including triplet Higgs scalars (Model
II). In both versions, a bi-doublet Higgs scalar $\Phi~ \equiv (1,2,2,0)$
 under ${\cal G}_{LR}$, gives mass to the charged fermions and
also a Dirac mass to the neutrinos\footnote{Note that in sec.~\ref{gfmass}, in case of SM, 
the charged fermions acquired masses due to the Higgs mechanism eqn.(\ref{electronmass}) but 
the neutrino
was massless as there was no right-handed neutrino. Presence of a right-handed neutrino
in the left-right symmetric model changes the perspective.}.
In an $SO(10)$ GUT, this bi-doublet $\Phi$ belongs to the
representation ${\bf 10, 120}$ or ${\bf 126}$. Usually a ${\bf
10}$ representation is chosen. However, for correct
fermion mass relations \cite{gj}, a ${\bf \overline {126}}$ representation
containing the field $\Phi' \equiv \{15,2,2\}$ under the group
$SU(4)_C \times SU(2)_L \times SU(2)_R \equiv {\cal G}_{PS}$
is often also included. 

The main differences between Models I and II lie in the Higgs
sector and the generation of neutrino masses. Lepton number violation in these models arises
from the Higgs scalars that break the $B-L$ symmetry and hence
the left-right symmetry. In Model I, the left-right symmetric group 
${\cal G}_{LR}$ is broken by an $SU(2)_R$ doublet Higgs scalar $\chi_R
\equiv (1,1,2,-1)$ when its neutral component acquires a $vev$
$\langle \chi^\circ_R \rangle \sim v_R$.  Left-right parity
implies the presence of an $SU(2)_L$ doublet Higgs scalar $\chi_L
\equiv (1,2,1,-1)$. The $vev$ of the neutral component of this
field,  $\langle \chi^\circ_L \rangle \sim v_L$, in addition to
$\langle \Phi \rangle$, breaks the electroweak symmetry.  

In Model II, an $SU(2)_R$ triplet Higgs scalar $\bar {\Delta}_R
\equiv (1,1,3, 2)$ breaks the left-right symmetric group ${\cal
G}_{LR}$. When the neutral component acquires a $vev$,
$\langle{\bar{\Delta}}^\circ_R \rangle \sim v_R$, it gives
Majorana masses to the right-handed neutrinos breaking lepton
number by two units. When the bi-doublet Higgs scalar $\Phi$
breaks the electroweak symmetry, this leads to the small see-saw
neutrino mass \cite{see-saw}.  Due to left-right parity, there is also an
$SU(2)_L$ triplet Higgs scalar $\bar {\Delta}_L \equiv (1,3,1,
2)$. Although these scalars have a mass at the parity
breaking scale $M_R$, the $vev$ of the neutral component of this
field is extremely tiny and can give small Majorana masses to
the left-handed neutrinos leading to a new type of see-saw mechanism.

Models I and II have the same symmetry breaking chain:
\begin{eqnarray}
SO(10) & \stackrel{210 ~(M_U)}{\longrightarrow} &  SU(3)_C \times
SU(2)_L \times SU(2)_R \times U(1)_{B-L}  \nonumber \\
&\stackrel{16 ~{\rm or} ~126 ~(M_R)}{\longrightarrow}&SU(3)_C
\times SU(2)_L \times U(1)_Y  \nonumber \\ &\stackrel{10
~(M_Z)}{\longrightarrow}&SU(3)_C \times U(1)_Q \nonumber
\end{eqnarray}

At the GUT scale, the symmetry is broken by the vacuum
expectation value  of a ${\bf 210}$-dimensional representation of
$SO(10)$.  The ${\bf 210}$ has a singlet under the  subgroup ${\cal
G}_{PS}$, i.e., $\{1,1,1\}$, which is odd under parity. When this
field acquires a $vev$, $SO(10)$ is broken to  ${\cal G}_{PS}$
and D-parity is also spontaneously broken (i.e., $g_{2L} \neq g_{2R}$).
To keep D-parity intact at this level we have to look elsewhere.
The $SO(10)$ {\bf 210} also contains a \{15,1,1\} under  ${\cal
G}_{PS}$ which is D-parity even. This is the field to which the
$vev$ must be ascribed to get the desired symmetry breaking to
${\cal G}_{LR}$ while keeping D-parity intact.

The left-right symmetry, $\lr$, is broken by the $vev$ of the
fields $F + \bar F$, where $F$ is a ${\bf {16}(\equiv\bf {\Gamma})}$-dimensional
representation for Model I and a ${\bf 126}$-dimensional
representation for Model II. Finally, the electroweak symmetry
breaking takes place by the $vev$ of a {\bf 10}-plet of $SO(10)$.
In the minimal models under consideration, there are no other
Higgs representations.

The breakdown \cite{Slansky:1981yr} of the 16 multiplet of $SO(10)$ under the 
$SU(4)_C\times SU(2)_L\times SU(2)_R$ is 
\be
16 \equiv (4,2,1)\oplus(\bar{4},1,2).
\ee

This 16-multiplet, thus, in addition with the left-handed particles also contains the left-handed 
anti-particles (equivalently the right-handed particles). 

A 16-dimensional multiplet can be written as
\ba
\S = \left(\br{c}\psi_0 \\ \psi_j \\ \psi_{jk} \er \right)
\ea
where, in the language of $SU(5)$ representations

\be
\S (16) \equiv \psi_0(1) \oplus \psi_{j}(\bar{5}) \oplus \psi_{ij}(10)  
\ee
with, 
\ba
\psi_0 = N^c,~~\psi_j\equiv L+d^c \equiv \left(\br{c} d^c_1 \\ d^c_2 \\ d^c_3 \\ e^- \\  {\n}_e \er \right),~~
\psi_{ij} \equiv (Q+U^c+E^c)\equiv \left(\br{ccccc} 0 & u^c_3 & -u^c_2 & u_1  &d_1\\ 
                                                      & 0     &  u^c_1 & u_2  &d_2\\ 
                                                      &       &  0     & u_3  &d_3\\
                                                      &       &        & 0    &e^+\er \right).
\ea
 
Let us consider the following symmetry breaking chain
\begin{eqnarray}
SO(10) & \stackrel{210 ~(M_U)}{\longrightarrow} &  SU(3)_C \times
SU(2)_L \times SU(2)_R \times U(1)_{B-L}  \nonumber \\
&\stackrel{16 ~{\rm or} ~126 ~(M_R)}{\longrightarrow}&SU(3)_C
\times SU(2)_L \times U(1)_Y  \nonumber \\ &\stackrel{10
~(M_Z)}{\longrightarrow}&SU(3)_C \times U(1)_Q \nonumber
\end{eqnarray}

An obvious question comes in our mind is what is the normalisation factor for the $B-L$ 
quantum number, c.f. the case of the hypercharge quantum number in sec.~\ref{su5gut}. 
Let us define different quantum numbers for all the members of a 16-dimensional multiplet 
in the $SU(2)_L\times SU(2)_R\times U(1)_{B-L}\times SU(3)_C$ group notation as:
\ba
Q_L \equiv \left(\br{c} u \\ d \er\right)_L \equiv (2,1,{1\over 3},3) \subset {\bf 16}, ~~~~~  
\psi_L \equiv \left(\br{c} \nu \\ e \er\right)_L \equiv (2,1,-1,1) \subset {\bf 16}, \nonumber\\
Q_R \equiv \left(\br{c} u \\ d \er\right)_R \equiv (1,2,{1\over 3},3) \subset {\bf 16}, ~~~~~  
\psi_R \equiv \left(\br{c} \nu \\ e \er\right)_R \equiv (1,2,-1,1) \subset {\bf 16}. \label{eq13}\ea

For these particles we have, \\

$Tr\left(k^2({B-L\over 2})^2\right)=k^2 \left(3({1\over 6})^2 + 3({1\over 6})^2 + 3({1\over 6})^2 + 3({1\over 6})^2 + (-{1\over 2})^2 + (-{1\over 2})^2  + (-{1\over 2})^2 + (-{1\over 2})^2 \right)$ \\
\be
 = k^2{4\over 3} \hspace*{10.4cm}
\ee
where, the factor 3 arises due to colour while $`k$' is the normalisation factor to be determined.
In comparison with the Trace of the $SU(2)$ generator, which is 
\ba
Tr\left({I_{3L}^2\over 2}\right) = 3(\frac{1}{4} + \frac{1}{4}) + \frac{1}{4} + \frac{1}{4} = 2
\ea
 we have ${4\over 3}k^2 = 2$ \emph{i.e.} $k = \sqrt{3\over 2}$.

\subsection{Renormalisation group equations}\label{rgesm}

The renormalisation group, in quantum field theory, tells us how  
different couplings evolve with energy. But before discussing
the renormalisation group equations (RGE) an obvious question 
is: what is renormalisation \cite{QFT}? In QFT, Green function is a most important
thing to be calculated. In perturbative QFT these quantities are divergent. The 
systematic way to remove these divergences is known as renormalisation.
There are different ways to cancel these infinities.
In order to renormalise the theory we need a reference point which is also arbitrary. Different 
choices of this reference point lead to different sets of parameters
 for the theory, but physics should not depend on the arbitrary choice of the reference point and 
be invariant. This invariance leads
to the \emph{renormalisation group}. In quantum field theory it is a useful method to examine
the behaviour of physics at a different scale knowing the same at some other scale. Thus,
measuring the observables in a low energy experiment one can compare with the values predicted
from a theory at a higher scale, \emph{e.g} at the GUT scale and certify about the correctness
of the theory. In the standard model, variations of the gauge coupling constants with energy
are given by the following renormalisation group equations (RGEs)
\be
16{\pi}^2 E {dg_i \over dE} = b_i g_i^3 = \b_{SM}(g_i)
\ee
where $i$ stands for $U(1)_Y,~SU(2)_L~{\rm and}~SU(3)_C$ and the right-hand-side
is known as the $\b$-function of the corresponding coupling\footnote{This equation is valid for the 
lowest one-loop order in perturbations theory. At higher orders $\CO(g^5)$ 
terms arise.}. One can write this equation as
\be
{d \over {d lnE}}{{\a} _i}^{-1}(E) =  -{b_i \over 2\pi}.
\ee
where, $\a _i= {{g^2_i}\over {4\pi}}$.

Using the measured values of these coupling constants at the scale $M_Z$ as the initial values
one can solve these equations as, 
{\vskip-1cm
\be
{\a}_i^{-1}(E) = {\a}_i^{-1}(M_Z) - {b_i \over 2\pi} {\rm ln}{E \over M_Z}.
\label{gevol}
\ee

In the above equations the co-efficients, $b_i$, can be calculated for any
$SU(N)$ group as 
\be
b_i = -{11\over 3}C_2(G) + {2\over 3}n_fC_2(R) + {1\over 3}n_sC_2(R)
\label{bSM}
\ee
where $C_2(R)$ is the quadratic Casimir operator for the representation R while $C_2(G)$
is that for the adjoint representation. These Casimir operators are discussed below. In 
the above equation $n_f$ is the number of chiral fermions and $n_s$ is the number of complex
scalars contributing to the $\b$-function\footnote{For a more general formula which includes 
two loop contributions one should look at Ref.\cite{Jones}.}.

The generators of a gauge group obey the following rules
\be
Tr[t_R^at_R^b]= C(R)\d^{ab},
\label{rel1}
\ee

and 
\be
\sum_a t_R^at_R^a = C_2(R).\bf{1}
\label{rel2}
\ee

where, the proportionality constant $C_2(R)$ is the quadratic Casimir operator  
for the particular representation. One can easily show that the quadratic Casimir operator 
is related with the factor $C(R)$ via
\be
C_2(R)d(R) = C(R)r
\label{rel3}
\ee

where, $r$ is the number of generators $(= N^2 -1)$ of the $SU(N)$ gauge group, equivalent to the 
dimension of the \emph{adjoint representation}, and $d(R)$ is the dimension of the representation R.

According to the convention used in eqn.(\ref{su2gen}), the $SU(2)$ generators follow the relation
\be
Tr[{\tau ^a\over 2}{\tau ^b\over 2}]={1\over 2}\d^{ab}.
\label{su2genp}
\ee

As stated earlier the bigger GUT $SU(N)$ group will be chosen in such a way that it
will contain the $SU(2)$ as a subgroup. The generators of the $SU(N)$ will also follow the
same normalisation condition -- eqn.(\ref{su2genp}) -- and, thus, we have $C(R) = {1\over 2}$
in the fundamental representation.  
Immediately eqn.(\ref{rel3}) implies that for $R=N$, \emph{i.e} for the fundamental representation 
the quadratic Casimir operator is $C_2(N) = {{N^2-1}\over {2N}}$. For the adjoint representation
$C_2(G) = N$. For the $U(1)$ gauge group these values will be $C_2(G)=0$ and $C_2(R)= C(R)=(Y/2)^2$. 

\begin{figure}[htb]
  \centering
  \includegraphics[width=0.5\textwidth,height=0.6\textheight,angle=270]
  {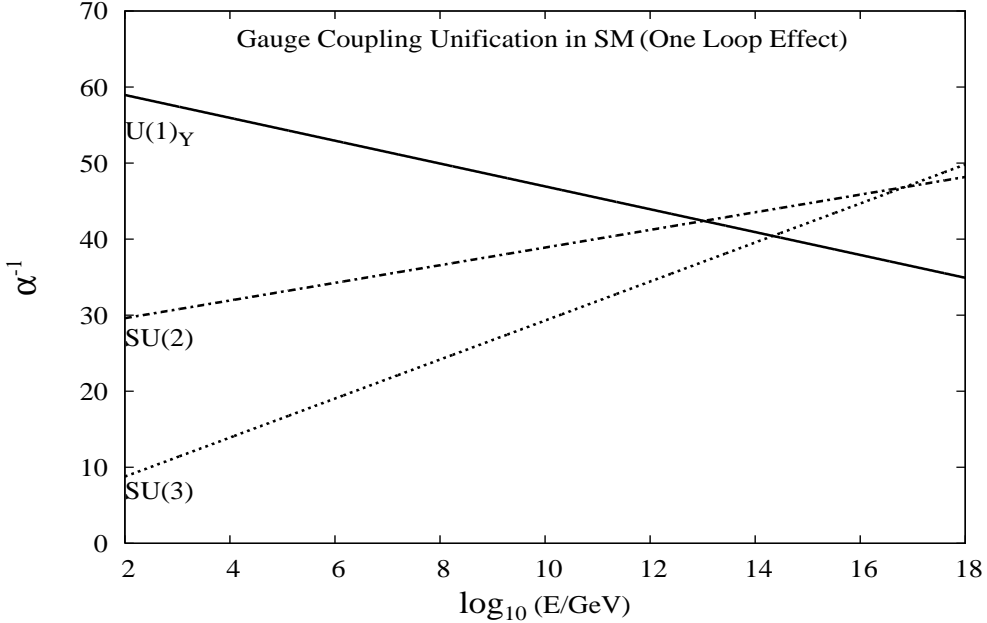}
   \caption{\sf \small{Evolution of the gauge couplings in the standard model}}
  \label{f:gsm}
\end{figure}
So, for the standard model, considering the contribution of all the particles 
listed in Table \ref{particle} one has for the three different co-efficients for the gauge groups 
 $U(1)_Y$, $SU(2)_L$ and  $SU(3)_C$ 
\ba
\left(\br{c} b_{Y}\\ b_{2L}\\ b_{3C}\er\right)=\left(\br{c} {41\over 10}\\
-{19 \over 6}\\ -7\er\right).
\label{coefsm}
\ea

where, the GUT normalisation factor ${3\over 5}$ is already multiplied to calculate the
co-efficient for the $U(1)_Y$ gauge group. Using these values of $`b$'  
one can find the evolution of the gauge couplings with energy from eqn(\ref{gevol}) as
depicted in Fig. \ref{f:gsm} upto one loop contribution only. 

It shows that all three 
standard model gauge couplings are trying to unify at some higher scale $\sim 10^{15}$ GeV, 
comparable to the predicted value of $M_G$ from the proton decay limits.
 Although in this
case they are not unifying exactly, they do so in the supersymmetric scenario.

\section{Supersymmetry}\label{sec:susy}
Supersymmetry is a space-time symmetry which relates the bosonic degrees of freedom to 
the fermionic degrees of freedom \cite{books, Martin, Ellis:2007wa}. The beautiful idea of 
supersymmetry helps to solve the gauge hierarchy problem (\ref{gaugehier}). 
The one loop radiative correction for the Higgs mass due to scalar particles in the loop  is  
\be
\delta {\rm m}_H^2 = {\lambda _S \over {16\pi ^2}}[\Lambda ^2 - 2{\rm m}_S^2{\rm ln}{\Lambda\over {M_S}} - ...],
\ee
where $\lambda _S$ is the corresponding coupling for the term in the Lagrangian, viz. 
 $-\lambda _S H^2S^2$. The same correction for a fermion-antifermion pair in loop takes the 
form
\be
\delta {\rm m}_H^2 = {|\lambda _f|^2 \over {8\pi ^2}}\left[-\Lambda^2 +3{\rm m}_f^2{\rm ln}(\Lambda/{\rm m}_f)+...\right], 
\ee
with $\lambda _f$ the coefficient of the term $-\lambda _f Hf\bar f$ in the Lagrangian.

So, we see that a conspiracy between 
the bosonic and fermionic degrees of freedom can solve the hierarchy problem. If we  
postulate that \emph{corresponding to each chiral fermion there should be  a complex scalar 
and vice versa with the condition 
\be
\lambda _S= |\lambda _f|^2,
\label{ss}
\ee
which follows in a supersymmetric theory}, then the quadratic correction can be erased.

In a supersymmetric transformation a boson changes to a fermion and vice versa. Thus, if
$Q$ is the generator of this transformation then
\be
Q|boson\rangle \equiv |fermion\rangle, ~~~{\rm and}~~Q|fermion\rangle \equiv |boson\rangle.
\ee
Hence, the generator Q has to be fermionic in nature with spin angular momentum $1/2$
 which is why supersymmetry is a space-time transformation.

The irreducible representation in which a particle and its superpartner will be accommodated 
is known as the supermultiplet. The number of bosonic and fermionic degrees of freedom are 
equal in each supermultiplet. There are different types of supermultiplets-- the simplest one
is the \emph{ chiral or matter} supermultiplet. It contains a chiral Weyl spinor and a complex 
scalar field, both of them are of two degrees of freedom. The \emph{gauge or vector supermultiplet} 
is the one in which a massless spin-1 vector gauge boson (degrees of freedom 2) is kept with it's 
fermionic superpartner, a massless spin-1/2 Majorana fermion, known as the \emph{gaugino}. The 
Minimal Supersymmetric Standard Model (MSSM), thus, contains the standard model particles, 
Table \ref{particle}, and their corresponding superpartners. The particles in the MSSM are listed in Table \ref{ssp}.

\begin{table}[]
\begin{center}
\begin{tabular}{|c|c|c|c|c|} \hline
\multicolumn{2}{|c|}{Names} 
& spin 0 & spin 1/2 & $SU(3)_C ,\, SU(2)_L ,\, U(1)_Y$
\\  \hline\hline
squarks, quarks & $Q$ & $({\stilde u}_L\>\>\>{\stilde d}_L )$&
 $(u_L\>\>\>d_L)$ & $(\>{3},\>{2}\>,\>{1\over 6})$
\\
 & $\overline u$
&${\stilde u}^*_R$ & $u^\dagger_R$ & 
$(\>{\overline 3},\> { 1},\> -{2\over 3})$
\\ & $\overline d$ &${\stilde d}^*_R$ & $d^\dagger_R$ & 
$(\>{\overline 3},\> {1},\> {1\over 3})$
\\  \hline
sleptons, leptons & $L$ &$({\stilde \nu}\>\>{\stilde e}_L )$&
 $(\nu\>\>\>e_L)$ & $(\>{1},\>{2}\>,\>-{1\over 2})$
\\
& $\overline e$
&${\stilde e}^*_R$ & $e^\dagger_R$ & $(\>{1},\> {1},\>1)$
\\  \hline
Higgs, higgsinos &$H_1$ &$(H_1^+\>\>\>H_1^0 )$&
$(\stilde H_1^+ \>\>\> \stilde H_1^0)$& 
$(\>{1},\>{2}\>,\>+{1\over 2})$
\\ &$H_2$ & $(H_2^0 \>\>\> H_2^-)$ & $(\stilde H_2^0 \>\>\> \stilde H_2^-)$& 
$(\>{1},\>{2}\>,\>-{1\over 2})$
\\  \hline
 & & spin 1/2 & spin 1 & \\ \hline
gluino, gluon &  & $\stilde g$&  $g$ & $(8,1,0)$ \\ \hline

winos, W-bosons &  & $\stilde{W}^\pm$, $\stilde {W}^0$ &${W^\pm}$, ${W^0}$ & $(1,3,0)$ \\ \hline

bino, B-boson &  & $\stilde{B}^0$ & ${B^0}$  & $(1,1,0)$ \\ \hline
\end{tabular}
\end{center}
\caption{\sf \small{Supersymmetric partners with the Standard Model members}}
\label{ssp}
\end{table}

Note that in Table \ref{ssp} MSSM 
requires two Higgs doublets for the following good reasons. Firstly, to keep the 
theory free from triangle gauge anomalies we need two Higgs scalar doublet. Since, the 
condition for a theory to be free from gauge anomalies is 
\be
\sum_{\rm Weyl~fermions}{\rm Tr}\left(({y\over 2})^3\right) = \sum_{\rm Weyl~fermions}{\rm Tr}\left(T^2_3{y\over 2}\right)=0.
\ee

The standard model itself was anomaly free, but supersymmetric extension of the SM brings
 one Weyl spinor, namely the Higgsino $\stilde{H_1}$ with hypercharge $+{1\over 2}$. This will 
generate anomalies.
 So if we add a new Higgs multiplet $(H_2,~ \stilde{H_2})$ with hypercharge
 opposite to that of the $H_1$ multiplet, then the anomaly will be cancelled again. 

Secondly, to make the up-type quarks massive we used $\Phi^*$ in sec.~\ref{gfmass}. 
Analyticity of the superpotential forces a field of definite chirality only and hence the use
of the complex conjugate of a field is disallowed. The two complex scalar doublets are
\begin{eqnarray}
&& {H_1} =\left(\begin{array}{c} {H_1}^0\\ {H_1}^- \end{array}\right),
~~~~~ {H_2}=\left(\begin{array}{c} {H_2}^+\\ {H_2}^0
\end{array}\right), \label{higgs}
\end{eqnarray}
After spontaneous symmetry breaking the minimum of $V_0$ involves the
following two vevs: $\langle H_1^0\rangle = v_1$ and $\langle H_2^0\rangle = v_2$. 
The combination $v = \sqrt{v_1^2+v_2^2} = 
(\sqrt{2} G_F)^{-1/2} \simeq 246$ GeV sets the Fermi scale.
These two different vevs will contribute to the up- and down-type quark masses 
respectively. The ratio of these two vevs,
\be
\tan\beta = \frac{v_2}{v_1}
\ee
is a very useful parameter for the discussion of the supersymmetric phenomenology.

 In a supermultiplet both the particle and its superpartner 
are included, so for exact supersymmetry both the members should have the same mass. 
If the superpartners were of same masses, they would have been already detected in experiments.
So far, none of the superpartners is observed. So,  
supersymmetry is a broken symmetry. On the other hand, Yuakwa-type coupling constants 
for the particles and corresponding antiparticles are already fixed by eqn.(\ref{ss}) due to
supersymmetry. The supersymmetry, thus, will be broken \emph{softly}; that means the 
coefficients of supersymmetry breaking couplings should be of mass dimension less 
than four and positive in order to cure the gauge hierarchy problem between the 
electroweak scale and a higher scale like, GUT or Planck scale. There are various models
 to predict the mass spectra for the MSSM scenario. In general, it is assumed that the origin
of supersymmetry breaking is at some higher scale and that all the superpartner masses
will be around the scale $M_S\sim {\rm TeV}$, known as the \emph{SUSY scale}.

\subsection{Gauge unification in SUSY}\label{rgemssm}

The one-loop renormalisation group equations for the MSSM case is 
\be
16{\pi}^2 E {dg_i\over dE} = b_i g_i^3 + \Theta(E-M_S) (\stilde{b_i}-b_i)g_i^3 ={\b}_{MSSM}(g_i),
\label{bMSSM1}
\ee
where, $\Theta$ is a step function, used due to the fact that the standard model is
valid upto scale $M_S$ after which supersymmetry will come into play. The standard model 
b-coefficients\footnote{Exact b-coefficients between $M_Z$ and $M_S$ are given in eqn. (\ref{smplus}).} $`b_i$' are given by eqn.(\ref{bSM}) while the MSSM b-coefficients are:
\ba
\left(\br{c} {\stilde b_{Y}}\\{\stilde b_{2L}}\\{\stilde b_{3C}}\er\right)=\left(\br{c} {33\over 5}\\
1 \\ -3\er\right).
\label{coefmssm}
\ea
\begin{figure}[htbp]
  \centering
  \includegraphics[width=0.5\textwidth,height=0.6\textheight,angle=270]
  {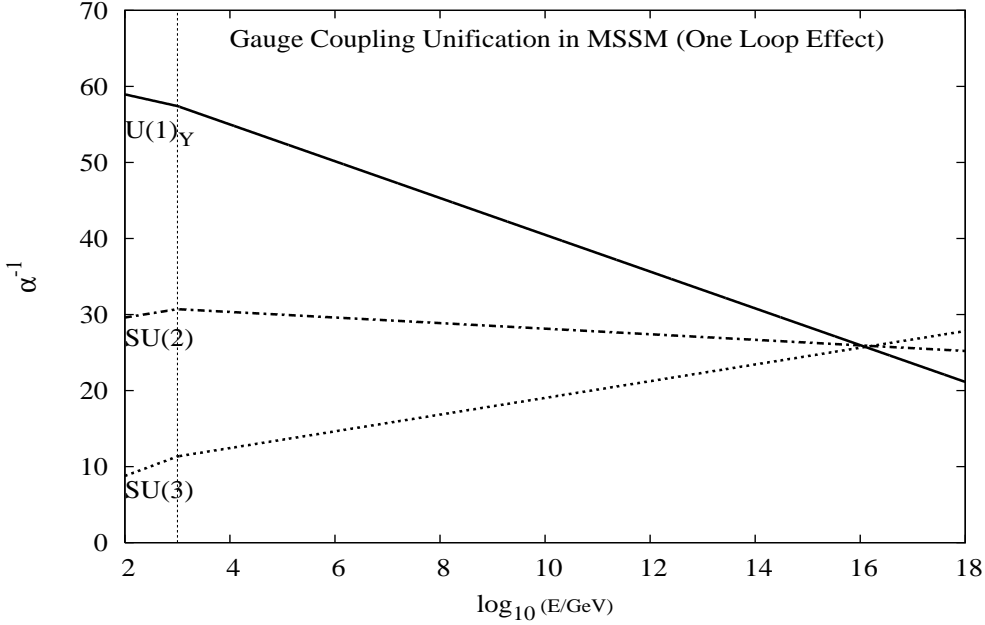}
   \caption{\sf \small{Evolution of the gauge couplings in the supersymmetric scenario}}
  \label{f:gsusy}
\end{figure}
These are the contributions due to the whole supermultiplet \emph{i.e} both from the SM 
particles and the corresponding superpartners. As a result not to overcount the contributions from
the SM particles, the standard model b-coefficients $`b_i$' are subtracted out from the second term
 in eqn.(\ref{bMSSM1}). The evolution of the gauge couplings, thus, is given by
\ba
\a _i^{-1}(E) = \a _i^{-1}(M_Z)-{b_i \over 2\pi}ln{M_S \over M_Z}-\Theta(E-M_S)\left({{\stilde{b_i}- b_i}\over{2\pi}}\right){\rm ln}{E \over M_S}. 
\ea
The evolution of the gauge couplings $\a _i(={g^2_i\over 2\pi})$ is depicted in Fig. \ref{f:gsusy}. We have already seen that supersymmetry can solve the gauge hierarchy problem,
 but in addition Fig. \ref{f:gsusy} shows that it's particle content is such that it also 
gives a very good unification of the standard model gauge couplings. Also, the unification,
 like in the SM case, is at such a high scale then it will not conflict\footnote{Note, this huge
mass of the $X,Y$ boson will not protect proton from decaying via dimension-5 operator, which is
discussed in Ref.~\cite{dim5pdk}.} with the proton life time predicted in sec.~\ref{su5gut}.


\section{Extra dimensions}

In the SM we have seen that the hierarchy problem is arising due to the huge ratio of the
Planck scale, ${M_{Pl}}$, or the GUT scale, ${M_{G}}$, to the
electroweak scale. As discussed in the previous section, supersymmetry provides a 
beautiful way to solve this hierarchy problem. In that case, the supersymmetric particles
are situated around the TeV scale. Actually to solve the hierarchy problem if we incorporate 
any new physics it should appear around that scale to address the huge ratio. More recently, 
a new kind of physics, Extra Dimension (ED), was introduced in particle physics. If we can 
distinguish a fermion from a bosonic particle by measuring the spin of 
of the particle at the Large Hadron Collider (LHC) or the International Linear Collider (ILC), 
then we can have a distinct signature of the physics of extra dimension from that of 
supersymmetry.  

Historically, this idea was first introduced by Kaluza and Klein in 1920, to 
unify the electromagnetic interaction with the gravitational one by generating the photon
from the extra components of the five-dimensional metric. Nowadays in 
a more popular and fundamental theory, namely, string theory, it is common to use more 
than one 
space dimension, as the theory is consistent only in the extra-dimensional scenario.
There are many open questions about the extra dimension, \emph{e.g}, what would be
nature of the extra dimension, what is the size of it and many more. A huge number
of phenomenological studies have been pursued in this subject in this decade. Let us 
have a closer look on some of these.

Let us consider a massless particle in a 5d Cartesian co-ordinate system, where Lorentz
invariance holds. The square, thus, of the 5d momentum gives us 
\be
p^2=0=g^{AB}p_Ap_B= p_0^2-{ \vec p}^2 \pm p_5^2, 
\ee
as
$g_{AB}={\rm diag}(1,-1,-1,-1,\pm 1)$. This implies that the four-dimensional
mass square of the particle given by 
\be
m^2=p_{\mu}p^{\mu}=p_0^2-{\vec p}^2=\mp p_5^2,
\label{fifthm}
\ee 
becomes negative if we consider the extra dimension as time-like \cite{Rizzo}.
\vskip1cm
\begin{figure}[thb]
\hskip 2.0cm
\includegraphics[width=0.4\textwidth,height=0.3\textheight,angle=0]
{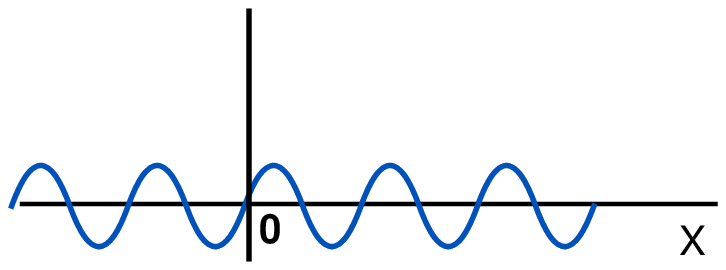}
\vskip -8.23cm
\hskip 10.0cm
\includegraphics[width=0.4\textwidth,height=0.3\textheight,angle=0]{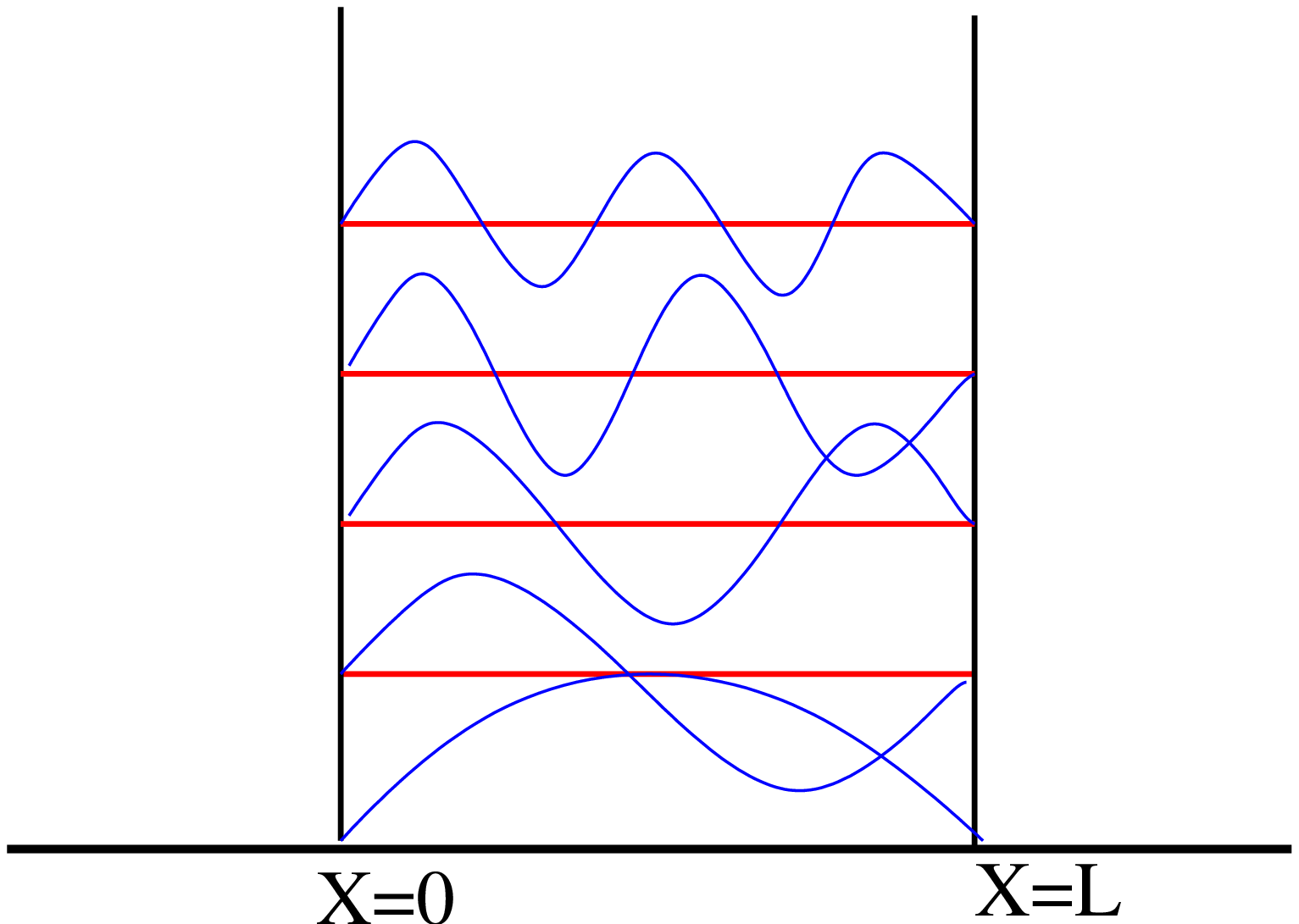}
\vskip1cm
\caption{\sf \small {Free particle wave function (left), wave function in a compact space (right)}}
\label{f:wave}
\end{figure}
 Thus it's velocity will exceed the velocity of light in vacuum and lead to a problem: the
{\it tachyon} state. So in this discussion we will consider a {\it space-like} co-ordinate
as the extra dimension which will be compactified on a circle $S^1$ or $S^1/Z_2$ orbifold,
for one extra-dimensional scenario, with radius of compactification R.

Before going into the more detailed discussion of extra dimensions let us 
recall the well known quantum mechanical one-dimensional box of size L. 
As we know,  
the solution of a particle moving along the $x$-direction with momentum $p$ is given 
by $\sim Ae^{ipx}+Be^{-ipx}$,
 here $x$ is infinitely long, $i.e$, the physical system is not compact and the 
particle momentum $p$ takes continuous values from 
$-\infty~{\rm to}~+\infty$. Let's go to a bound system. Suppose 
the potential is infinite outside the box $0\leq x\leq \pi R$; $L=\pi R$ is the 
length of the box, while it is zero inside of the box. With the proper boundary 
condition that the wave function vanishes at the boundary, the solution takes 
the form $\psi = \sin{nx\over R}$ and the momentum of the particle 
is given by $p={n\over R}$, where $n$ can take any integer value. Due 
to the compactness of the $x$-dimension, the corresponding momentum $p_x$ becomes quantized. 
In the five-dimensional scenario where the extra space direction $y$ is 
compactified in a similar way, the corresponding quantized fifth component of momentum is 
given by $p_5={n\over R}$. 
Hence a particle which is massless in its zero mode, in the excited states, according to
eqn.(\ref{fifthm}), acquire a mass $m_n={n\over R}$. This implies a large number of massive states 
whose mass is inversely proportional to the dimension of the box \cite{Rizzo, apel}.  

\subsection{Scalar particle in ED}\label{sed}

In addition to the four space-time co-ordinates $x({\rm {\vec x, t}})$, 
let us denote the extra space-type co-ordinate by $y$, compactified on
a circle or radius R. Thus, the Lagrangian of a free complex scalar $\Phi(x,y)$with 
mass m will be a function of both $x~{\rm and}~y$ co-ordinates with a condition that
the field at $y=2\pi R$ will match with that at $y=0$, $i.e$ it has a 
periodicity of $2\pi R$ along the $y$ direction. So one can expand it in 
a Fourier series as 
\be
\Phi(x,y) =  \frac{1}{\sqrt{2\pi R}}\Phi_0(x) 
  +   \sum_{n=1}^\infty \frac{1}{\sqrt{\pi R}} 
  \left[ \Phi_n^+(x)\cos\left(\frac{ny}{ R}\right) 
   +  \Phi_n^-(x)\sin\left(\frac{ny}{R}\right)\right]. 
 \label{5scalar}
\ee


The five-dimensional action is given by
 \ba  
\hspace*{-3cm} S^5[\Phi] = \frac{1}{2}\int\! d^4x\ dy\  
 \left\{(\partial^A\Phi)^{\dagger}(\partial_A\Phi)  - m^2 \Phi^{\dagger}\Phi\right\} \nonumber
 \label{5action}
 \ea

with A = 0,1,2,3,5.

With the use of eqn.(\ref{5scalar}) if we replace the 
scalar field $\Phi$ 
and integrate out the extra dimension $y$ then
the action will correspond to a large number Kaluza-Klein (KK) modes as
 \ba 
 S^4[\Phi] &=& \frac{1}{2}\int\! d^4x 
 \left\{(\partial^\mu\Phi_0)^{\dagger}(\partial_\mu \Phi_0) - m^2 \Phi_0^{\dagger}\Phi_0\right\}
+ \sum_{n=1}^\infty \frac{1}{2}\int\! d^4x 
 \left\{(\partial^\mu\Phi_n^+)^{\dagger}(\partial_\mu \Phi_n^+) - m_n^2 {\Phi_n^+}^{\dagger}{\Phi_n^+}\right) \nonumber\\
& +&  \sum_{n=1}^\infty \frac{1}{2}\int\! d^4x 
 \left\{(\partial^\mu\Phi_n^-)^{\dagger}(\partial_\mu\Phi_n^-) - 
 m_n^2{\Phi_n^-}^{\dagger}{\Phi_n^-}\right\},
 \ea

where the n-th KK mode mass is given as 
\be
{\rm m}_n^2 = {\rm m}^2 + \frac{n^2}{R^2}.
\ee

In four-dimensional effective theory, thus, in addition to the
zero mode field, we are getting two different sets -- one is even and
another odd under the transformation $y \ra -y$ -- of fields when the extra 
space dimension is compactified on the circle $S^1$.

\subsection{Fermion particle in ED}\label{fed}

In some models only the scalar bosons are allowed to access
the extra dimensions while the fermions are kept in a fixed point
of the extra dimension, called ``brane''. In such cases the above 
compactification
is quite natural but what happens if we intend to allow the fermions
as well to access the extra dimension? Do we have the 
same set of Kaluza-Klein modes for the fermionic fields or something else?

Let us consider a fermion $\S$ in the five-dimensional field, where the 
extra space dimension is compactified the way we discussed in sec.~\ref{sed}. 
The five-dimensional spinor can be written as a two component four-dimensional spinor
 \ba 
 \S = \left(\br{c} \psi_R\\ \psi_L\er\right)~.
 \label{f5d0}
 \ea 

Note that in the five-dimensional field theory, one can construct the five $\G^A$ matrices 
with A=0,1,2,3,5, from the usual four-dimensional ones as follows:

\be
\G^\mu = \g^\mu ~~~ {\rm and}~~ \G^5 = i \g_5. 
\label{gm5d}
\ee

In 5d the fifth component of the $\G^A$ is constructed from the $\g_5$ matrix, which
is used, in four-dimensions, to define the chiral operator $P_{R/L} = (1\pm\g_5 )$. So,
in five-dimensions, and it is true for any odd number of dimensions, there is
no chiral operator. To be clear, in eqn.(\ref{f5d0}) the subscripts $L$ and $R$
are just two component notations only.

Like the scalar field discussion in sec.~\ref{sed}, let us 
consider the action for a massless fermion $\S$ as 
\ba
S &=& \int ~ d^4x dy~~ i\bar\S\G^A\partial_A\S \nonumber \\
 &=& \int ~d^4x dy~~  \left( i\bar\S \g ^\m \P _\m\S + \bar\S\g ^5\P _y\S
 \right).
\label{fs5d}
\ea

Due to the symmetry of the fermion field at the point $y=0$ and $y=2\pi R$, we can have the Fourier 
expansion of the field as
\be
\psi_{L/R}(x,y)  = 
{1\over{\sqrt{2\pi R}}}\psi^0_{L/R}(x)
  +   \sum_{n=1}^\infty {1\over{\sqrt{\pi R}}} 
  \left\{\psi^{+n}_{L/R}(x)\cos\left({ny\over R}\right)
  +  \psi^{-n}_{L/R}(x)\sin\left({ny\over R}\right)\right\}.
\label{f5d}
\ee

Once we put these fermions -- eqn.(\ref{f5d}) -- in the above action eqn.(\ref{fs5d}) we 
end up with a few phenomenological problems. 

For example, let us use the zero mode term in eqn.(\ref{f5d}), then we have 
\ba
S_{\rm zero~mode} &=& \int d^4x \int_0^{2\pi R} ~ dy~~  \left({1\over{\sqrt{2\pi R}}} (\bar\psi^0_L +\bar\psi^0_R)i \g ^\m \P _\m 
{1\over{\sqrt{2\pi R}}}(\psi^0_L + \psi^0_R) \right). \nonumber \\
 &\sim & \int ~d^4x  \left\{ \bar\psi ^0_L i \g ^\m \P _\m \psi_L + \bar\psi ^0_R i \g ^\m \P _\m \psi_R \right\}. 
\ea

Thus, for each massless field in five-dimension we are having two massless zero modes in the
four-dimensional effective theory. The four-dimensional fermion is thus vector like in nature. It 
is well-known that fermions in the SM are chiral in nature, the left chiral part 
transforms as a doublet under $SU(2)$ gauge transformation and the  right chiral part transforms 
trivially. 
If the dimensional reduction doubles the state can we regain our chiral nature of the fermion 
in its zero mode? 

To regain the chiral nature we have to compactify on an $S^1/Z_2$ orbifold instead of a circle. 
The expansions of different kind of field for the $S^1/Z_2$ orbifold will be 
discussed in sec.~\ref{ued}. In that case, although the higher KK modes of the chiral 
fermion behave as vector but the zero mode remains a chiral one.

\subsection{Vector gauge bosons in ED}\label{vbed}
Let us consider a vector gauge boson $A_M$ with $M=0,1,2,3,5$ in five-dimensional scenario, where
$A_\m$ are the usual four vector bosons while the extra component $A_5$ will be a scalar\footnote{The coupling of the $A_5^{(n)}$ states to
fermions involve $\gamma_5$ and so, strictly, they are pseudoscalars.}. The Lagrangian of such a field is given by
\ba 
{\CL}_{5d} &=& -{1\over 4} F_{MN}F^{MN} 
\ea
where $F_{MN} = \P _M A_N - \P_N A_M$.

The compactification will be like in the two previous sections 
sec.~\ref{sed} and sec.~\ref{fed}. In the same way, the periodicity of the field at $y=0$ 
and $y=2\pi R$ implies,
 \be 
 A_M(x,y) = {1\over{\sqrt{2\pi R}}}A_M^{(0)}(x)
 + \sum_{n=1}{1\over{\sqrt{\pi R}}}  
  \left\{ A_M^{n}(x){\rm cos}\left({ny\over R}\right) + 
  \tilde A_M^{n}(x){\rm sin}\left({ny\over R}\right)\right\}.
 \ee


To discuss different components of an extra-dimensional vector gauge boson, let
us assume a photon field in five-dimensional quantum electrodynamics (QED). In five dimensions
this massless photon has five components. In the effective four-dimensional
theory one should expect a tower of four-component photon fields and a tower of 
adjoint scalars. Although the zero mode of the photon has to be massless but the excitations for
 $n \not= 0$ will be massive due to the KK contribution ($\sim n/R$). In QED there is no 
spontaneous symmetry breaking. The extra longitudinal degree of 
freedom for each of these massive KK gauge bosons will be obtained by 
absorbing the adjoint scalar of the same level. In five-dimensional QED, thus, we will 
have KK modes of the photon field only, but no KK modes of any adjoint scalar.

The corresponding SM scenario is quite complex. As besides the usual 
electroweak mass the weak gauge bosons acquired masses from the KK contribution also,
 so the usual unphysical components of the weak scalar doublet will no more be the 
Goldstone bosons. In reality, the KK modes of these unphysical fields will mix with 
the corresponding KK modes of ${W_5^\pm}_n$ and ${Z_5}_{n}$, to form three Goldstone 
modes $G^0_{n}$, $G^\pm_{n}$ and three physical scalar fields $a^0_n$ and $a^\pm_n$ \cite{buras}.
With increasing KK number, the contributions of ${W_5^\pm}_n$ and ${Z_5}_{n}$ dominate the
Goldstone boson modes while the unphysical components of the weak scalar doublet will, now,
become main part of three physical scalar fields $a^0_n$ and $a^\pm_n$. In addition with
these real scalars we will also have usual higgs boson $h$ and its KK excitations $h_n$. 

  Before going into the 
discussion of the universal extra dimension and the field expansion in the $S^1/Z_2$ orbifold
let us have a brief discussion on how the large extra dimension scenario can explain the 
gauge hierarchy problem.

\subsection{ADD model and solution of the gauge hierarchy problem}
The main motivation of 
introducing extra dimensions into particle physics was to explain the 
huge hierarchy between the electroweak scale $\sim 100~{\rm GeV}$ and the Planck 
scale $\sim 10^{19}~{\rm GeV}$. The model we discuss in this subsection is the one of Large Extra 
Dimension (LED), introduced by Arkani-Hamed, Dimopoulos, Dvali (ADD) \cite{add}. 
According to this model there exist $n$ extra spatial dimensions of radius $R$, which are 
accessed by only gravity while all other standard model particles are constrained at a 
particular point of these extra space dimensions. 
To fulfill the requirement it is assumed that the higher dimensional Planck 
mass $M_{Pl}^{(4+n)}$ is equal to the four-dimensional electroweak mass $m_{EW}$ thus evading
the vexing hierarchy.
Using Gauss's law in $(4+n)$ dimensions, the gravitational potential between two
masses ${\rm m}_1$ and ${\rm m}_2$ separated by a distance $r>>R$ is given by 
\be
V(r) \sim {{{\rm m}_1{\rm m}_2}\over{M_{Pl(4+n)}^{(n+2)}R^n}}{1\over r} .
\ee

Thus the four-dimensional effective $M_{Pl}$ is 
\be
M_{Pl}^2 =  M_{Pl(4+n)}^{2+n} R^n.
\label{mrel}
\ee

But, as stated above, our requirement is $M_{Pl(4+n)}~{\rm will~ be} \sim {\rm m}_{EW}$. Thus replacing the
same in the above eqn.(\ref{mrel}), we have,
\be
M_{Pl}^2 =  {\rm m}_{EW}^{2+n}R^n,
\ee
implies,
\be
R \sim 10^{{30\over n} -17} cm \times \left( {1{\rm TeV}\over{{\rm m}_{EW}}} \right) ^{1+{2\over n}} .
\label{R}
\ee

Thus, the requirement that $M_{Pl}^{(4+n)}$ will be equal to the electroweak scale $\sim TeV$ 
implies $R=10^{13}$ cm for $n=1$, instantly excluded due to the huge deviation from
Newtonian gravity.
 
At the time the model was proposed, Newtonian gravity was precisely checked upto 1mm. 
For $n=2$ from the above formula eqn.(\ref{R}) we have $R=1$ mm. Thus models with at 
least two extra dimensions with a size of a millimeter can explain the gauge hierarchy problem. 

\subsection{Universal Extra Dimension}\label{ued}

A model in which all the standard model particles are allowed to 
access the extra dimensions
is known as the \emph{Universal Extra Dimension} (UED) model also known as the \emph{ACD model} 
after its proposers Appelquist, Cheng and Dobrescu \cite{acd}. 

 Construction-wise it is very similar to the ADD model,
but as in this case, in addition to gravity, scalars, 
fermions and vector gauge bosons are also accessing the 
extra dimension so, as discussed earlier, it should be compactified on 
an orbifold $S^1/Z_2$ instead of a circle.  This
orbifold is nothing but equivalent to the compactification
on a circle of radius $R$ with a $Z_2$ symmetry - identifying
$y \to -y$, where $y$ denotes the fifth compactified coordinate.
The orbifolding is crucial in generating {\em chiral} zero modes
for fermions.  

The motivations of universal extra dimensions are quite speculative. Besides providing 
viable dark matter candidates the six-dimensional theory can explain from anomaly 
cancellation why we have only three generations \cite{anomaly-can}. 
Only three generation of fermions can remove the $SU(2)$ global gauge anomaly. 
Another good feature about universal extra dimensions is to provide a natural way
to explain the long life time for the proton \cite{pdecay}. They
could lead to a new mechanism of supersymmetry breaking 
\cite{antoniadis1}, address the fermion mass hierarchy in an
alternative way, provide a cosmologically
viable dark matter candidate \cite{Servant:2002aq}, stimulate power
law renormalization group running \cite{dienes,blitzkrieg}, admit
substantial evolution of neutrino mixing angles defined through an
effective Majorana neutrino mass operator \cite{Deandrea:2006mh},
etc.
The interesting point is that in this
case the discrete symmetry which removes operators providing dangerous contributions to
the proton decay is not imposed externally but is an essential ingredient for the theory.

With the compactification, as defined, on an orbifolding $S^1/Z_2$ for the five-dimensional
scenario the expansion of the five-dimensional gauge bosons, scalars and fermions with
the proper use of boundary conditions are given by

\begin{eqnarray}
\label{fourier}
A_{\mu}(x,y)&=&\frac{\sqrt{2}}{\sqrt{2\pi
R}}A_{\mu}^{(0)}(x)+\frac{2}{\sqrt{2\pi
R}}\sum^{\infty}_{n=1}A_{\mu}^{(n)}(x)\cos\frac{ny}{R},~~~~
A_5(x,y) = \frac{2}{\sqrt{2\pi
R}}\sum^{\infty}_{n=1}A_5^{(n)}(x)\sin\frac{ny}{R}, \nonumber\\
\phi(x,y)&=&\frac{\sqrt{2}}{\sqrt{2\pi
R}}\phi^{(0)}(x)+\frac{2}{\sqrt{2\pi
R}}\sum^{\infty}_{n=1}\phi^{(n)}(x)\cos\frac{ny}{R}, \nonumber \\
\mathcal{Q}_{i}(x,y)&=&\frac{\sqrt{2}}{\sqrt{2\pi 
R}}\bigg[{\pmatrix{u_i\cr d_i}}_{L}(x)+\sqrt{2}\sum^{\infty}_{n=1}\Big[
\mathcal{Q}^{(n)}_{iL}(x)\cos\frac{ny}{R}+
\mathcal{Q}^{(n)}_{iR}(x)\sin\frac{ny}{R}\Big]\bigg], \\
\mathcal{U}_{i}(x,y)&=&\frac{\sqrt{2}}{\sqrt{2\pi
R}}\bigg[u_{iR}(x)+\sqrt{2}\sum^{\infty}_{n=1}\Big[
\mathcal{U}^{(n)}_{iR}(x)\cos\frac{ny}{R}+
\mathcal{U}^{(n)}_{iL}(x)\sin\frac{ny}{R}\Big]\bigg], \nonumber\\
\mathcal{D}_{i}(x,y)&=&\frac{\sqrt{2}}{\sqrt{2\pi
R}}\bigg[d_{iR}(x)+\sqrt{2}\sum^{\infty}_{n=1}\Big[
\mathcal{D}^{(n)}_{iR}(x)\cos\frac{ny}{R}+
\mathcal{D}^{(n)}_{iL}(x)\sin\frac{ny}{R}\Big]\bigg], \nonumber
\end{eqnarray}

where $i=1,2,3$ are generation indices. Above, $x (\equiv x^{\mu})$ denotes
the first four coordinates, and as mentioned before, $y$ is the compactified
coordinate. The complex scalar field $\phi (x,y)$ and the gauge boson
$A_\mu(x,y)$ are $Z_2$ even fields with their zero modes identified with the
SM scalar doublet and a SM gauge boson respectively. On the contrary, the
field $A_5(x,y)$, which is a real scalar transforming in the adjoint
representation of the gauge group, does not have any zero mode. The fields
$\mathcal{Q}$, $\mathcal{U}$, and $\mathcal{D}$ describe the 5-dimensional
quark doublet and singlet states, respectively, whose zero modes are
identified with the 4-dimensional chiral SM quark states. The KK expansions of
the weak-doublet and -singlet leptons will be likewise and are not shown for
brevity.

Similar to the supersymmetric R-parity in this 
UED scenario we have Kaluza-Klein parity, in short KK-parity, which
 is conserved. This KK-parity is defined as 
\be
KK-{\rm parity} = (-1)^n,
\ee
with $n$ as the KK number of the corresponding states. Thus in any KK-parity conserving
process the lightest Kaluza-Klein particle (LKP) state, with $n=1$, cannot decay to the standard 
model particles and will be a good example of dark matter.
	
\subsection{Bounds on the Universal Extra Dimension}\label{bued}
During the discussion of LED scenario, we put some bound on the 
extra dimension based on the Newtonian gravitational interaction. In that
case all the SM particles were constrained to be at a single point of the
extra dimension so we did not care about the electro-weak or any other
interactions. In the case UED scenario all the SM particles can access
the extra dimensions. On the other hand weak interaction is perfectly
measured up to a length $\sim {\rm m}_{EW}^{-1}$, much smaller than the
LED bound 1mm, so we expect that the bound on the universal extra dimensions 
should be much more constraint. Constraints on 
this scenario from $g-2$ of the muon \cite{nath}, flavour changing
neutral currents \cite{buras,chk,desh}, $Z \to b\bar{b}$ decay
\cite{santa}, the $\rho$ parameter \cite{acd,appel-yee}, other
electroweak precision tests \cite{ewued}, implications from hadron
collider studies \cite{collued}, etc. imply that $R^{-1}~\gtap~300$
GeV.  A recent inclusive $\bar{B} \to X_s \gamma$ analysis sets a
stronger constraint $R^{-1}~\gtap~600$ GeV \cite{Haisch:2007vb}.

This thesis is devoted mainly to explore some of the distinctive characterisicts of new physics
beyond the standard model. In chapter 2 we have presented the 
power law evolution of gauge, Yukawa and quartic couplings in the universal extra-dimensional scenario. We have also noted in this chapter that if supersymmetry is found at 
the LHC then UED will be out of reach of any future collider experiment. 
How the upper limit of the lightest 
supersymmetric neutral Higgs mass will be relaxed if supersymmetry is embeded in the extra 
dimension is discussed in chapter 3. In chapter 4, using three different approaches
we have achieved a low intermediate left-right symmetry breaking scale in the supersymmetric 
$SO(10)$ Grand Unified Theory. Remaining within the SM in chapter 5 we have analysed the triquark 
state using $SU(6)$ unitary scalar factors and derived tree level pentquark masses. In chapter 6 
we have presented the summary and conclusions of the work.

\chapter{ Power law scaling in Universal Extra Dimension
scenarios }
 
\section{Introduction} 
In section \ref{rgesm} we have seen that in the standard model, the gauge 
couplings (Yukawa and quartic scalar couplings as well) run logarithmically 
with the energy scale. The gauge couplings do not all meet at a point, 
they tend to unify near $10^{15}$ GeV. Such a high scale is beyond the 
reach of any present or future experiments. Instead of this logarithmic
running, if the gauge couplings were running exponentially or 
with a definite power of energy then we could have a lower GUT scale. Extra
dimensions is such a scenario which will lead to a power law running of 
the gauge couplings due to the large number of Kaluza-Klein states. 
Different KK modes will contribute the same way, as the zeroth mode does,
to the gauge coupling evolution once we cross their corresponding threshold 
energies. The cumulative 
effect of this leads to a power law running of the gauge couplings. 

Here we will work in a one UED scenario, where a flat extra
dimension is compactified on an $S^1 /Z_2$ orbifold as discussed
earlier. With this compactification on an orbifolding 
$S^1/Z_2$ for the five-dimensional scenario the expansion of the five- 
dimensional gauge bosons, scalars and the fermions are already given in sec.~\ref{ued}. 
We examine the cumulative contribution of the KK states to the 
renormalisation group (RG) evolution of the gauge, Yukawa and 
quartic scalar couplings. Our motive, here, is to extract any subtle 
features that
emerge due to the KK tower induced power law running of these couplings in
contrast to the usual logarithmic running of the standard 4-dimensional
theories, and whether they set any limit on parameters for the sake of
theoretical and experimental consistency.

Let us now clarify the technical meaning of RG running in a higher
dimensional context. This has been extensively discussed in
\cite{dienes} in a general context, and here we merely reiterate it to
put our specific calculations into perspective. Like all other
extra-dimensional models, from a 4-dimensional point of view, the UED
scenario too is non-renormalisable due to the infinite multiplicity of
the KK states\footnote{For a study of ultraviolet cutoff sensitivity
in different kinds of TeV scale extra-dimensional models, see
\cite{db}.}. So $`$running' of couplings as a function of the energy scale 
$E$
ceases to make sense. What we should say is that the couplings receive finite
quantum corrections whose size depend on some explicit
cutoff\footnote{The beta functions are coefficients of the
divergence $1/\epsilon$ in a 4-dimensional theory. Here, a second
kind of divergence appears when the finite beta functions get
corrections from each layer of KK states which are summed over.
This summation is truncated at a scale $\Lambda$.} $\Lambda$. The
corrections originate from the following number of KK states\footnote{Since, the energy of the n-th KK-mode is given by $E = {n\over R}$, where $R$ is the radius of compactifaction.}
\ba
 \int ^\L _{1\over R} dn = R \int ^\L _{1\over R} dE = (\L R - 1)
\ea

which lie between the scale $R^{-1}$ where the first KK states
are excited and the cutoff scale $\Lambda$. The couplings will
have a power law dependence on $\Lambda$ as a result of the KK
summation.  This cutoff is interpreted as the scale where a
paradigm shift occurs when some new renormalisable physics
underlying our effective non-renormalisable framework surfaces.

\section{Renormalisation Group Equations} 
We now lay out the strategy followed to compute the RG correction
to the couplings from the KK modes. The first step is obviously
the calculation of the contribution from a given KK level which
has both $Z_2$-even and -odd states. Three points are noteworthy
and should be taken into consideration during this step:
\begin{enumerate} 
\item While the zero mode fermions are chiral as a result of
orbifolding, the KK quarks and leptons at a given level are
vector-like. 
\item The fifth compotent of the gauge bosons are ($Z_2$ odd)
scalars, but in
the adjoint representation of the gauge group. Such states are not encountered
in the SM context.
\item The KK index $n$ is conserved at each tree level vertex.
\end{enumerate}
The first step KK excitation occurs at the scale $R^{-1}$ (modulo the zero
mode mass). Up to this scale the RG evolution is logarithmic, controlled by
the SM beta functions. Between $R^{-1}$ and $2 R^{-1}$, the running is still
logarithmic but with beta functions modified due to the first KK level
excitations, and so on. Every time a KK threshold is crossed, new resonances
are sparked into life, and new sets of beta functions rule till the next
threshold arrives. The beta function contributions are the same for each of
the $(\Lambda R - 1)$ KK levels, which, in effect, can be summed. After this, the
scale dependence is not logarithmic any more, it shows power law behaviour, as
illustrated by Dienes {\em et al} in \cite{dienes2}. This illustration shows
that if $\Lambda R \gg 1$, then to a very good accuracy the calculation
basically boils down to computing the number of KK states up to the cutoff
scale. For one extra dimension up to the energy scale $E$ this number is $S =
ER$, and $E^{\rm max} = \Lambda$. Then if $\beta^{\rm SM}$ is a generic SM
beta function valid during the logarithmic running up to $R^{-1}$, beyond that
scale one should replace it as\footnote{We refer the readers to eqns.~(2.15)
and (2.21) of Ref.~\cite{dienes2}, and the subtleties leading to these
equations in the context of gauge couplings, to have a feel for our
Eq.~(\ref{master}).}
\begin{equation} 
\label{master}
\beta^{\rm SM} \to \beta^{\rm SM} + \Theta (E-\frac{1}{R}) ~(S - 1) \tilde{\beta}, 
\end{equation}
where $\tilde{\beta}$ is a generic contribution from a single KK
level. The function $\Theta (E-\frac{1}{R})$ is used just to state that this
UED $\b$-function will only occur after an energy scale $E\geq {1\over R}$. 
Irrespective of whether we deal with the `running' of gauge, Yukawa, or
quartic scalar couplings, the structure of Eq.~(\ref{master}) would continue
to hold. Clearly, the $S$ dependence reflects power law running. How this
master formula (\ref{master}) enters diagram by diagram into the evolution of
the above couplings in the UED scenario constitutes the main part of
calculation in this chapter.

\subsection{Gauge couplings} 
While considering the evolution of the gauge couplings, we first write
$\tilde{\beta}_i^g = \tilde{b}_i g_i^3$. The calculation of $\tilde{b}_i^g$
would proceed via the same set of Feynman graphs which give the SM
contributions $b_i^{\rm SM}$ but now containing the KK internal lines.  The
key points to remember are the presence of adjoint scalars and doubling of KK
quark and lepton states due to their vectorial nature.

  
\begin{center}
\begin{figure}[thb]
\hspace*{2cm}\psfig{figure=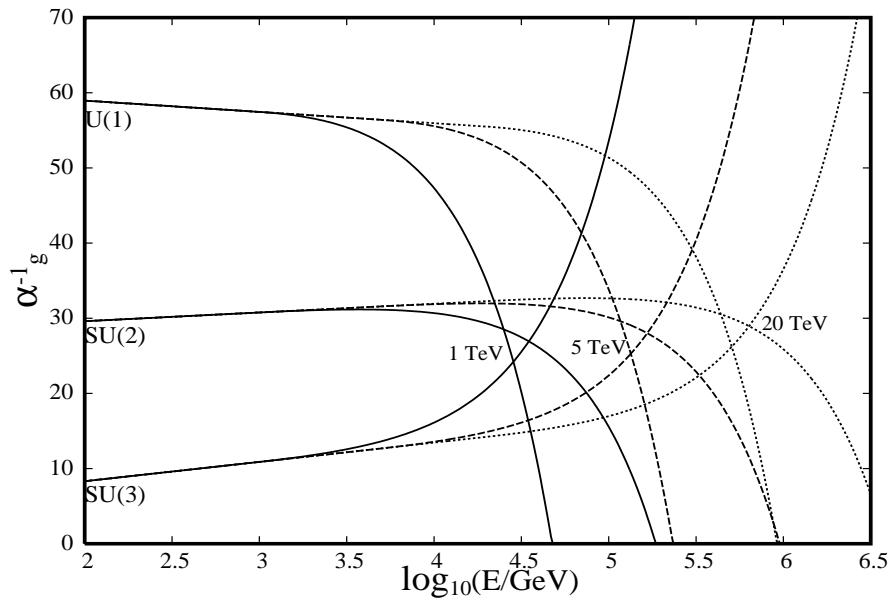,width=12cm,height=8cm,angle=270}
\caption{\sf \small {Evolution of gauge couplings for UED with $R^{-1}$ = 1,
5, and 20 TeV. For each of the three couplings, $\alpha_i \equiv g_i^2/4\pi$.}}
\label{f:gunif}
\end{figure}
\end{center}

We obtain
\begin{equation}
\tilde{b}_1 = \frac{81}{10}, ~~ \tilde{b}_2 = \frac{7}{6}, ~~ 
\tilde{b}_3 = - \frac{5}{2},
\label{ued}
\end{equation}   
where the $U(1)_Y$ beta function is appropriately normalised. Just to
recall, the corresponding SM numbers are given in eqn.(\ref{coefsm}). We have 
plotted the evolution of gauge couplings in UED
for $R^{-1} =$ 1, 5, and 20 TeV in Fig.~\ref{f:gunif}. The running is
fast, as expected, and the couplings nearly meet around\footnote{The
issue of proton stability in such low scale unification scenarios has
been dealt in \cite{pdecay}.}  30, 138 and 525 TeV,
respectively.  It is not hard to provide an intuitive argument for
such low unification scales and how they vary with $R$: roughly
speaking, $\Lambda R$ is order $\ln (M_{\rm GUT}/M_W) \sim
\ln(10^{15})$, where $M_{\rm GUT}$ is the 4-dimensional GUT scale,
i.e. the effect of a slow logarithmic running over a large scale is
roughly reproduced by a fast power law sprint over a short track.  The
other striking feature reflected in Fig.~\ref{f:gunif} is that the
$SU(2)$ gauge coupling ceases to be asymptotically free: the dominance
of the KK matter sector over the gauge part in $\tilde{b}_2$ severely
challenges the $SU(2)$ asymptotic freedom.  In contrast, the negative
sign of $\tilde{b}_3$ causes a precipitous drop in the $SU(3)$ gauge
coupling with energy.

\subsection{Yukawa couplings} 
\begin{center}
\begin{figure}[thb]
\hspace*{2cm}\psfig{figure=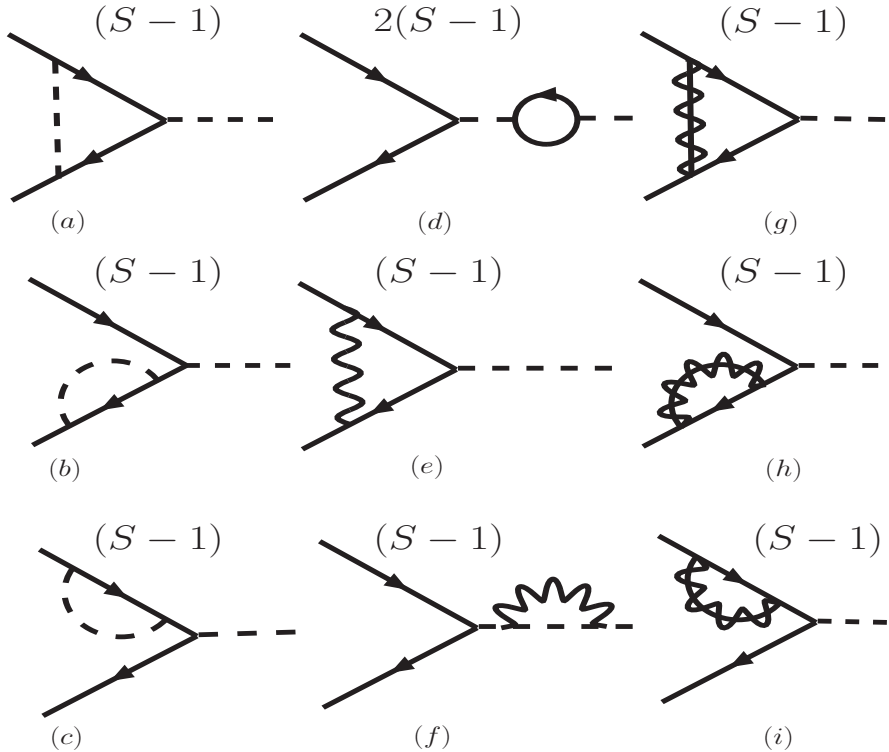,width=12cm,height=10cm,angle=0}
\caption{\sf \small {Diagrams contributing to Yukawa coupling
evolution in the Landau gauge. Solid (broken) lines correspond to
fermions (SM scalars), while wavy lines (wavy+solid lines) represent
ordinary gauge bosons (fifth components of gauge bosons).}}
\label{f:yuk}
\end{figure}
\end{center}
The Feynman diagrams that contribute to the power law evolution of
Yukawa couplings (in Landau gauge) are shown in Fig.~\ref{f:yuk}.  The
contributions come from the pure SM states, their KK towers, and from
the adjoint representation scalars\footnote{A subtle feature is worth
noticing. In four dimensions, the calculational advantage of working
in Landau gauge is that some diagrams give vanishing
contributions. The argument breaks down in a higher dimensional
context. More explicitly, consider the Figs.~\ref{f:yuk}h and
\ref{f:yuk}i. These graphs proceed through the exchange of adjoint
$A_5$ scalars and yield non-vanishing contributions. The corresponding
figures with $A_\mu$ exchange are absent because they give null
results in the Landau gauge.}.  The last two contributions, as the
master formula (\ref{master}) indicates, have an overall
proportionality factor $(S-1)$. As we examine contributions from
individual KK states, we see that due to the argument of fermion
chirality, not in all diagrams do the cosine and sine mode states both
{\em simultaneously} contribute.  This accounts for a relative factor
of 2 between the two types of diagrams.  For example, in
Fig.~\ref{f:yuk}a the fermionic KK modes can only come from cosine
expansions, whereas in Fig.~\ref{f:yuk}d both cosine and sine fermion
modes contribute. This is why Fig.~\ref{f:yuk}a has a multiplicating
factor $(S-1)$, while for Fig.~\ref{f:yuk}d the factor is $2(S-1)$.
Whereever $A_5$ is involved as an internal line, the associated KK
internal fermions necessarily come from sine expansion, e.g. in
Figs.~\ref{f:yuk}g, \ref{f:yuk}h and \ref{f:yuk}i.  The above book-keeping has been done for
individual graphs and the proportionality factors have been mentioned
for each diagram in Fig.~\ref{f:yuk}. The Yukawa RG equations (beyond
the threshold $R^{-1}$) can be written as ($t = \ln E$):
\begin{equation}
16 \pi^2 \frac{dy_f}{dt} = \beta^{\rm SM}_{y_f} +   \Theta (E-\frac{1}{R}) ~ \beta^{\rm UED}_{y_f},
\label{eq:gen}
\end{equation}

\begin{center}
\begin{figure}[thb]
\hskip 2.0cm
\psfig{figure=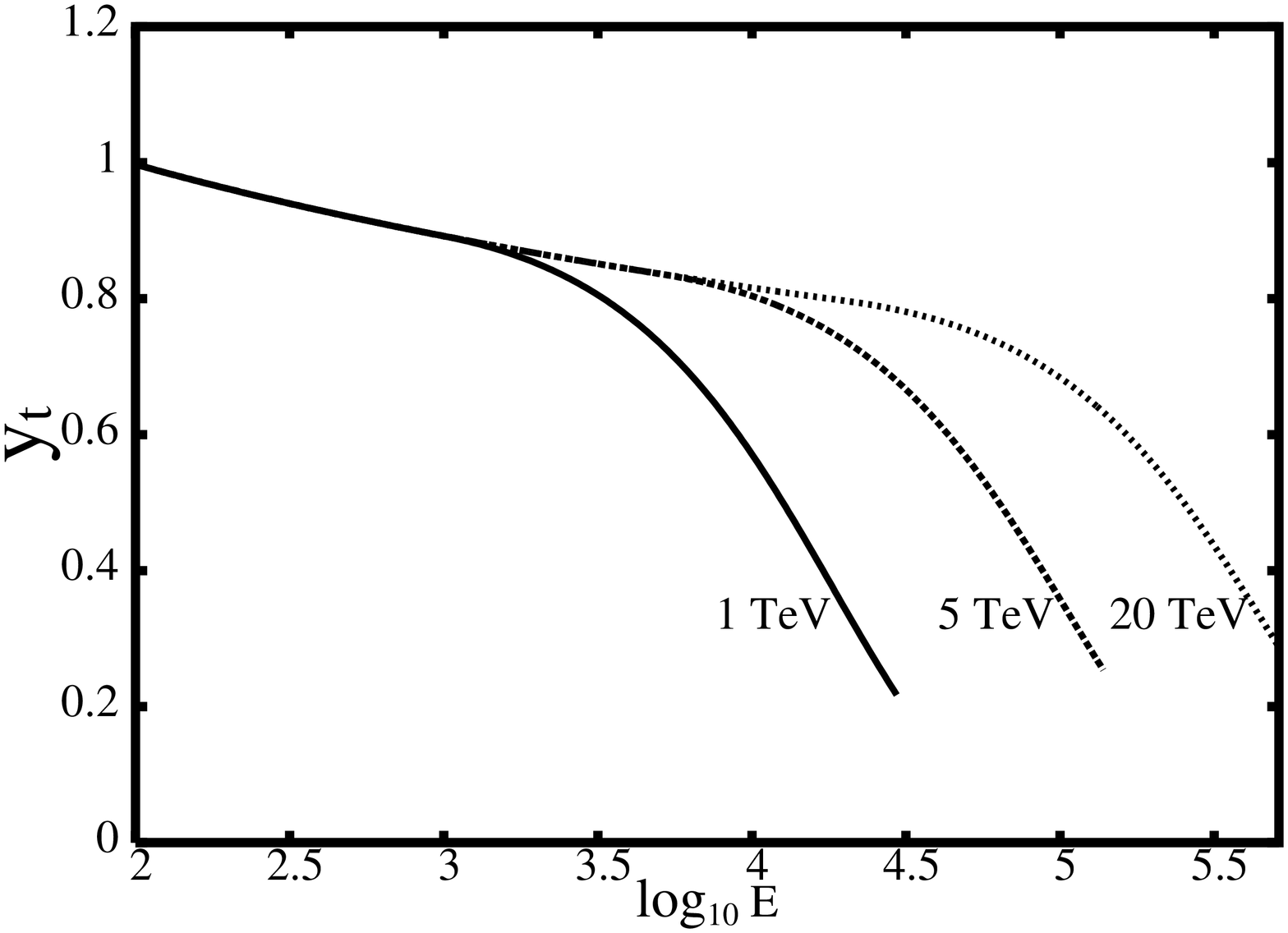,width=6.2cm,height=6.2cm,angle=270}
\vskip -6.23cm
\hskip 8.0cm
\hskip 1.50cm
\psfig{figure=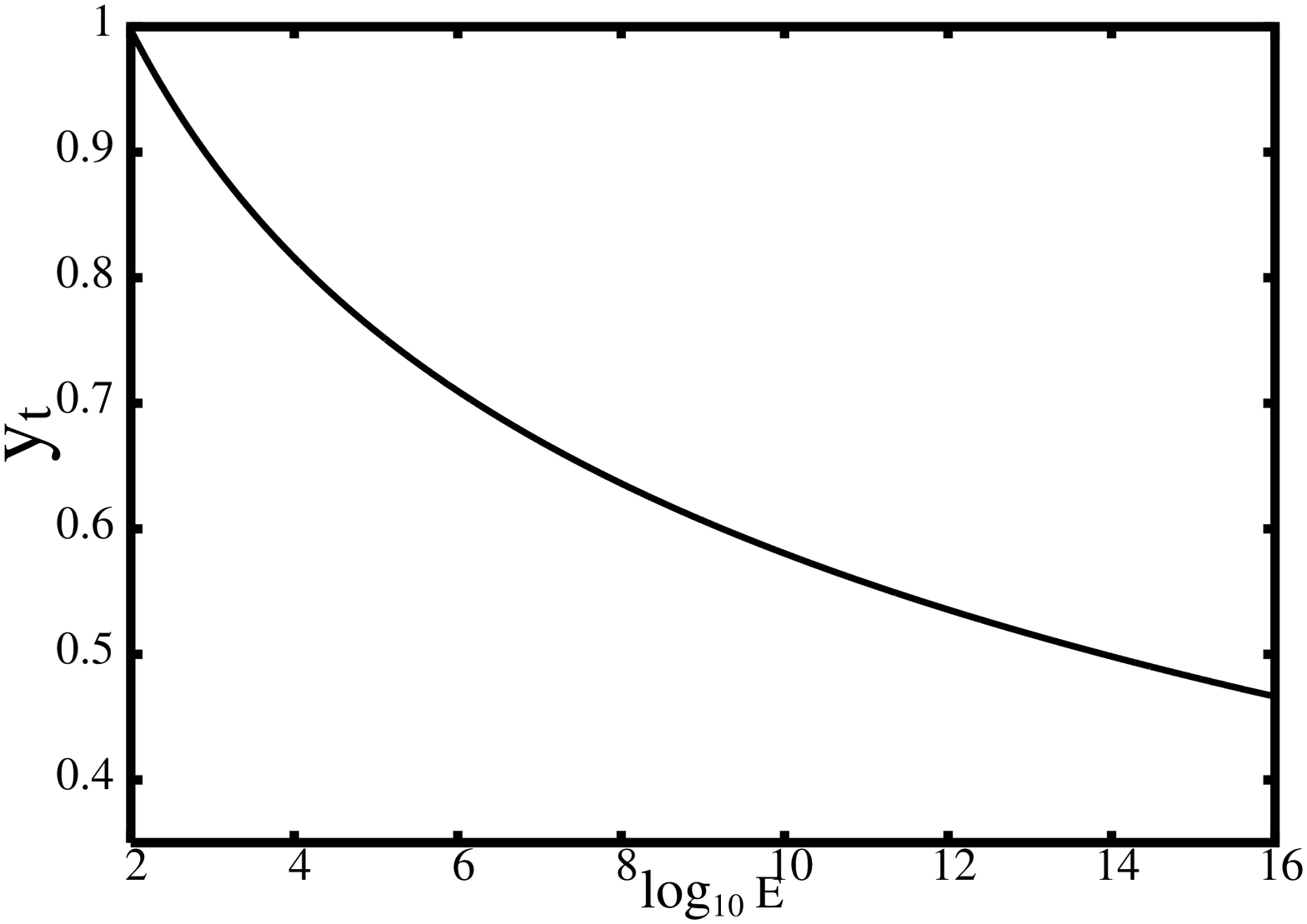,width=6.2cm,height=6.2cm,angle=270}
\caption{\sf \small {Evolution of the top quark Yukawa coupling in the UED
scenario (left panel) and (b) the SM (right panel). UED evolution is shown for
three different values of $R^{-1}$ and the curves are terminated at the
corresponding unification scales.}}
\label{f:tyuk}
\end{figure}
\end{center}

where $f$ generically stands for the up/down quarks or leptons. The SM
beta functions $\beta^{\rm SM}_{y_f}$ can be found e.g. in
\cite{sumathi}.

The UED contributions to the beta functions
$\beta^{\rm UED}_{y_{l,u,d}}$ are given by:
\vbox{
\begin{eqnarray}
\beta^{\rm UED}_{y_{l}} &=& (S -
1)\;\left[-(\frac{21}{8}g_2^2 +\frac{129}{40}g_1^2 )
  +\frac{3}{2} y_l^2 \right]\;y_l + 2(S -
1)\;\left[Y_l + 3 Y_u + 3 Y_d \right]\;y_l ,\nonumber \\
\beta^{\rm UED}_{y_{u}} &=& (S - 1)\;\left[-(12 g_3^2 +
  \frac{21}{8}g_2^2 +\frac{9}{8}g_1^2 ) + \frac{3}{2}(y_u^2 -
  y_d^2)\right]\;y_u + 2(S -
1)\;\left[Y_l + 3 Y_u + 3 Y_d \right]\;y_u ,   \\
\beta^{\rm UED}_{y_{d}} &=& (S - 1)\;\left[-(12 g_3^2 +
  \frac{21}{8}g_2^2 + \frac{9}{40}g_1^2 ) + \frac{3}{2}(y_d^2 -
  y_u^2)\right]\;y_d + 2(S - 1)\;\left[Y_l + 3 Y_u + 3 Y_d \right]\;
y_d , \nonumber
\end{eqnarray}
}
with $Y_l = \sum_{l} y_l^2$, $Y_d = \sum_{d} y_d^2$, and $Y_u =
\sum_{u} y_u^2$. 
To illustrate how the power law dependence of Yukawa couplings quantitatively
compares and contrasts with their 4-dimensional logarithmic running, we have
exhibited in Fig.~\ref{f:tyuk} the behaviour of the top-quark Yukawa
coupling in the two cases. 

Another consequence of unification in many models is a prediction
of the low energy value of $m_b/m_\tau$. This ratio, unity at
the unification scale, at low energies takes the values 4.7, 4.2,
and 3.9 for $1/R$ = 1, 5, and 20 TeV, respectively. Admittedly,  $m_b$ is 
on the high side; a limitation which perhaps may be attributable
to the one-loop level of the calculation.

\subsection{Quartic scalar coupling and the Higgs mass}
The one-loop diagrams through which the KK modes contribute to the power law
running of the quartic scalar coupling $\lambda$ (in Landau gauge) are shown
in Fig.~\ref{f:lamb}.  As clarified before in the case of Yukawa running, the
extra factor of 2 in front of $(S - 1)$ for some graphs indicates that
cosine and sine KK modes {\em both} contribute only to those graphs. The
evolution equation can be written as
\begin{equation} 
16 \pi^2 \frac{d\lambda}{dt} = \beta^{\rm SM}_{\lambda} +
 \Theta (E-\frac{1}{R}) ~\beta^{\rm UED}_{\lambda} 
\end{equation}

The expressions for $\beta^{\rm SM}_{\lambda}$ can be found e.g.~in
\cite{chengli}.  The UED beta functions are given by
\vbox{
\begin{eqnarray}
\beta^{\rm UED}_{\lambda} &=&
(S - 1)\left[ 3 g_2^4 + \frac{6}{5}g_2^2 g_1^2 + \frac{9}{25}
g_1^4 -3 \lambda (3 g_2^2 +
\frac{3}{5}g_1^2) + 12 \lambda^2 \right] \nonumber \\
&+& 2(S - 1)\left[  4 \left(Y_l + 3 Y_u +
    3 Y_d \right) \lambda -4\sum_{l,u,d}\left(y_l^4 + 3 y_u^4 + 3 y_d^4
  \right) \right].  
\end{eqnarray}
}

\begin{center}
\begin{figure}[thb]
\hspace*{3cm}\psfig{figure=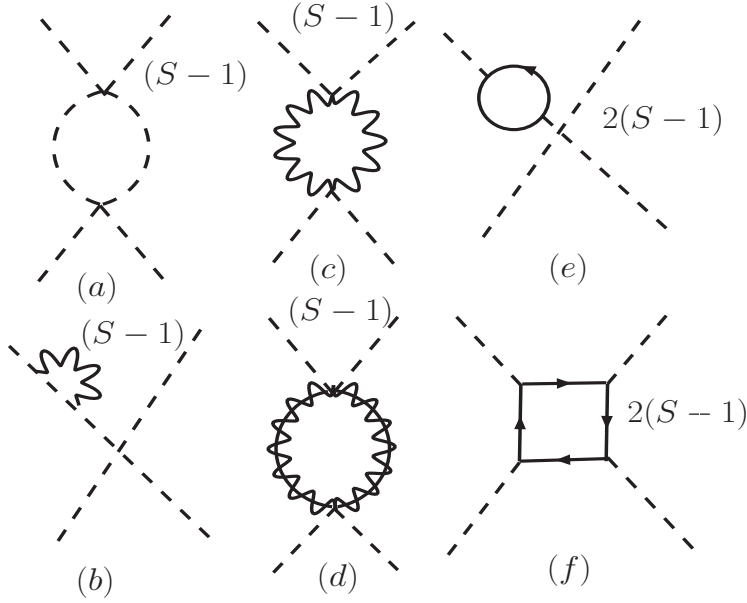,width=10cm,height=8cm,angle=0}
\caption{\sf \small {Diagrams contributing to quartic scalar
coupling evolution. The conventions are the same as in Fig. \ref{f:yuk}.}}
\label{f:lamb}
\end{figure}
\end{center}

\begin{center}
\begin{figure}[thb]
\hspace*{2cm}\psfig{figure=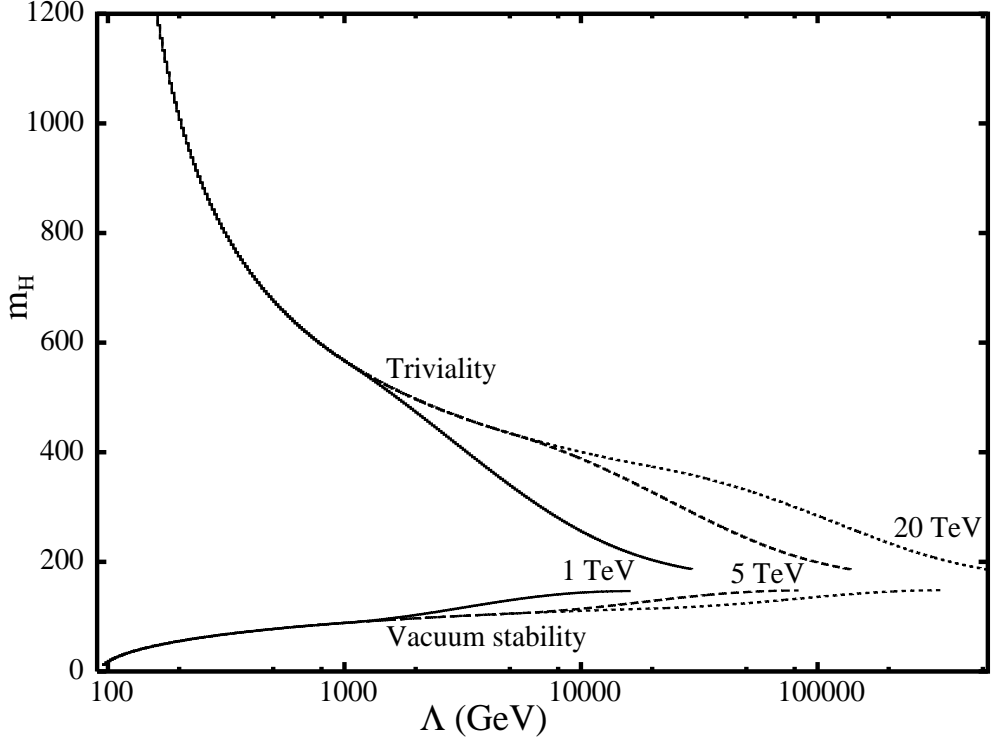,width=14cm,height=10cm,angle=270}
\caption{\sf \small {Bounds on the Higgs mass at the electroweak scale in the
UED scenario from the triviality and vacuum stability conditions for $R^{-1}$
= 1, 5, and 20 TeV. See text for details.}}
\label{f:higgs}
\end{figure}
\end{center}

 The evolution of $\lambda$ has interesting bearings on the Higgs mass. In
the standard 4-dimensional context, bounds on the Higgs mass have been placed
on the grounds of `triviality' and `vacuum stability' \cite{trivia}. What do
they imply in the UED context? The `triviality' argument requires that
$\lambda$ stays away from the Landau pole, i.e. remains finite, all the way to
the cutoff scale $\Lambda$. The condition that $1/\lambda(\Lambda) > 0$ can be
translated to an upper bound on the Higgs mass ($m_H$) at the electroweak
scale when the cutoff of the theory is $\Lambda$. This has been plotted in
Fig.~\ref{f:higgs} (the upper curves) for three different values of $R$.  A
given point on that curve (for a given $R$) corresponds to a maximum allowed
$m_H$ at the weak scale; for a larger $m_H$ the coupling $\lambda$ becomes
infinite at some scale less than $\Lambda$ and the theory ceases to be
perturbative.  Clearly, this $m_H^{\rm max}$ varies as we vary the cutoff
$\Lambda$ . The argument of `vacuum stability' relies on the requirement that
the scalar potential be always bounded from below, i.e. $\lambda (\Lambda) >
0$. This can be translated to a lower bound $m_H^{\rm min}$ at the weak
scale. The lower set of curves in Fig.~\ref{f:higgs} (for three values of
$R^{-1}$) represent the `vacuum stability' limits, the region below the curve
for a given $R$ being ruled out. Recalling that the cutoff is where the gauge
couplings tend to unify, we observe that the Higgs mass is limited in the
narrow zone
\begin{equation}
148 ~ \ltap~ m_H ~\ltap~ 186 ~{\rm GeV}
\label{higgsued} 
\end{equation}
in all the three cases, for a zero mode top quark mass of 174.2
GeV. Admittedly, our limits are based on one-loop corrections
only. That the upper and lower limits are insensitive to the choice of
$R$ is not difficult to understand, as what really counts is the
number of KK states, given by the product $\Lambda R$, which, as
mentioned before, is nearly constant, order $\ln (10^{15})$.  The
limits in Eq.~(\ref{higgsued}) are very close to what we obtain in the
SM at the one-loop level, namely $147 ~\ltap ~ m_H^{\rm SM} ~
\ltap 189$ GeV (see also \cite{Kielanowski:2003jg}, where one-loop SM results
have been derived\footnote{The SM two-loop limits are \cite{trivia}: $145
~\ltap ~ m_H^{\rm SM} ~ \ltap~ 168$ GeV for $m_t = 174.2$ GeV.}).

\subsection{Supersymmetric UED}  
What happens if we take the supersymmetric (SUSY) version of UED?  A
5-dimensional $N=1$ supersymmetry when perceived from a 4-dimensional context
contains two different $N=1$ multiplets forming one $N=2$ supermultiplet.  For
a comprehensive analysis, we refer the readers to \cite{dienes}.  There are
two issues that immediately concern our analysis. First, unlike in the
non-SUSY case, the Higgs scalar in a chiral multiplet will now have both even
and odd $Z_2$ modes on account of degrees of freedom counting consistent with
supersymmetry. Also, there will be two such $N=2$ chiral supermultiplets to
meet the requirement of supersymmetry. Second, in the RG evolution two energy
scales will come into play. The first of these is the supersymmetry scale,
called $M_S$, which we take to be 1 TeV. Beyond $M_S$, supersymmetric
particles get excited and their contributions must be included in the RG
evolution. The second scale is that of the compactified extra dimension $1/R$,
which we take to be larger than $M_S$.

The gauge coupling evolution must now be specified for three
different regions. The first of these is when $E < M_S$ where the
SM with the additional scalar doublet\footnote{SUSY requires two
complex scalar doublets.}  beta functions are in control. In this region:
\begin{equation}
{b}_{1o} = \frac{21}{5}, ~~ {b}_{2o} = -\frac{10}{3}, ~~ 
{b}_{3o} = -7 .
\label{smplus}
\end{equation}
Once $M_S$ is crossed and up until $1/R$, we also have the superpartners of
the SM particles pitching in with their effects. The contributions of the SM
particles and their superpartners together are given by:
\begin{equation}
{b}_{1s} = \frac{33}{5}, ~~ {b}_{2s} = 1, ~~ 
{b}_{3s} = -3 .
\label{susy1}
\end{equation}
Finally, when the KK-modes are excited ($E > 1/R$) one has 
further contributions from the individual modes: 
\begin{equation}
\tilde{b}_1 = \frac{66}{5}, ~~ \tilde{b}_2 = 10, ~~ 
\tilde{b}_3 = 6 .
\label{susy2}
\end{equation}   
Thus, beyond $1/R$, the total contribution is given by 
\begin{equation} 
b_i^{\rm tot} = b_{io}  + \Theta (E-M_S) ~(b_{is} - b_{i0}) + \Theta (E-
\frac{1}{R}) ~(S - 1)  ~\tilde{b}_i, 
\end{equation}

Not unexpectedly, for the SUSY UED case, gauge unification is possible.  We
observe that the introduction of this plethora of KK excitations of the SM
particles and their superpartners radically changes the beta functions; so
much so, that the gauge couplings tend to become non-perturbative before
unification is achieved. For clarity, we make the argument more explicit
below. First, from Eqs.~(\ref{susy1}) and (\ref{susy2}) we note that the
dominance of the KK matter over the KK gauge parts is so overwhelming that the
$SU(3)$ beta function ($\tilde{b}_3$) beyond the first KK threshold ceases to be
negative any longer. The other two gauge beta functions, which were already
positive with contributions from zero mode particles plus their superpartners,
become even more positive. So the curves for all the three gauge couplings
would have the same sign slopes once the KK modes are excited. As a result,
with increasing energy the three curves for $\alpha_g^{-1}$ would dip with a
power law scaling fast into a region where the couplings themselves become too
large at the time they meet.  Therefore, in order that all of them remain
perturbative during the entire RG evolution, the onset of the KK dynamics has
to be sufficiently delayed. This requirement imposes $R^{-1}~\gtap~5.0 \times
10^{10}$ GeV.  In effect, this implies that the twin requirements of a
SUSY-UED framework as well as perturbative gauge coupling unification pushes
the detectability of the KK excitations well beyond the realm of the LHC.

\section{Conclusions and Outlook}
As the LHC is getting all set to roar in 2008, expectations are
mounting as we prepare ourselves to get a glimpse of new and
unexplored territory. New physics of different incarnations,
especially supersymmetry and/or extra dimensions, are crying out for
verification. How does the landscape beyond the electroweak scale
confront the evolution of the gauge, Yukawa and scalar quartic
couplings? Will there be a long logarithmic march through the desert
all the way to $10^{(16-17)}$ GeV, or is a power law sprint 
awaiting us with a stamp of extra dimensions? In which way does the
latter quantitatively differ from the former has been the subject of
our investigation in the present chapter. We observe the following
landmarks that characterise the extra-dimensional running:
\begin{enumerate}
\item
The orbifolding renders some subtle features to the RG running in UED.  Due to
the conservation of KK number at tree level vertices, the $Z_2$ even and odd
KK states selectively contribute to different diagrams. While some diagrams
are forbidden, there are new diagrams originating from adjoint scalar
exchanges. In the present chapter we have performed a diagram by
diagram book-keeping leading to the evolution equations.

\item Low gauge coupling unification scales can be achieved without
introducing non-perturbative gauge couplings. The unification scale depends on
$R$, and is approximately given by $\Lambda \sim (25-30)/R$.

\item The `triviality' and `vacuum stability' bounds on the Higgs mass have
  been studied in the context of power law evolution. This limits the
  Higgs mass in the range $148 ~ \ltap m_H ~\ltap 186$ GeV at the
  one-loop level. The corresponding SM limits at the one-loop level
  are not very different.

\item If low energy SUSY is realised in Nature, then the requirement of
  perturbative gauge coupling unification pushes the inverse radius of
  compactification all the way up to $\sim 10^{10}$ GeV. Thus if
  superpartners of the SM particles are observed at the LHC, the
  nearest KK states within the UED framework are predicted to lie
  beyond the boundary of any observational relevance.
\end{enumerate}

It should be admitted that even if TeV scale extra-dimensional theories are
established, the spectrum might be more complicated than what UED
predicts. The confusion is expected to clear up at least when the low-lying KK
states face appointment with destiny within the first few years of the LHC
run. Our intention in the present chapter has been to choose a simple
framework to study power law evolution.  Flat extra-dimensional models are
particularly handy as they provide equispaced KK states which allow an elegant
handling of internal KK summation in the loops. UED is an ideal
test-bed to conduct this study as it has been motivated from various angles
and subjected to different phenomenological tests.


\chapter{ Extra-dimensional relaxation of the upper limit 
of the lightest supersymmetric neutral Higgs mass}

 

 
 
 

 
\section{Introduction} 
The most general symmetries of local relativistic quantum field
theories include supersymmetry, briefly discussed in sec.-\ref{sec:susy}, 
a phenomenological version
\cite{books} of which is awaiting a final judgement within the next
few years as the LHC turns on. Indeed, one of
the most coveted targets of the LHC is to capture the Higgs boson, and
supersymmetry, admitting chiral fermions together with their scalar
partners in the same representations, tacitly provides a rationale for
treating the Higgs as an elementary object
\cite{Ellis:2007wa}. Furthermore, through the removal of the
quadratic divergence that plagues the ordinary Higgs mass,
phenomenological supersymmetry has emerged as a leading candidate of
physics beyond the standard model.  A key signature of the
minimal version of supersymmetry is that the lightest Higgs boson mass
obeys an upper bound $\sim$ 135 GeV, see sec.-\ref{NhiggsRad:mssm}, -- a 
prediction which
will be put to test during the LHC run. Now, supersymmetry is an
integral part of string theory which attempts to provide a quantum
picture of all interactions.  Since string theory is intrinsically a
higher dimensional theory, a reanalysis of some 4-dimensional (4d)
supersymmetric wisdom in the backdrop of extra dimensions might
provide important clues to our search strategies.  As we know the 
main motivation of LHC is to find the Higgs boson besides examining the new physics. 
 In this chapter we 
address the following question which we believe is extremely timely:
{\em What is the upper limit of the lightest CP-even neutral Higgs
mass if the minimal supersymmetric standard model (MSSM) is embedded
in extra dimensions} \cite{Bhattacharyya:2007te}?  We consider the 
embedding first in one and then
in two extra dimensions.

Let us first discuss why this is an important issue. Recall that MSSM
has two Higgs doublet superfields ($H_1$ and $H_2$), and supersymmetry
does not allow the scalar potential to have independent quartic
couplings. Gauge interactions generate them through supersymmetry
breaking $D$-terms and the effective quartic interactions are written
in terms of the gauge couplings.  This makes the Higgs spectrum
partially predictive, in the sense that at the tree level the lightest
neutral Higgs ($h$) weighs less than ${\rm m}_Z$ (${\rm m}_h^2 < {\rm m}_Z^2
\cos^2 2\beta$, where $\tan \beta$ is the ratio of two vacuum expectation
values (VEVs)), see sec.-\ref{Nhiggs:tree}. However, ${\rm m}_h$ receives quantum 
corrections, see sec.-\ref{NhiggsRad:mssm}, which, due
to the large top quark Yukawa coupling and for heavy stop squarks, can
become as large as $\Delta {\rm m}_h^2 \sim (3G_F {\rm m}_t^4/\sqrt{2}\pi^2) \ln
({\rm m}_{\tilde{t}}^2/{\rm m}_t^2)$, where ${\rm m}_{\tilde{t}}$ is an average stop
squark mass \cite{radcorr1,radcorr2}.  The upper limit on ${\rm m}_h$ is
then pushed to around 135 GeV for squark mass in the ${\cal{O}}$(TeV)
range. Notice that the non-observation of the Higgs boson at LEP2 has
already set a lower limit ${\rm m}_h > 114.5$ GeV
\cite{Barate:2003sz,Schael:2006cr}, which is satisfied only if a
sizable quantum correction elevates the Higgs mass beyond the tree
level upper limit of ${\rm m}_Z$. This implies (i) lower values of $\tan
\beta$, which is usually chosen in the range $1 < \tan\beta <
{\rm m}_t/{\rm m}_b$, are disfavoured, and (ii) the squark mass ${\rm m}_{\tilde{t}}$
has to be in the TeV range, which also sets the scale of a generic
soft supersymmetry breaking mass $M_S$.  The MSSM prediction of a
light Higgs is also in line with the indication coming from
electroweak precision tests that the neutral Higgs should weigh below
199 GeV\footnote{This indirect upper limit as well as the LEP2 direct
search lower limit of ${\rm m}_h >$ 114.5 GeV apply, strictly speaking, for
the SM Higgs. However, in the $`$decoupling limit' of the MSSM (large
${\rm m}_A$ leading to full-strength $ZZh$ coupling), which is the region of
interest in the present chapter, the above limits continue to hold.}
\cite{Alcaraz:2006mx}.  The so called `little hierarchy' problem then
arises out of an order of magnitude mass splitting between the Higgs
and the superparticles.

Adding a gauge singlet superfield ($N$) in the MSSM spectrum and
coupling it with $H_{1,2}$ via the superpotential $\lambda N H_1 H_2$
helps to ease the tension. Not only does this next to minimal version
of supersymmetry (the so called NMSSM \cite{Ellwanger:1999ji}) help to
address the `$\mu$ problem', it also generates a tree level quartic
coupling in the scalar potential which modifies the tree level upper
limit on ${\rm m}_h$ through ${\rm m}_h^2 < {\rm m}_Z^2 \cos^2 2\beta [1 + 2\lambda^2
\tan^2 2\beta/(g^2+g'^2)]$ (see
\cite{Drees:1988fc}). Assuming $\lambda$ to be in the perturbative regime,
i.e., $\lambda \sim g,g'$, one basically obtains a new contribution
$\sim {\rm m}_Z^2 \sin^2 2\beta$ to the tree level ${\rm m}_h^2$. This way the low
$\tan\beta$ regime can be revived. Since many supersymmetric couplings
depend on $\tan\beta$, search strategies alter in a significant way if
the disfavoured low $\tan\beta$ region is thus
resurrected\footnote{Low $\tan\beta$ is preferred by electroweak
baryogenesis as well
\cite{ewb}.}${}^,$\footnote{The constraint arising from perturbativity of
couplings can be evaded if the Higgs is charged under an
asymptotically free gauge group \cite{Batra:2003nj}.}.

In this chapter we adopt a different approach which also revives the low
$\tan\beta$ region. We stick to the MSSM particle content, but embed
it in a higher dimension compactified at the inverse TeV scale
\cite{dienes}. Although we argued in the beginning that string theory
provides a rationale for linking the two ideas, namely, supersymmetry
and extra dimension, establishing any rigourous connection between the
two at the level of phenomenological models is still a long shot. Here
we take a $`$bottom-up' approach: we first outline what has already been
studied in the phenomenological context of TeV scale extra-dimensional
scenarios, and then illustrate what we aim to achieve in this chapter.
\begin{enumerate}
\item 
As in the previous chapter, we will discuss the
physics with one extra dimension first (with inverse radius of 
compactification around a TeV) but without supersymmetry and later 
for completeness we will also glance at the scenario with two extra 
dimensions.  A typical model is the one UED scenario where all 
particles access the extra dimension. Constraint, in general, on 
this scenario, discussed in the sec-\ref{bued}, is $R^{-1}~\gtap~300$
GeV while a recent inclusive $\bar{B} \to X_s \gamma$ analysis sets a 
stronger constraint $R^{-1}~\gtap~600$ GeV \cite{Haisch:2007vb}.

\item 
Our object of interest is a supersymmetric theory (e.g. MSSM) but
embedded in a higher dimension. Here we ask the following question:
{\em What would be the shift in the Higgs mass due to radiative
effects induced by extra dimensions}? The kind of scenario, here, we 
will consider is not the exact UED scenario but a \emph{modified 
universal extra dimension} (mUED)\footnote{If all the three matter
generations are bulk fields, then the theory become non-perturbative
too soon, unless $1/R > 5.0 \times 10^{10}$ GeV \cite{blitzkrieg}.}
 where the SM bosons along with their 
superpartners access the higher dimensional bulk in addition with only 
one generation (here, the third one) of the SM fermions together with their superpartners. 
From a 4d
perspective, all the states which access the bulk will have
Kaluza-Klein (KK) towers. The zero modes, i.e., those states which do
not have any momenta along the extra coordinates, are identified with
the standard 4d MSSM spectra. Now, not only the top quark and the stop
squarks would contribute to the radiative correction to $m_h^2$, their
KK partners would do so as well. As it turns out, the radiative
correction driven by the KK states has the same sign as the one from
the zero modes. As a result, $\Delta {\rm m}_h^2$ becomes larger and thus
the upper limit on ${\rm m}_h$ is pushed to higher values beyond the usual
4d MSSM limit of around 135 GeV. As we shall see, in the absence of any
left-right scalar mixing, the new contribution coming from KK modes is
to a good approximation proportional to $R^2 ({\rm m}_{\tilde{t}}^2 -
{\rm m}_t^2)/n^2$. This fits our intuition that the KK contribution falls
with higher KK modes and vanishes both when $R \to 0$ and in the limit
of exact supersymmetry. We can interpret the result in two
ways. Either, we take large $\tan\beta$ and ${\cal{O}}$(TeV) squark
mass that yielded the 4d supersymmetry limit $\sim$ 135 GeV, in which case
the new upper limit shoots up by several tens of GeV. Or, we may admit
lower $\tan\beta$ and/or accommodate lighter zero mode squarks which were
hitherto disfavoured in the 4d context.  Either way, the Higgs
phenomenology gets an interesting twist which is intuitively
comprehensible and analytically tractable, owing largely due to the
fact that we are here dealing with only {\em one additional}
parameter, namely, the radius of compactification. Moreover, the top
quark mass which appears with fourth power in the expression of
$\Delta {\rm m}_h^2$ is now known to a precision better than ever (${\rm m}_t =
170.9 \pm 1.8$ GeV
\cite{Brubaker:2006xn}).
\end{enumerate} 

As mentioned before, we have considered the embedding of 4d supersymmetry in
one as well as two extra dimensions. We shall see that qualitatively the 
KK contributions to the radiative corrections 
of ${\rm m}_h$ from 5d and 6d theories are similar, the quantitative
estimates differ due to the different density of KK states in the
two cases. In 5d, the KK states are spaced as $n/R$ (modulo their
zero mode masses) where $n$, an integer, is the KK number,
whereas in 6d, a similar expression holds except $n^2 \Rightarrow
j^2+k^2$, where $j$ and $k$ are two different sets of KK numbers
corresponding to the two compactified directions.

Section \ref{rev:mssm} is basically a review of the standard derivation of
the upper limit of the lightest neutral Higgs in conventional 4d
MSSM in the effective potential approach. This paves the way, in
 the next section, to upgrade the above derivation for accommodating
contributions from the KK modes of the top quark and squarks in
5d and 6d scenarios. In section \ref{result:nhiggs}, we shall comment on the
numerical impact of the higher KK modes on the lightest neutral
Higgs mass and its consequences. We shall draw our conclusion in
the final section of the chapter.

\section{MSSM and the neutral Higgs spectrum}\label{rev:mssm}
\subsection{Tree level mass relations}\label{Nhiggs:tree}
Let us start with the discussion what we started in sec.-\ref{sec:susy} that
in supersymmetry we have two complex scalar doublets as
\begin{eqnarray}
&& {H_1} =\left(\begin{array}{c} {H_1}^0\\ {H_1}^- \end{array}\right),
~~~~~ {H_2}=\left(\begin{array}{c} {H_2}^+\\ {H_2}^0
\end{array}\right), \label{higgs}
\end{eqnarray}
whose $SU(2)\times U(1)$ quantum numbers are (2,$-$1) and (2,$+$1)
respectively. $H_1^0$ couples with down-type quarks and charged leptons, while
$H_2^0$ couples with up-type quarks. This guarantees natural suppression of
flavour-changing neutral currents in the limit of exact supersymmetry. Now, out of the eight degrees of freedom contained in the two Higgs
doublets three are absorbed as the longitudinal modes of the $W$ and the
$Z$ bosons, while the remaining five modes appear as physical states. Of
these five states, two are charged $(H^\pm)$ and three are neutral ($h, H,
A$). Our present concern is the neutral sector of which ($h,H$) are
CP-even, while $A$ is CP-odd.
  
The tree level potential involving these two doublets 
is given by 
\begin{equation}
V = m_1^2|H_1|^2 + m_2^2|H_2|^2 + m_{12}^2(H_1H_2 + {\rm h.c}) +
{1 \over 8}g_2^2(H_2^{\dagger}\sigma^a H_2 +H_1^{\dagger}\sigma^a H_1)^2 
+{1 \over 8}{g_1}^2(|H_2|^2-|H_1|^2 )^2 , \label{pot1}
\end{equation}
where ${\rm m}_1^2$, ${\rm m}_2^2$ and ${\rm m}_{12}^2$ are soft supersymmetry breaking mass
parameters, $g_2$ and $g_1$ are the $SU(2)$ and $U(1)$ gauge couplings, and
$\sigma^a$ $(a=1,2,3)$ are the Pauli matrices. Note that the quartic coupling
is related to the gauge couplings. The part involving the neutral fields is
given by
\begin{equation}
V_0 = {\rm m}_1^2|H_1^0|^2 + {\rm m}_2^2|H_2^0|^2 - {\rm m}_{12}^2(H_1^0 H_2^0 + {\rm h.c}) +
{1 \over 8}(g_2^2+g_1^2)(|H_1^0|^2-|H_2^0|^2)^2 . \label{pot2}
\end{equation}
After spontaneous symmetry breaking the minimum of $V_0$ involves the
following two VEVs: $\langle H_1^0\rangle = v_1$ and $\langle
H_2^0\rangle = v_2$.  The combination $v = \sqrt{v_1^2+v_2^2} =
(\sqrt{2} G_F)^{-1/2} \simeq 246$ GeV sets the Fermi scale. Let us
define shifted neutral Higgs fields as
\ba
H_1^0 \ra H_1^0 + {1\over {\sqrt{2}}} (S_1 +iP_1) \\ \nonumber
H_2^0 \ra H_2^0 + {1\over {\sqrt{2}}} (S_2 +iP_2) 
\ea
The mass matrix square for the CP-odd sector is given by

\begin{eqnarray}
\cal{M}_{\rm Im}^{\rm 2}& = & {{{\P}^{\rm 2}V}\over{{\P}P_{\rm i}{\P}P_{\rm j}}}\\ \nonumber
                & = & \left(\begin{array}{cc} {\rm m}_1^2 + {1\over 4}(g_1^2 +g_2^2)(v_1^2 -v_2^2) & {\rm m}_{12}^2 \\ {\rm m}_{12}^2 &  {\rm m}_2^2 - {1\over 4}(g_1^2 +g_2^2)(v_1^2 -v_2^2) \end{array}\right)
 =  {\rm m}^2_{12}\left(\begin{array}{cc} {v_2\over v_1} & 1 \\  1 &  {v_1\over v_2} \end{array}\right).
 \label{imhiggs}
\end{eqnarray}

As it\footnote{In deriving the second step we have used the minimization condition of the conition of the potetial,
\be
{\P V\over{{\P} H_{\rm 1}^0}} = {\P V\over{{\P} H_{\rm 2}^0}} = 0 
\ee
at the minima of the potential \emph{i.e} at $H_1^0=v_1$ and $H_2^0=v_2$.}
 is a singular matrix\footnote{A singular matrix is matrix with \emph{zero} determinant}, 
so one of the eigenvalues will be zero which is the mass of the neutral massless Goldstone boson $G^0$ 
\be
{\rm m}_{G^0} = 0
\ee
while other one is the CP-odd Higgs boson $A$ whose mass simply is given by the $`$ trace' of
the matrix as
\begin{eqnarray}  
\label{imtreemass}
{\rm m}_A^2 & = & \frac{2 {\rm m}_{12}^2}{\sin 2\beta} ~
\end{eqnarray} 
where,
\be
{\sin 2\beta} = {{2v_1v_2}\over{v_1^2+v_2^2}}.
\ee

The CP-even neutral sector mass is
\begin{eqnarray}
\cal{M}_{\rm Re}^{\rm 2} = {{{\P}^{\rm 2}V}\over{{\P}S_{\rm i}{\P}S_{\rm j}}} = \left(\begin{array}{cc} 2{\rm m}_1^2 + {1\over 2} (g_1^2 +g_2^2)(3v_1^2 -v_2^2) & -2{\rm m}_{12}^2 - (g_1^2 +g_2^2)v_1v_2 \\ -2{\rm m}_{12}^2 -(g_1^2 +g_2^2)v_1v_2 & 2{\rm m}_1^2 + {1\over 2}(g_1^2 +g_2^2)(3v_2^2 -v_1^2)  \end{array}\right) \label{rehiggs}
\end{eqnarray}
and we get CP-even neutral Higgs masses as
\begin{eqnarray}  
\label{retreemass}
{\rm m}^2_{h,H} & = & {1\over 2}
\left[{\rm m}_A^2 + {\rm m}_Z^2 \mp \sqrt{({\rm m}_A^2 + {\rm m}_Z^2 )^2 - 
4 {\rm m}_A^2 {\rm m}_Z^2 \cos^2 2\beta} \right] , 
\end{eqnarray} 
where, by definition, $h$ is the lighter of the two CP-even Higgs. 

Using these relations we can have a useful sum rule just by adding $m_h^0$ and $m_H^0$ from 
eqn.-(\ref{retreemass})
\begin{eqnarray}
\label{sumrule}
& & {\rm m}_h^2 + {\rm m}_H^2  =  {\rm m}_A^2 + {\rm m}_Z^2.
\end{eqnarray}

On the otherhand, from eqn.(\ref{retreemass}), we have 
\begin{eqnarray}  
\label{big1}
{\rm m}^2_{H} &=& {1\over 2}\left[({\rm m}_A^2 + {\rm m}_Z^2) 
+({\rm m}_A^2 + {\rm m}_Z^2) \left\{1 - {{4{\rm m}_A^2{\rm m}_Z^2\cos^2 2\beta}\over{({\rm m}_A^2 + {\rm m}_Z^2)^2}}\right\}^{1\over 2} \right] ,\\ \nonumber 
&=& {\rm m}_A^2 + {\rm m}_Z^2 - {{{\rm m}_A^2{\rm m}_Z^2\cos^2 2\beta}\over{{\rm m}_A^2 + {\rm m}_Z^2}}.
\end{eqnarray}

Thus we have,
\begin{eqnarray}  
\label{big2}
{\rm m}^2_{H} = {\rm m}_A^2 + {\rm m}_Z^2 
\left\{ 1 -  {{\cos^2 2\beta} \over {1+{{\rm m}_Z^2\over{{\rm m}_A^2}}}}\right\}, ~~{\rm implies,}
~~ {\rm m}^2_{H} >{\rm m}^2_A 
\end{eqnarray}
or,
\begin{eqnarray}  
\label{big2}
{\rm m}^2_{H} = {\rm m}_Z^2 + {\rm m}_A^2 
\left\{ 1 -  {{\cos^2 2\beta} \over {1+{{\rm m}_A^2\over{{\rm m}_Z^2}}}}\right\}, ~~{\rm implies,}
~~ {\rm m}^2_{H} >{\rm m}^2_Z. 
\end{eqnarray}

Again, just by multiplying $m_h^0$ with $m_H^0$ from eqn.-(\ref{retreemass}) we have,

\begin{eqnarray}  
\label{big3}
{\rm m}^2_{h}{\rm m}^2_{H} = {\rm m}_Z^2{\rm m}_A^2{\cos^2 2\beta}.
\end{eqnarray}

So, we have the relation,

\begin{eqnarray}  
\label{big3}
{\rm m}^2_{h} = {{\rm m}_A^2\over {\rm m}^2_{H}}{\rm m}_Z^2{\cos^2 2\beta}~~{\rm implies,}~~
 {\rm m}^2_{h} < {\rm m}_Z^2{\cos^2 2\beta} < {\rm m}_Z^2, ~~ {\rm since,}~~{\rm m}^2_{H} >{\rm m}^2_A;  
\end{eqnarray}
and,
\begin{eqnarray}  
\label{big3}
{\rm m}^2_{h} = {{\rm m}_Z^2\over {\rm m}^2_{H}}{\rm m}_A^2{\cos^2 2\beta}~~{\rm implies,}~~
 {\rm m}^2_{h} < {\rm m}_A^2{\cos^2 2\beta} < {\rm m}_A^2, ~~ {\rm since,}~~{\rm m}^2_{H} >{\rm m}^2_Z.  
\end{eqnarray}

Thus, finally, in the tree level we have the inequality
\begin{eqnarray}
\label{inequal} 
& & {\rm m}_h  \leq  {\rm min}~({\rm m}_A, {\rm m}_Z) |\cos 2\beta| \leq {\rm min}~({\rm m}_A, {\rm m}_Z) , 
\end{eqnarray}
i.e., at the tree level (i) the lighter of the two CP-even Higgs ($h$) weighs
less than ${\rm m}_Z$, and (ii) the CP-odd Higgs ($A$) is heavier than $h$ but
lighter than $H$.

\subsection{Radiative corrections}\label{NhiggsRad:mssm}
We shall now discuss how the above tree level relations are affected
by quantum loops \cite{radcorr1,radcorr2}. We shall confine our
discussion on the correction to ${\rm m}_h$ only, and that too at the
one-loop level. We note two important points:
\begin{enumerate}
\item Radiative corrections to ${\rm m}_h$ are dominated by the top quark Yukawa
coupling ($h_t$) and the masses of the stop squarks ($\tilde{t}_1$,
$\tilde{t}_2$). For large values of $\tan \beta$, the
contributions from the $b$-quark sector also assume significance.
We shall ignore loop contributions mediated by lighter quarks
or the gauge bosons.
\item The tree level Higgs mass is protected by supersymmetry. In the limit of
exact supersymmetry, the entire quantum correction vanishes. So
radiative corrections to ${\rm m}_h$ will be controlled by $M_S$.
\end{enumerate}

Three different approaches have been adopted in the literature to
calculate the radiative corrections to ${\rm m}_h$: (i) effective potential
technique, (ii) direct diagrammatic calculations, and (iii)
renormalisation group (RG) method, assuming $M_S\gg {\rm m}_Z$ and fixing
the quartic coupling proportional to $(g^2+g'^2)$ at that scale and
then evolving down to weak scale. In this chapter, we shall follow the
effective potential approach primarily for the sake of conveniently
including the effect of new physics later.

We first start with an RG-improved tree level potential $V_0(Q)$ which
contains running masses ${\rm m}_i^2 (Q)$ and running gauge couplings
$g_i(Q)$.  The full one-loop effective potential is now given by
\begin{equation}
V_1(Q) = V_0(Q) + \Delta V_1(Q) 
\label{epa1}, 
\end{equation}
where, in terms of the field dependent masses $M(H)$, 
\begin{equation}
\Delta V_1(Q) = {1 \over {64\pi^2}}{\rm Str} M^4(H) 
\left \{\ln{M^2(H)\over {Q^2}} - {3 \over 2} \right \}.
\label{epa2} 
\end{equation}
The $Q$-dependence of $\Delta V_1(Q)$ cancels against that of $V_0(Q)$ making
$V_1(Q)$ independent of $Q$ up to higher loop orders.  The supertrace in
Eq.~(\ref{epa2}), defined through
\begin{equation}
{\rm Str} f({\rm m}^2) = \sum_i (-1)^{2J_i} (2J_i+1) f({\rm m}_i^2), 
\end{equation}
has to be taken over all members of a supermultiplet
and where ${\rm m}_i^2 \equiv {\rm m}_i^2(H)$ is the field-dependent mass
eigenvalue of the particle $i$ with spin $J_i$. 
As an example, the contribution from the chiral multiplet containing the top
quark and squarks is given by 
\begin{equation}
\label{epatop}
\Delta V_t = {3 \over {32\pi^2}}
\left\{ {\rm m}_{\tilde t_1}^4 \left(\ln{{\rm m}_{\tilde
    t_1}^2\over{Q^2}}-{3\over2}\right)
      + {\rm m}_{\tilde t_2}^4 \left(\ln{{\rm m}_{\tilde
          t_2}^2\over{Q^2}}-{3\over2}\right)
      -2 {\rm m}_t^4 \left(\ln{{\rm m}_t^2\over{Q^2}}-{3\over2}\right)\right\}, 
\end{equation}
where the overall factor of 3 comes from colour.  Note that ${\rm m}_{\tilde{t}_i}$
and ${\rm m}_t$ in Eq.~(\ref{epatop}) are field dependent masses.  Even though $h_b
\ll h_t$, the contribution from the bottom supermultiplet turns out to be
numerically significant in the large $\tan\beta$ region.  $\Delta V_b$
can be written analogously to $\Delta V_t$ with the appropriate
replacements of top and stop masses by bottom and sbottom masses
respectively.

We now explicitly write down the field dependent mass terms. This
simply means a replacement of $v_i$ by $H_i^0$ ($i=1,2$) wherever
$v_i$ appear in the expression of masses. The field dependent top and
bottom quark masses are given by
\begin{equation} 
\label{mthmbh}
{\rm m}_t^2(H) = h_t^2 |H_2^0|^2 ~;~ {\rm m}_b^2(H) = h_b^2 |H_1^0|^2. 
\end{equation}
The field dependent stop and sbottom squark mass matrices are written as 
\begin{eqnarray}
&& {M_{\tilde t}^2} (H) =
\left(\begin{array}{cc} 
{\rm m}_Q^2 + h_t^2|H^0_2|^2 & h_t(A_t H^0_2+\mu{H_1^0}^*)\\  
h_t(A_t{H^0_2}^* +\mu H_1^0) & {\rm m}_U^2 + h_t^2|H^0_2|^2   
\end{array}\right), 
\label{tsquark}
\end{eqnarray} 
and 
\begin{eqnarray}
&& {M_{\tilde b}^2} (H) =
\left(\begin{array}{cc} 
{\rm m}_Q^2 + h_b^2|H^0_1|^2 & h_b(A_b H^0_1+\mu{H_2^0}^*)\\  
h_b(A_b{H^0_1}^* +\mu H_2^0) & {\rm m}_D^2 + h_b^2|H^0_1|^2   
\end{array}\right).  
\label{bsquark}
\end{eqnarray} 
In Eqs.~(\ref{tsquark}) and (\ref{bsquark}) ${\rm m}_Q$, ${\rm m}_U$ and ${\rm m}_D$ are
soft supersymmetry breaking masses, $A_t$ and $A_b$ are trilinear soft
supersymmetry breaking mass dimensional couplings, and $\mu$ is the
supersymmetry preserving mass dimensional parameter connecting $H_1$
and $H_2$ in the superpotential. We take both trilinear and the $\mu$
couplings to be real. We have neglected the $D$-term contributions which
are small, being proportional to gauge couplings. The squark masses
appearing in Eq.~(\ref{epatop}) are obtained from the diagonalisation
of Eq.~(\ref{tsquark}).

We now consider the radiative correction to the CP-odd scalar mass
matrix. The one-loop corrected mass matrix square, obtained by taking double
derivatives of the full potential with respect to the pseudo-scalar
excitations, can be written as
\begin{eqnarray}
{\cal{M}}^2_{({\rm odd})} = 
\left(\begin{array} {cc} {\rm tan}\beta & 1 \\ 
1 & {\rm cot}\beta \end{array} \right) ({\rm m}_{12}^2 + \Delta) .
\label{cpodd1}
\end{eqnarray}
The radiative corrections generated as a consequence of
supersymmetry breaking are contained in $\Delta = \Delta^{\rm
t} + \Delta^{\rm b}$, which is given by 
\begin{equation} 
\Delta^{\rm t(b)} = - {3\over{32\pi^2}} {h_{t(b)}^2\mu A_{t(b)}
\over\left[{{\rm m}^2_{{\tilde t_1}({\tilde b_1})} -
{\rm m}^2_{{\tilde t_2}({\tilde b_2})}}\right]}
\left[f\left({\rm m}^2_{{\tilde t_1}({\tilde b_1})}\right)-
f\left({\rm m}^2_{{\tilde t_2}({\tilde b_2})}\right)\right]
\end{equation}
where 
\begin{equation}
f({\rm m}^2) = 2{\rm m}^2\left(\ln{{\rm m}^2\over{Q^2}} - 1 \right) .
\end{equation}
The zero eigenvalue corresponds to the massless Goldstone boson which is eaten
by the $Z$ boson. The massive state is the pseudo-scalar $A$ whose radiatively
corrected mass square is given by 
\begin{equation}
{\rm m}_A^2 = \frac{2({\rm m}_{12}^2 + \Delta)} {\sin 2\beta} . 
\label{cpodd2}
\end{equation}
The $Q$-dependence of ${\rm m}_A$ cancels in Eq.~(\ref{cpodd2}) up to
one-loop order. In any case, we shall treat the radiatively corrected
${\rm m}_A$ as an input parameter.

Now we are all set to calculate the radiative corrections in the neutral
CP-even mass eigenvalues. The one-loop corrected mass matrix square is  
obtained by taking double derivatives of the full potential
with respect to the scalar excitations and is given by 
\begin{eqnarray}
\label{msqeven}
{\cal{M}}^2_{({\rm even})}  =  
\left(\begin{array} {cc} 
{\rm m}_Z^2 \cos^2 \beta + {\rm m}_A^2 \sin^2 \beta & 
-({\rm m}_A^2 + {\rm m}_Z^2) \sin\beta \cos\beta \\ 
-({\rm m}_A^2 + {\rm m}_Z^2) \sin\beta \cos\beta &  
{\rm m}_Z^2 \sin^2 \beta + {\rm m}_A^2 \cos^2 \beta
\end{array}\right) 
+ {3\over{4\pi^2 v^2}}
\left(\begin{array}{cc} \Delta_{11} & \Delta_{12} \\ 
\Delta_{12} & \Delta_{22}\end{array}\right) ,
&&  \label{cpeven}
\end{eqnarray}
where $\Delta_{ij} = \Delta^{\rm t}_{ij} + \Delta^{\rm b}_{ij}$. The
individual $\Delta_{ij}$'s are explicitly written below: 
\begin{eqnarray}
\label{delta}
\Delta_{11}^t &=& {{\rm m}_t^4\over{{\sin}^2\beta}}\left(\mu (A_t+\mu {\rm
  cot}\beta)\over{{\rm m}_{\tilde t_1}^2 - {\rm m}_{\tilde t_2}^2}\right)^2g({\rm m}_{\tilde
  t_1}^2,m_{\tilde t_2}^2) ,\nonumber \\
\Delta_{12}^t &=& {{\rm m}_t^4\over{{\sin}^2\beta}}{\mu (A_t+\mu {
\cot}\beta)\over{{\rm m}_{\tilde t_1}^2 - {\rm m}_{\tilde t_2}^2}}\left[{
\ln}{{{\rm m}_{\tilde t_1}^2}\over{{\rm m}_{\tilde t_2}^2}}+{A_t(A_t+\mu {
\cot}\beta)\over{{\rm m}_{\tilde t_1}^2 - {\rm m}_{\tilde t_2}^2}} g({\rm m}_{\tilde
t_1}^2,{\rm m}_{\tilde t_2}^2)\right] ,\nonumber\\
\Delta_{22}^t& =& {{\rm m}_t^4\over{\sin^2\beta}}\left[\ln{{{\rm m}_{\tilde
t_1}^2}{{\rm m}_{\tilde t_2}^2}\over{{\rm m}_t^4}} + {2A_t(A_t+\mu {
\cot}\beta)\over{{\rm m}_{\tilde t_1}^2 -{\rm m}_{\tilde t_2}^2}} {\ln}{{{\rm m}_{\tilde
t_1}^2}\over{{\rm m}_{\tilde t_2}^2}} + 
\left(A_t(A_t+\mu {\cot}\beta)\over{{\rm m}_{\tilde t_1}^2 -
{\rm m}_{\tilde t_2}^2}\right)^2g({\rm m}_{\tilde t_1}^2,{\rm m}_{\tilde t_2}^2)\right] , 
\nonumber\\
\Delta_{11}^b& =& {{\rm m}_b^4\over{{\cos}^2\beta}}\left[{\ln}{{{\rm m}_{\tilde
b_1}^2}{{\rm m}_{\tilde b_2}^2}\over{{\rm m}_b^4}} + {2A_b(A_b+\mu {
\tan}\beta)\over{{\rm m}_{\tilde b_1}^2 -{\rm m}_{\tilde b_2}^2}} {\ln}{{{\rm m}_{\tilde
b_1}^2}\over {\rm m}_{\tilde b_2}^2} + 
\left(A_b(A_b+\mu \tan\beta)\over{{\rm m}_{\tilde b_1}^2 -
{\rm m}_{\tilde b_2}^2}\right)^2g({\rm m}_{\tilde b_1}^2,{\rm m}_{\tilde b_2}^2)\right] ,
\nonumber \\
\Delta_{12}^b &=& {{\rm m}_b^4\over{{\cos}^2\beta}}{\mu (A_b+\mu {
\tan}\beta)\over{{\rm m}_{\tilde b_1}^2 - {\rm m}_{\tilde b_2}^2}}\left[{
\ln}{{{\rm m}_{\tilde b_1}^2}\over{{\rm m}_{\tilde b_2}^2}}+{A_b(A_b+\mu {
\tan}\beta)\over{{\rm m}_{\tilde b_1}^2 - {\rm m}_{\tilde b_2}^2}} g({\rm m}_{\tilde
b_1}^2,{\rm m}_{\tilde b_2}^2)\right] , \\
\Delta_{22}^b &=& {{\rm m}_b^4\over{{\cos}^2\beta}}\left(\mu (A_b+\mu {
\tan}\beta)\over{{\rm m}_{\tilde b_1}^2 - {\rm m}_{\tilde b_2}^2}\right)^2 g({\rm m}_{\tilde
b_1}^2,{\rm m}_{\tilde b_2}^2) .\nonumber 
\end{eqnarray}
where  
\begin{equation}
g({\rm m}_1^2,{\rm m}_2^2)=2 - {{{\rm m}_1^2+{\rm m}_2^2}\over{{\rm m}_1^2-{\rm m}_2^2}}\ln{{\rm m}_1^2\over {\rm m}_2^2} .
\end{equation}
Two points deserve mention at this stage: 
\begin{enumerate}
\item While the leading log contribution appears in $\Delta_{22}$ for the top
sector, the same appears in $\Delta_{11}$ for the bottom sector. This
happens because the right-handed top quark couples to $H_2$ while the
right-handed bottom quark couples to $H_1$. In the absence of any
left-right scalar mixing, these leading logs are the only radiative
contributions.
\item Ignoring the left-right scalar mixing, the radiative shift to the Higgs
mass square coming from the top-stop sector turns out to be $\Delta
{\rm m}_h^2 = (3/4\pi^2 v^2) \Delta^{t}_{22} \sin^2\beta \sim (3
{\rm m}_t^4/2\pi^2 v^2)
\ln ({\rm m}_{\tilde{t}}^2/{\rm m}_t^2)$, where ${\rm m}_{\tilde{t}} = \sqrt{{\rm m}_{\tilde{t}_1}
{\rm m}_{\tilde{t}_2}}$ is an average stop mass. This is the expression we
quoted in the Introduction.
\end{enumerate} 

\section {Radiative corrections due to extra dimensions}\label{NhiggsRad:1d}

We now discuss the supersymmetric version of the theory. A 5d $N=1$
supersymmetry from a 4d perspective appears as two $N=1$
supersymmetries forming an $N=2$ theory. For the details of the
hypermultiplet structures of this theory, we refer the readers to
\cite{dienes}. Our concern in this chapter is to calculate the
radiative contribution to ${\rm m}_h$ coming from the KK partners of
particles and superparticles. We now proceed through the
following steps.
\begin{enumerate}  
\item Let us first recall that the $N=2$ supersymmetry prohibits
any bulk Yukawa interaction involving three chiral multiplets. The
Yukawa interaction is considered to be localised at a brane, like $-
(h_{t5}/\Lambda^{3/2})\int d^4x~ \int dy~
\delta(y) \int d^2\theta~ ({\cal{H}}_2{\cal{Q}}{\cal{T}}~+~{\rm
h.c.})$, where the residual supersymmetry is that of $N=1$,
$h_{t5}$ is a dimensionless Yukawa coupling in 5d and $\Lambda$
the cutoff scale. This localisation has a consequence in the
counting of KK degrees of freedom that contribute to the Higgs
mass radiative correction. The delta function ensures that those
fields which accompany the sine function after Fourier
decomposition do not sense the Yukawa interaction.

\item As in the case of 4d (zero mode) supersymmetry, here too
the dominant effect arises solely from the third generation
quark superfields, only that now we have to include the contributions
from their KK towers. We shall continue to ignore contributions 
from the gauge interactions or those from the first two quark
families, as they are not numerically significant. As we are 
working in the mUED scenario, where only one generation accesses the bulk and 
the other two are confined to a brane, then the validity of the theory extends 
further, allowing even a perturbative gauge coupling unification, we checked,
around $E \sim 40/R$. The five-dimensional Fourier decomposition of the corresponding
superfields from the eqn.~(\ref{fourier}).

\item In our scheme $M_S$ and $R$ are independent parameters, although
we take them to be of the same order\footnote{This is in contrast to
other higher dimensional supersymmetric scenarios in which both the
superpartner masses and the scale of electroweak symmetry breaking
arising from quantum loops are set by $1/R$, where $R$ is the distance
between the brane at which top quark Yukawa coupling is localised and
the brane where supersymmetry is broken \cite{arkbar}. Higher order
finiteness of the Higgs mass, where supersymmetry is broken in the
bulk by Scherk-Schwarz boundary conditions \cite{ss}, has been
discussed in \cite{quiros}.}. Towards the end of sec-\ref{result:nhiggs}, we
briefly remark on the numerical implications of any possible connection between $M_S$ and $R$.

\item The KK equivalent of Eq.~(\ref{epatop}), which captures the KK
contribution arising from the top quark chiral hypermultiplet, is then
given by 
\begin{equation}
\label{kkepatop}
\Delta V_t^n = {3 \over {32\pi^2}}
\left[{\rm m}_{\tilde t_1^n}^4 \left(\ln{{\rm m}_{\tilde
    t_1^n}^2\over{Q^2}}-{3\over2}\right)
      + {\rm m}_{\tilde t_2^n}^4 \left(\ln{{\rm m}_{\tilde
          t_2^n}^2\over{Q^2}}-{3\over2}\right)
      -2 {\rm m}_{t^n}^4
\left(\ln{{\rm m}_{t^n}^2\over{Q^2}}-{3\over2}\right)\right] , 
\end{equation}
where the field dependent KK masses are given by ${\rm m}_{\tilde
t_{1}^n}^2 = {\rm m}_{\tilde t_{1}}^2 + n^2/R^2$, ${\rm m}_{\tilde t_{2}^n}^2
= {\rm m}_{\tilde t_{2}}^2 + n^2/R^2$, and ${\rm m}_{t^n}^2 = {\rm m}_t^2 +
n^2/R^2$. The field dependence is hidden inside the zero mode
masses, as illustrated in Eqs.~(\ref{mthmbh}), (\ref{tsquark})
and (\ref{bsquark}). The corresponding contribution triggered by
the bottom quark hypermultiplet,  $\Delta V_b^n$, can be written
{\em mutatis mutandis}.

\item We now calculate the KK loop contribution to the neutral
scalar mass matrix. The procedure will be exactly the same as
that followed  for the 4d MSSM scenario in the previous section.
Since we are going to treat the radiatively corrected physical
${\rm m}_A$ as an input parameter, we concentrate only on the CP-even
mass matrix.  We first take another look at the expressions of
the different $\Delta_{ij}$, assembled in Eq.~(\ref{delta}),
calculated in the context of the 4d MSSM. The prefactors like
${\rm m}_t^4$ or ${\rm m}_b^4$ originated by the action of double
differentiation on the field dependent squark or quark masses.
Recall that the squark and quark masses are (quadratically)
separated by the soft supersymmetry breaking mass-squares which
are {\em not} field dependent.  So, irrespective of whether we
double-differentiate the squark or quark masses we get either the
top or bottom quark Yukawa coupling\footnote{This also indicates
that by fixing the first and second generation matter superfields
at the brane we have not made any numerically serious compromise
as otherwise their contributions would have been adequately
suppressed on account of their small Yukawa couplings.}.  In the
same way, the KK mass-squares are separated from the zero mode
mass-squares by a field independent quantity $n^2/R^2$.
Therefore, the expressions for $(\Delta_{ij})^n$, the radiative
corrections from the $n$th KK level, continue to have the zero
mode quark masses ${\rm m}_t^4$ or ${\rm m}_b^4$ as prefactors, but now the
arguments of the other functions contain the corresponding KK
masses.
\end{enumerate} 

We are now all set to write down the expressions for different
$(\Delta_{ij})^n$ for $n \neq 0$. They are given by 
\begin{eqnarray}
\label{deltan}
(\Delta_{11}^t)^n &=& {{\rm m}_t^4\over{{\rm sin}^2\beta}}\left(\mu (A_t+\mu
{\rm cot}\beta)\over{{\rm m}_{\tilde t_1^n}^2 - {\rm m}_{\tilde
t_2^n}^2}\right)^2g({\rm m}_{\tilde t_1^n}^2,{\rm m}_{\tilde t_2^n}^2) ,\nonumber \\
(\Delta_{12}^t)^n &=& {{\rm m}_t^4\over{\sin^2\beta}}{\mu (A_t+\mu 
\cot\beta)\over{{\rm m}_{\tilde t_1^n}^2 - {\rm m}_{\tilde t_2^n}^2}}\left[\ln
{{{\rm m}_{\tilde t_1^n}^2}\over{{\rm m}_{\tilde t_2^n}^2}}+{A_t(A_t+\mu \cot
\beta)\over{{\rm m}_{\tilde t_1^n}^2 - {\rm m}_{\tilde t_2^n}^2}}g({\rm m}_{\tilde
t_1^n}^2,{\rm m}_{\tilde t_2^n}^2)\right] , \nonumber\\
(\Delta_{22}^t)^n& =& {{\rm m}_t^4\over{{\rm sin}^2\beta}}\left[{
\ln}{{{\rm m}_{\tilde t_1^n}^2}{{\rm m}_{\tilde t_2^n}^2}\over{{\rm m}_{t^n}^4}} +
{2A_t(A_t+\mu \cot\beta)\over{{\rm m}_{\tilde t_1^n}^2 -{\rm m}_{\tilde
t_2^n}^2}} \ln {{{\rm m}_{\tilde t_1^n}^2}\over{{\rm m}_{\tilde
t_2^n}^2}} + \left(A_t(A_t+\mu
\cot\beta)\over{{\rm m}_{\tilde t_1^n}^2 - {\rm m}_{\tilde
t_2^n}^2}\right)^2g({\rm m}_{\tilde t_1^n}^2,{\rm m}_{\tilde t_2^n}^2) \right] ,
\nonumber\\
(\Delta_{11}^b)^n& =& {{\rm m}_b^4\over{{\rm cos}^2\beta}}\left[
\ln{{{\rm m}_{\tilde b_1^n}^2}{{\rm m}_{\tilde b_2^n}^2}\over{{\rm m}_{b^n}^4}} +
{2A_b(A_b+\mu \tan\beta)\over{{\rm m}_{\tilde b_1^n}^2 -{\rm m}_{\tilde
b_2^n}^2}} \ln{{{\rm m}_{\tilde b_1^n}^2}\over{{\rm m}_{\tilde
b_2^n}^2}} + \left(A_b(A_b+\mu
\tan\beta)\over{{\rm m}_{\tilde b_1^n}^2 - {\rm m}_{\tilde
b_2^n}^2}\right)^2g({\rm m}_{\tilde b_1^n}^2,{\rm m}_{\tilde b_2^n}^2)\right] ,
\nonumber \\
(\Delta_{12}^b)^n &=& {{\rm m}_b^4\over{\cos^2\beta}}{\mu (A_b+\mu 
\tan\beta)\over{{\rm m}_{\tilde b_1^n}^2 - {\rm m}_{\tilde b_2^n}^2}}\left[
\ln{{{\rm m}_{\tilde b_1^n}^2}\over{{\rm m}_{\tilde b_2^n}^2}}+{A_b(A_b+\mu 
\tan\beta)\over{{\rm m}_{\tilde b_1^n}^2 - {\rm m}_{\tilde b_2^n}^2}}g({\rm m}_{\tilde
b_1^n}^2,{\rm m}_{\tilde b_2^n}^2)\right] ,\\
(\Delta_{22}^b)^n &=& {{\rm m}_b^4\over{\cos^2\beta}}\left(\mu (A_b+\mu
\tan\beta)\over{{\rm m}_{\tilde b_1^n}^2 - {\rm m}_{\tilde
b_2^n}^2}\right)^2g({\rm m}_{\tilde b_1^n}^2,{\rm m}_{\tilde b_2^n}^2) .\nonumber 
\end{eqnarray} 
Now we have to add the $(\Delta^t)^n$ and $(\Delta^b)^n$ matrices to
the one-loop corrected (from zero modes only) mass matrix in
Eq.~(\ref{msqeven}), sum over $n$, and then diagonalise to obtain the
eigenvalues ${\rm m}_h^2$ and ${\rm m}_H^2$. The KK radiative corrections decouple
in powers of $(R^2/n^2)$. To provide intuition to the expressions in
Eq.~(\ref{deltan}), we display below the approximate formulae for
$(\Delta^t)^n$ in leading powers of $(R^2/n^2)$:
\begin{eqnarray}
\label{apdeltan}
(\Delta_{11}^t)^n & = & - \frac{1}{6}
\left(\frac{R^4}{n^4}\right)
\frac{{\rm m}_t^4}{\sin^2\beta} \left[ \mu(A_t+\mu\cot\beta) \right]^2 ,
\nonumber \\ (\Delta_{12}^t)^n & = & \left(\frac{R^2}{n^2}\right)
\frac{{\rm m}_t^4}{\sin^2\beta} \mu(A_t+\mu\cot\beta) , \\
(\Delta_{22}^t)^n & = & \left(\frac{R^2}{n^2}\right)
\frac{{\rm m}_t^4}{\sin^2\beta} \left[({\rm m}_{\tilde t_1}^2 + {\rm m}_{\tilde t_2}^2 -
2{\rm m}_t^2) + 2 A_t(A_t+\mu\cot\beta)\right] . \nonumber 
\end{eqnarray} 
Similar expressions for $(\Delta^b)^n$ can be written, with
appropriate replacements like ${\rm m}_t \leftrightarrow {\rm m}_b$, $\cot\beta 
\leftrightarrow \tan\beta$, etc. So, in the absence of any left-right
scalar mixing, the KK contribution to $\Delta {\rm m}_h^2$ is controlled by
$R^2 ({\rm m}^2_{\tilde t} - {\rm m}_t^2)/n^2$ and its higher powers.

{\bf Six-dimensional scenario:}~ For the 6d scenario we follow the
compactification on a chiral square, as done in \cite{sixd}, which
admits zero mode chiral fermions. The two extra spatial coordinates
$(y_1,y_2)$ are compactified on a square of side length $L$, such that
$0 < y^1,~y^2 < \pi R (\equiv L)$. The boundary condition is the
identification of the two pairs of adjacent sides of the squares such
that the values of a field at two identified points differ by a phase
($\theta$). Nontrivial solutions exist when $\theta$ takes four
discrete values $(n\pi/2)$ for $n=0,1,2,3$ and the zero modes appear
when $n=0$. What matters to our calculation in this chapter is the
structure of the KK masses, a generic pattern of which is given by
\begin{equation}
\label{kksixd} 
{\rm m}_{j,k}^2 = {\rm m}_0^2 + \frac{j^2+k^2}{R^2} ,  
\end{equation} 
where $j,k$ are integers such that $j \geq 0$ and $k \geq 0$.  We
display in Table \ref{sixdm} the KK mass spectrum (neglecting the zero
mode mass ${\rm m}_0$ for simplicity of presentation while in the actual
calculation we do keep it). The formalism we developed for 5d will
simply go through for 6d. More concretely, the structure of
Eqs.~(\ref{kkepatop}) and (\ref{deltan}) would remain the same in 6d,
only that one should now read $n \Rightarrow (j,k)$.  The numerical
impact in the two cases obviously differ, as we shall witness in the
next section\footnote{Admittedly, the 6d sum is logarithmically
sensitive to the cutoff. The low-lying KK states we include reflect
the dominant contribution to the Higgs mass shift. We thank Anindya
Datta for raising the 6d divergence issue.}.
\begin{table}
\begin{center}
\begin{tabular}{|l|c|c|c|c|c|c|c|c|c|} \hline
$(j,k)$&$1,0$&(1,1)&(2,0)&(2,1) or (1,2)&(2,2)&(3,0)&(3,1) or
(1,3)&(3,2) or (2,3)&(4,0)\\ \hline ${\rm m}_{j,k}$&1&$\sqrt 2$&2&$\sqrt
5$&$2\sqrt 2$&3&$\sqrt {10}$& $\sqrt {13}$&4 \\\hline
\end{tabular}
\caption{{\sf \small{ 6d scenario mass spectrum in $(1/R)$
units, neglecting the zero mode mass}.}}
\label{sixdm}
\end{center}
\end{table}

\section{Results}\label{result:nhiggs}
In this section we explore the consequences of the
extra-dimensional contributions to the Higgs mass encoded in the
exact one-loop expressions in Eq.~(\ref{deltan}). But to start
with, to get a feel for the numerical impact of the extra
dimensions, consider the scenario pared down to its bare minimum
by assuming that left-right scalar mixing ingredients are
vanishing, i.e., $\mu = A_t = A_b = 0$. This leads to two
degenerate stop squarks: ${\rm m}_{\tilde{t}}^2 = M_S^2 + {\rm m}_t^2$. Then,
for a moderate $\tan\beta$, 
\begin{equation} 
\label{approxmh}
\Delta {\rm m}_h^2 ~(n=0) \sim \frac{3 {\rm m}_t^4 }
{2\pi^2 v^2} \ln\left(1 + \frac{M_S^2}{{\rm m}_t^2}\right); ~~~
\Delta {\rm m}_h^2 ~(n \neq 0) \sim \frac{3 {\rm m}_t^4 }{2\pi^2 v^2} 
\frac{(M_S R)^2}{n^2}   .
\end{equation}

\begin{figure}[tbh]
  \centering
  \includegraphics[width=0.9\textwidth,height=0.32\textheight]
  {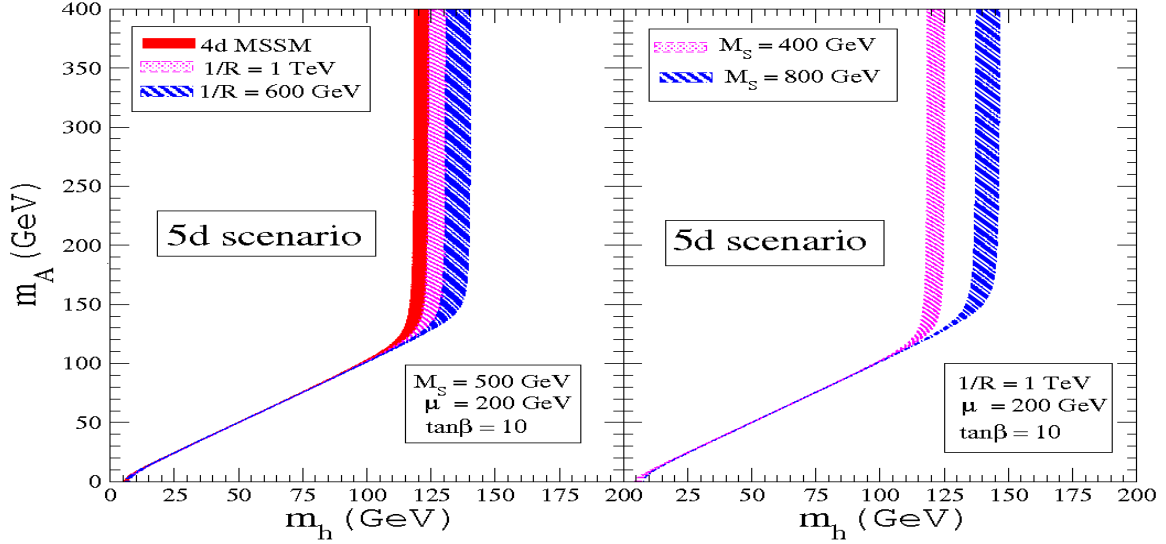} \caption{\sf \small{ The variation of ${\rm m}_h$ with ${\rm m}_A$ in the
  5d MSSM for different choices of the supersymmetry breaking scale
  ($M_S$) and the compactification radius ($R$). The width of each
  band corresponds to the variation of $A_t$ and $A_b$ in the range
  $(0.8 - 1.2)M_S$ (see text).}}
  \label{f:5dmhma}
\end{figure}
\begin{figure}[tbh]
  \centering
  \includegraphics[width=0.9\textwidth,height=0.32\textheight]
  {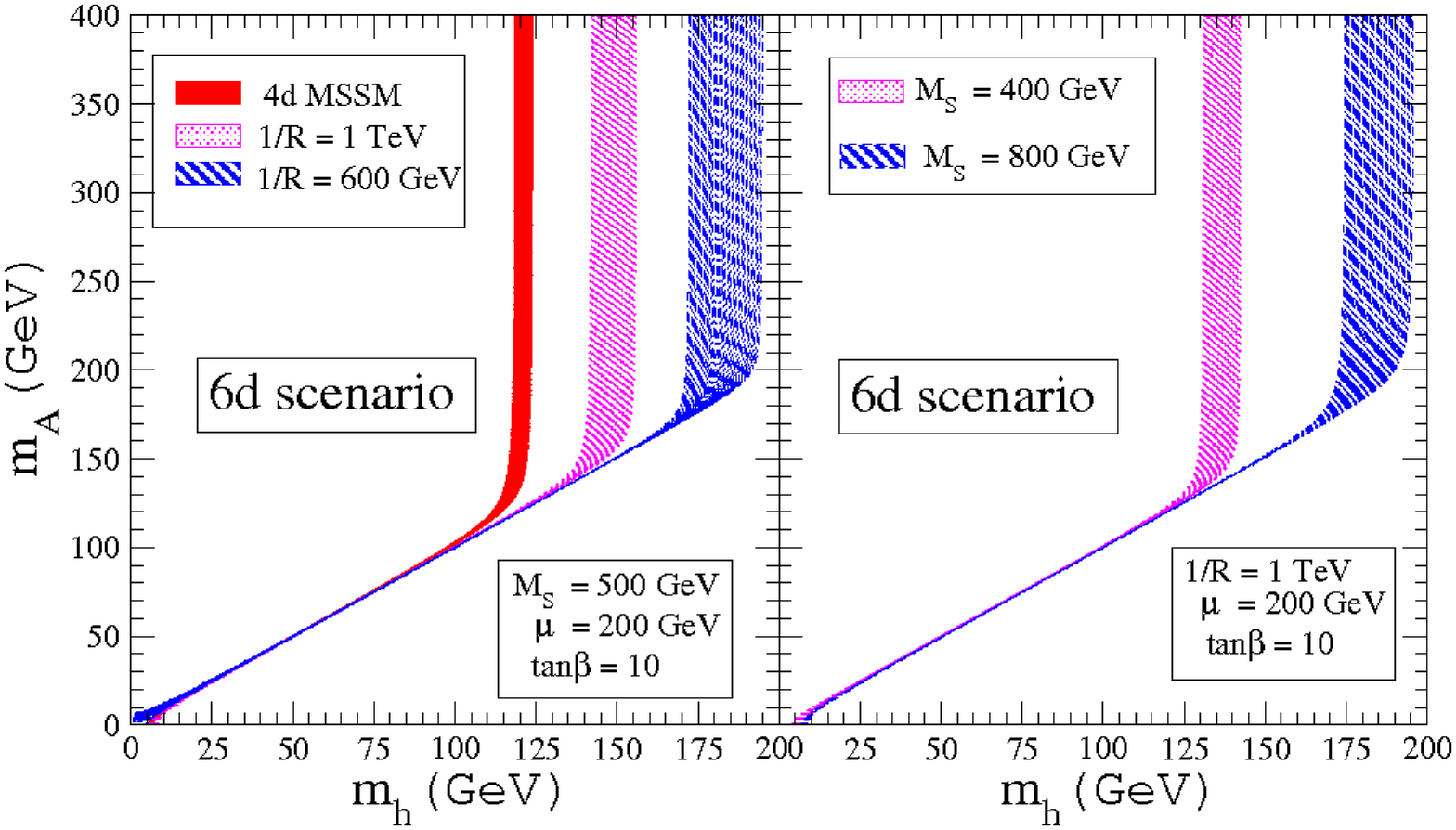}
   \caption{\sf \small {As in Fig. \ref{f:5dmhma} but for the 6d MSSM.}}
  \label{f:6dmhma}
\end{figure}


Indeed, non-zero trilinear and $\mu$ terms would complicate the
expressions, yet Eq.~(\ref{approxmh}) provides a good intuitive
feel for our results displayed through the different plots. The
expected decoupling of extra-dimensional effects in the $1/R
\rightarrow \infty$ limit is transparent in Eq. (\ref{approxmh}),
leaving the logarithmic dependence on the supersymmetry scale,
$M_S$.

As stressed already, the primary emphasis in this work is to
examine the effect of extra dimensions on the upper bound of
${\rm m}_h$. In 4d supersymmetry it is usual to choose the pseudoscalar Higgs
mass, ${\rm m}_A$, as a free parameter and exhibit ${\rm m}_h$ as its
function. This has been done for the extra-dimensional MSSM
models in Figs. \ref{f:5dmhma} (5d case) and \ref{f:6dmhma} (6d case). Let us discuss them
in turn.

In these and the subsequent figures, the parameters involved are
chosen as follows:\\
(a)~ ${\rm m}_Q = {\rm m}_U = {\rm m}_D \equiv M_S$, which is a common soft
supersymmetry breaking mass. Several values of 
$M_S$ have been chosen in the figures to depict its impact. \\
(b)~ The trilinear scalar couplings  $A_t$ and $A_b$ are varied in
the range $[0.8 - 1.2]~M_S$. This results in bands in the
figures. We have found that the results are not particularly
sensitive to $\mu$ and we hold it fixed at 200 GeV. Also, sign
flips in the trilinear couplings do not change the results.\\
(c)~ The stop and sbottom (zero mode) mass eigenvalues are
calculated from the diagonalisation of matrices in
Eqs.~(\ref{tsquark}) and (\ref{bsquark}) after setting the Higgs
fields to their VEVs. For a chosen value of $\tan\beta$ and $M_S$,
those eigenvalues will vary in a range in accord with the variation
of $A_t$ and $A_b$ stated above. \\ 
(d)~ Since we are interested in probing the upper limit of the
lightest Higgs, we maximize its {\em tree} level mass as much as
possible.  For displaying our results we have fixed $\tan\beta =
10$, a moderate value for which the tree level ${\rm m}_h$ is almost close 
to ${\rm m}_Z$.

In Fig.~\ref{f:5dmhma} we have displayed the result in the ${\rm m}_h$-${\rm m}_A$ plane
for only one extra dimension. The dependence of ${\rm m}_h$ on ${\rm m}_A$ in
the MSSM case is mimicked in the extra-dimensional case and $m_h$
settles at its upper limit for ${\rm m}_A$ greater than about 150 GeV.
In the left panel, $M_S$ has been fixed at 500 GeV.  As
anticipated, larger the value of $1/R$ smaller is the extra-
dimensional impact. The 4d MSSM case corresponds to $1/R \to
\infty$. The width of each band reflects the variation of the
trilinear parameters in the zone mentioned above.  For the chosen
supersymmetry parameters, the maximum value of ${\rm m}_h$ is a little
below 125 GeV for the 4d MSSM case while for the extra-dimensional
situation it is enhanced to above 135 (130) GeV for $1/R$ = 600
GeV (1 TeV). In the right panel, the dependence on $M_S$ is
exhibited holding $1/R$ at 1 TeV.  Clearly, a larger 
$M_S$ results in  bigger radiative corrections -- recall Eq.
(\ref{approxmh}) -- both from the zero mode as well as from the
KK modes. 

Fig.~\ref{f:6dmhma} is a 6d version of  Fig.~\ref{f:5dmhma}. While the pure 4d MSSM band
remains the same, the KK radiative effects are larger now due to
the denser KK spectrum in the 6d case, specified by two sets of
integers $j$ and $k$, as shown in Table \ref{sixdm}.  Quantitatively, for
an $1/R$ of 600 GeV (1 TeV), ${\rm m}_h$ can now be as heavy as 195
(155) GeV, to be compared with 125 GeV in 4d MSSM for these
parameter values.

\begin{figure}[tbh]
  \centering
  \includegraphics[width=0.9\textwidth,height=0.30\textheight]
  {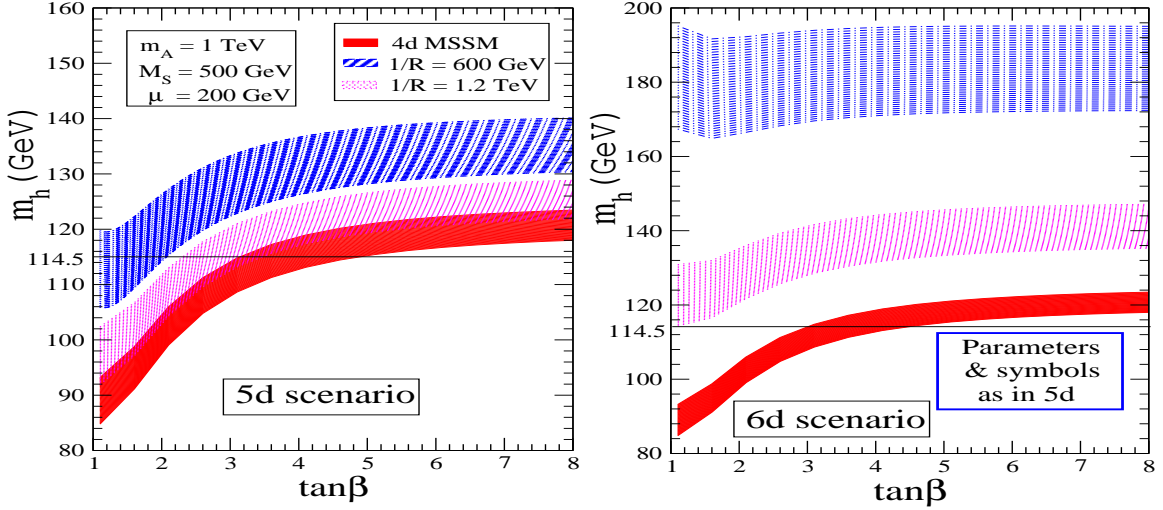} \caption{\sf\small{ The dependence of ${\rm m}_h$ on $\tan\beta$
  (zoomed for the low values) in 5d (left panel) and 6d (right panel)
  MSSM.  The width of each band corresponds to the variation of $A_t$
  and $A_b$ in the range $(0.8 - 1.2)M_S$.}}
  \label{f:tanB}
\end{figure}

As mentioned earlier, the current lower bound on ${\rm m}_h$ of 114.5
GeV excludes low values of $\tan \beta$ in the 4d MSSM. It is
expected that in the extra-dimensional scenarios some of this
excluded range of $\tan \beta$ will make it into the allowed
zone. In Fig.~\ref{f:tanB}, we have shown the variation of ${\rm m}_h$ with
respect to $\tan\beta$ (for low values) to illustrate this
effect. For the 5d case (left panel), $1/R$ of even 1.2 TeV
eases the tension somewhat while for $1/R$ of 600 GeV the effect
is very prominent.  For 6d (right panel), the extra-dimensional
contributions are further enhanced and the restriction on $\tan
\beta$ is essentially entirely lifted.  We should recall that
$\tan\beta$ enters in the Higgs couplings to other particles and
so the above result has significant bearing on collider searches
of supersymmetry.

\vspace*{1cm}
\begin{figure}[tbh]
  \centering
  \includegraphics[width=0.9\textwidth,height=0.30\textheight]
  {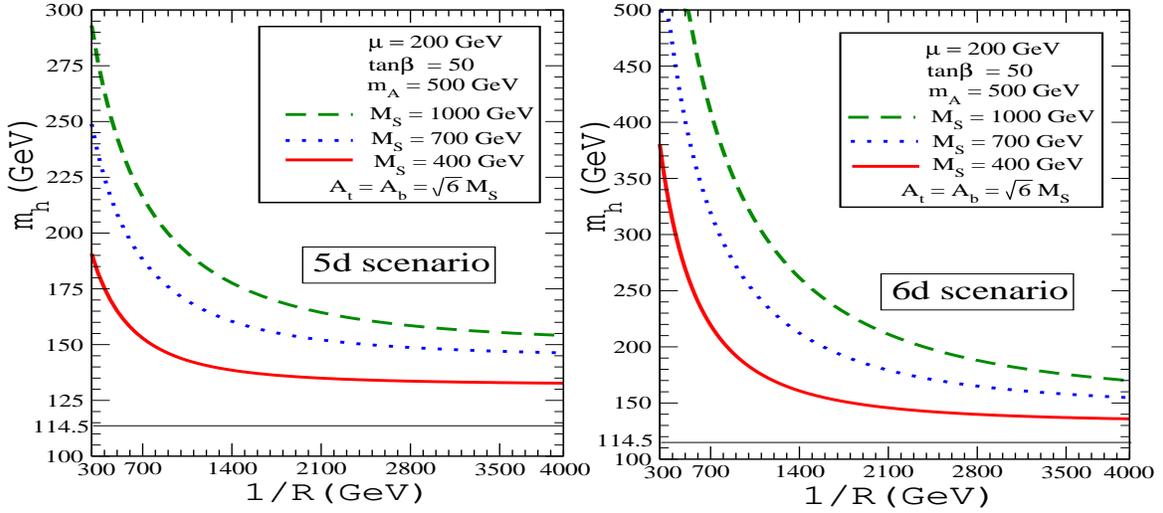} \caption{\sf \small{ The dependence of ${\rm m}_h$ on $1/R$
  for different choices of $M_S$ for 5d (left panel) and 6d (right
  panel) cases. The ratio $\sqrt{6}$ between $A_t (= A_b)$ and $M_S$
  maximises the trilinear contribution.} }
  \label{f:maxmh}
\end{figure}

So far, we have exhibited results for a few choices of the
compactification scale, $1/R$. Fig.~\ref{f:maxmh} demonstrates how the
KK-induced radiative correction depends on $1/R$ for the 5d (left
panel)  and  6d (right panel) scenarios. If the Higgs boson is
detected at the LHC then using these figures one can gain a
handle on $1/R$ dependent on the supersymmetry parameters like
$M_S$.  The decoupling behaviour as $1/R$ increases is in
agreement with expectation.

We have also studied, in passing, the possibility that the soft
supersymmetry breaking scale arises from compactification
(e.g. through the Scherk-Schwarz mechanism \cite{ss}). Let us suppose
$M_S = C/R$, where $C$ is an order one dimensionless constant. Since
we are interested in weak scale supersymmetry breaking, we keep $1/R$
around a few hundred GeV to a TeV. In this region, the radiative
correction roughly depends on $M_S$ and $R$ only through their product
($\equiv C$), and for a choice of $C \in [0.5-2.0]$, the upper limit
on the lightest Higgs mass turns out to be in the range ${\rm m}_h \in
(150-230)$ GeV (5d) and $(200-450)$ GeV (6d).

It may bear mentioning again that in these calculations we have
retained the loop conntributions from the $t$ and $b$ quarks only. The
other quarks and gauge bosons make negligible impact.  Also, we have
dealt only with real MSSM parameters and limited our studies up to
one-loop KK contributions. We have not, therefore, included either the
two-loop improvements of the 4d MSSM calculations or the numerical
effects of the phases associated with complex MSSM parameters in our
discussions (for a recent survey, see \cite{Heinemeyer:2007aq}).

\section{Conclusions}
One of the virtues for which supersymmetry stands out as a leading
candidate of physics beyond the SM is that it sets an upper bound
on the Higgs mass.
The lightest neutral Higgs mass, ${\rm m}_h$, could at most be ${\rm m}_Z$ at the
tree level, but is pushed further obeying a definite relation,
obtained from quantum corrections, involving ${\rm m}_h$, ${\rm m}_t$ and the stop
squark mass, ${\rm m}_{\tilde{t}}$. The sensitivity of this correction to
${\rm m}_{\tilde{t}}$ is only logarithmic. Consequently, a firm prediction
results, namely, that ${\rm m}_h ~\ltap ~135$ GeV in MSSM for
${\rm m}_{\tilde{t}}~\ltap ~{\cal{O}}$(1 TeV). This is regarded as a
critical test of supersymmetry and is naturally high on the
agenda of the upcoming LHC experiments. Here, we have probed
how much this upper limit could be relaxed, should the MSSM be embedded
in one ($S^1/Z_2$) or two ($T^2/Z_4$) extra dimensions. We highlight
our main findings:
\begin{enumerate}
\item The KK towers of the  top quark and stop squarks provide a
positive contribution to ${\rm m}_h^2$ raising it by several tens of GeV.
If we ignore left-right scalar mixing and assume moderate
$\tan\beta \sim (5-10)$, then using Eq.~(\ref{approxmh}) and
summing over all the KK modes, we obtain $\Delta {\rm m}_h^2 ({\rm KK})
\sim (60~{\rm GeV})^2 \times (M_S R)^2$. This is a 5d result.
Including the left-right scalar mixings, i.e., non-zero $\mu$ and
trilinear parameters, somewhat enhances the magnitude of the
correction (see Fig.~\ref{f:5dmhma}). As in the case of 4d MSSM, here too the size
of the correction is controlled by the large top Yukawa coupling.

\item If we consider a 6d theory with two extra dimensions
compatcified on a chiral square, whose motivations have been
mentioned earlier, the correction gets sizably enhanced (see
Fig.~\ref{f:6dmhma}), compared to 5d, due to a denser packing of KK states,
which are now fixed by two independent KK numbers.

\item Non-observation of a Higgs boson weighing below 114.5 GeV
disfavours low $\tan\beta$ in 4d MSSM. Some part of this 
region can be revived by extra-dimensional embedding (see
Fig.~\ref{f:tanB}).

\item The 4d MSSM relationship between the lightest neutral Higgs mass
and the stop squark mass is extremely profound in the sense that its
specific form does not depend on the supersymmetry breaking
mechanism. If supersymmetry is embedded in extra dimension(s) and,
with some cooperation from Nature, the KK states happen to be light
enough to mark their imprints on the LHC data recorder, then the
relationship between the stop mass and the Higgs mass alters in a
numerically significant way (see Fig.~\ref{f:maxmh}).  

\end{enumerate}




\newpage
\mbox{}
\chapter{ Low intermediate scales for leptogenesis in supersymmetric
 $SO(10)$ grand unified theories}
\section{Introduction}\label{intromprs}
An area where the standard model based on the group $SU(3)_C
\times SU(2)_L \times U(1)_{Y} \equiv {\cal G}_{std}$ merits
improvement is the origin of parity violation. The most natural
extension that addresses this issue is the left-right symmetric
model in which the gauge group is enlarged to $SU(3)_C \times
SU(2)_L \times SU(2)_R \times U(1)_{(B-L)} \equiv {\cal G}_{LR}$
\cite{lr}. Here, the left-handed fermions transform nontrivially
under $SU(2)_L$ and are singlet under $SU(2)_R$, while it is the
converse for the right-handed (RH) fermions.  It is then possible to
extend the definition of parity of the Lorentz group to all
particles and ensure that the theory is invariant under the
transformation of parity. Spontaneous breaking of the group
$SU(2)_R$ would trigger violation of parity in the low energy
theory. It is also possible to break the parity symmetry
spontaneously by the vacuum expectation value of a gauge
singlet scalar field which has odd parity \cite{dpar}.  In either
case, parity violation at low energy originates from some
spontaneous symmetry breaking at high energy.

Two experimental evidences for the existence of a theory beyond the
standard model are the baryon asymmetry of the universe observed
by the Wilkinson Microwave Anisotropy Probe (WMAP) \cite{wmap} and the light neutrino
mass as noted earlier. The see-saw mechanism can explain the light neutrino 
mass while the most viable mechanism to generate the matter antimatter asymmetry 
\emph{i.e} baryon asymmetry is the baryogenesis via leptogenesis through the \emph{sphaleron} 
processes. Actually, the tiny neutrino masses could be related with leptogenesis. 
In the SM $\n$ is massless, but by the see-saw mechanism with inclusion 
of a right-handed neutrino or triplet Higgs scalar or both one can generate the 
tiny neutrino mass. 
The lepton number violating decays of the right-handed neutrino or the triplet 
Higgs scalar at some large scale, on the otherhand, can generate a lepton 
asymmetry\footnote{The decay should  satisfy two necessary Sakharov conditions: 
$i)$ should have enough CP -violation and $ii)$ the decay satisfies the out-of-equilibrium 
condition.}, which is then converted into a baryon asymmetry of the universe.

As we discussed in sec.~\ref{so10gutM} the left-right symmetric 
extension of the standard model can emerge from a Grand Unified Theory based on the 
gauge group $SO(10)$.
There are two broad classes of minimal $SO(10)$ models: those
with only doublet Higgs scalars (Model I) and the conventional
left-right symmetric model including triplet Higgs scalars (Model
II). The main differences between Models I and II lie in the Higgs
scalar that breaks the left-right symmetry and the generation of
neutrino masses. Lepton number violation in these models arises
from the Higgs scalars that break the $B-L$ symmetry and hence
the left-right symmetry. The origin of leptogenesis is also
different in these two models.  There is a natural mechanism of
resonant leptogenesis in Model I (see below) while Model II has other advantages.

In Model I there is an extra singlet fermion, $S$, that combines
with the neutrinos and a new type of see-saw mechanism is
operational \cite{valle}. There are several interesting features
associated with this. The one relevant here is that the singlet
fermions can be almost degenerate with the neutrinos, leading to
resonant leptogenesis naturally in this scenario \cite{new}.
On the otherhand Model II is truly a renormalizable high scale SUSY SO(10)
theory of fermion masses and mixings.
 
From an analysis of gauge coupling unification, we have determined the 
scale of left-right symmetry breaking, which is intimately related to 
a successful prediction of leptogenesis in these models. An apparent 
obstacle arises in the following form: 
\emph{either these models do not allow any intermediate mass scales or
the intermediate left-right symmetry breaking scale comes out to
be large ($\sim 10^{15}$ GeV)}. To implement leptogenesis, on
the other hand, the left-right symmetry breaking scale has to be
much lower. We exhibit several alternate possibilities which may
provide a way out from this impasse.

The Majorana mass of right-handed neutrinos is given by $M_N \sim
\tilde{f} v_R$, where $\tilde{f}$ is the Yukawa coupling and $v_R$
is the \emph{vev} given to $\chi _R$, defined in sec. \ref{so10gutM}. The
right-handed neutrino mass-scale controls leptogenesis as well as
light neutrino masses and, in particular, a value around $10^{9}$
GeV or lower is favored by the `gravitino constraint' discussed
below.  Since $\tilde{f}$ does not affect the experimentally
measured charged fermion masses at low energies, one can assign
any value to it, leaving the left-right symmetry breaking scale
unrestricted. However, such a low RH neutrino mass is
likely to give too large contributions to the left-handed neutrino
masses through the see-saw mechanism,  contradicting experimental
observation. The main motivation of the see-saw mechanism was to
avoid arbitrarily small Yukawa couplings, so we shall assume the
value of $\tilde{f}$ to be of order unity \footnote{Here, for simplicity 
of discussion, we have considered 
$\tilde{f}$ to be multiplying a unit matrix in flavor space. The RH
neutrino masses can also be lowered through small eigenvalues, if
$\tilde{f}$ has a non-trivial matrix structure \cite{ji}}.

While considering leptogenesis in the minimal supersymmetric
$SO(10)$ GUTs, the potential problem \cite{grav} arising from the
overclosure of the universe by gravitinos (and its adverse influence on
the successful Big Bang Nucleosynthesis predictions) must be taken into
account.  This requires the reheating temperature, $T_{RH}$, to
be less than $\sim 10^8$~GeV.  Since leptogenesis takes place
just below the scale of left-right symmetry breaking, $M_R >
T_{RH}$ can make models inconsistent with the above or at least
unnatural.  However, Model I may still be consistent because it
offers the alternative of resonant leptogenesis.

Using renormalization group (RG) equations, in the following sections 
we examine for both Models whether gauge coupling unification at all 
allows a low left-right symmetry breaking scale which would make
successful leptogenesis viable. The simplicity of the minimal 
supersymmetric $SO(10)$ GUT allows several interesting predictions. 
With some standard assumptions
it is possible to determine the mass scales involved in the
symmetry breaking.  Below, we shall  show that one-loop
renormalization group evolution leads to left-right
symmetry breaking and unification scales,
\ba 
{M_R}^0 \simeq 1.3\times 10^{16}~ {\rm GeV},~~~~{M_U}^0 \simeq 2.9\times
10^{16}~ {\rm GeV} .
\ea
$M_R^0$ and $M_U^0$ are already very close. But, the situation
worsens when two-loop RG contributions are included and we find
that no intermediate scales are allowed at all below the
unification scale.  All this makes leptogenesis unnatural in this
class of models. We suggest some possible remedies \cite{Majee:2007uv}.


In this chapter we show how inclusion of GUT-threshold effects, 
gravitational corrections through $\rm {dim}.5$ operators, or presence of 
additional light fields near  $M_R$, can lower the intermediate scale, bringing it even
to the range of a few TeV in the doublet model. Thus, in this
model, the gravitino constraint can be easily satisfied
leading to successful resonant leptogenesis at low scales. In
addition, the signatures of right-handed gauge bosons,
($W_R^{\pm}$, $Z_R$), and new Higgs scalars can be tested at the LHC
and ILC.  In the triplet model, on the other hand, even though the GUT
threshold corrections are much larger, we derive a 
bound on the intermediate scale, $M_R > 10^{9}$ GeV arising out
of the requirement of perturbation theory to be valid, due to 
which the scale cannot be reduced further. With this lower bound
on $M_R$ the triplet model emerges genuinely as a high scale
supersymmetric theory for successful description of fermion
masses and mixings.

The chapter is organised in the following manner. In sec.~\ref{genform} we discuss
renormalization group equations and origins of threshold and
Planck scale effects. Discussing SUSY $SO(10)$ model and left-right symmetric
breaking in sec. \ref{minsusyso10} and \ref{lowscale} respectively, we will
show in sec.~\ref{techniq} how low intermediate scales are obtained in the 
doublet model and triplet model. The perturbative lower bound on $M_R$ 
is derived 
in sec.~\ref{sec:lb}. After making  brief remarks on fermion masses and light
scalars in the SUSY $SO(10)$ model in sec.~\ref{remarkfermion}, we summarise the
results and state our conclusions in sec.~\ref{summprs}.

\subsection{General formulation}\label{genform}
We have already discussed the renormalisation group superficially for the standard model 
and supersymmetric standard model in sec.~\ref{rgesm} and sec.~\ref{rgemssm} respectively.
Let us first gather all the RGE with two loop contributions. We have also mentioned how these
equations will change in the presence of some threshold correction or due to the Planck scale
effects\footnote{More detail will be given later in sec.~\ref{techniq}.}. The RG equations with 
one\footnote{ Here, we neglect the small logarithmic running between the 
electroweak scale ($M_Z$) and the SUSY scale ($M_S$) and, in  effect, set $M_S$ and $M_Z$ to be the same.}  intermediate scale, $M_R$, between $M_U$ and $M_Z$ are:
\ba
{1\over{\a _i}(\MZ)}&=&{1\over\a_i(\MR)}+{a_i\over
2\pi}\ln{\MR\over\MZ}+\T_i-\D_i,
\label{eqA.1}\\
{1\over\a_i(\MR)}&=&{1\over\a_i(\MU)}+{a'_i\over
2\pi}\ln{\MU\over\MR}+\T'_i-\D'_i-
\D_i^{(gr)}, \nonumber \\
\label{eqA.2}
\ea
\noindent
where $i$ runs over the different gauge couplings. Let us clarify different
notations used in the above two equations. In the R.H.S.
of eqns. (\ref{eqA.1}) and (\ref{eqA.2}), the second and third
terms represent one- and two-loop contributions, respectively,
with
\ba
\T_i &=&{1\over 4\pi}\sum_jB_{ij}\ln{\a_j(\MR)\over\a_j(\MZ)},\nonumber\\
\T'_i&=&{1\over 4\pi}\sum_jB'_{ij}\ln{\a_j(\MU)\over\a_j(\MR)},\nonumber\\
B_{ij}&=&{b_{ij}\over a_j},\, B'_{ij}={b'_{ij}\over a'_j}.\label{eqA.3}
\ea
\par\noindent
The one- and two-loop coefficients ($a_j, a'_j, b_{ij}, b'_{ij}$)
for specific scenarios are given later.   Between $M_Z$ and $M_R$
the indices $i,j \subset \std $ while above $M_R$ one has  $i,j
\subset \lr$.

 The $\D_i$ include SUSY threshold effects  and
intermediate scale threshold effects at $\MR$,
$$\D_i=\D^{(S)}_i+\D^{(R)}_i,$$ while $\D'_i$ includes the same at the unification scale $\MU$.
They are represented as \cite{carena, baer, langacker, parida1},
\ba
\D_i^{(S)} &=& \frac{1}{2\pi}\Sigma_{\a} b_i^\a
\ln{{M^{\a}}\over M_S} \equiv
\frac{b_i} {2\pi}\ln{{M_i}\over {M_S}}, \;\;\;
b_i =  \Sigma_{\a} b_i^\a,   \nonumber\\
\D_i^{(R)} &=& \frac{1}{2\pi}\Sigma_{\beta}  {c}_i^\beta
\ln{{M^{\beta}}\over M_R} \equiv
\frac{b'_i} {2\pi}\ln{{M_i}\over {M_R}}, \;\;\;
b'_i = \Sigma_{\beta}  {c}_i^\beta , \nonumber\\
\D'_i &=& \frac{1}{2\pi}\Sigma_{\gamma} {d}_i^\gamma
\ln{{M^{\gamma}}\over M_U} \equiv
\frac{b''_i} {2\pi}\ln{{M_i}\over {M_U}}, \;\;\;
b''_i = \Sigma_{\gamma} {d}_i^\gamma . ~~~~ \label{eqA.4} 
\ea

Here the indices $\a,\b$ and $\g$ signify the particle
components of $SO(10)$ representations spread around the  SUSY
scale  $M_S$, the $SU(2)_R\times U(1)_{B-L}$ breaking scale
$M_R$, and the $SO(10)$ breaking scale $M_U$, respectively.

 The definition of effective mass parameters at the SUSY
scale $M_S$ through the first of eqns. (\ref{eqA.4})  introduced by
Carena, Pokorski and Wagner~\cite{carena} has been generalised to
study GUT-threshold effects by Langacker and Polonsky
\cite{langacker} in SUSY $SU(5)$ and in ref.~\cite{parida1} to study
 intermediate breaking in SUSY $SO(10)$. The effective mass
parameters defined through these relations  are not arbitrary.
Logarithm of each of them is a well defined
linear combination of logarithms of actual  particle masses
(heavy or superheavy) spread around the respective thresholds.
Hence, in
principle, it is possible to express them
in terms of the parameters of the superpotential.~The actual
relationship would vary from model to model 
depending upon the type and number of representations used in
driving the spontaneous symmetry  breaking of SUSY $SO(10)$ to
the low energy theory.

In the absence of unnatural mass spectra,  the particles are
expected  to be a few times heavier or lighter than the
associated threshold scale which would  result in the effective
mass parameters bearing a similar relationship to that scale.\\

The term
$\D^{gr}_i$ represents the effect of $\rm {dim}.5$-operators
which may be induced at the Planck scale as \cite{shafi, parpat},
\ba 
\D_i^{(gr)} &=& -{{\epsilon}_i\over {\a}_G}, i = BL, 2L, 2R ,3C
.\label{eqA.4b}
\ea

These operators modify the boundary condition at $\MU$ as ,
\ba
&&\a_{2L}(\MU)(1+\ep_{2L})=\a_{2R}(\MU)(1+\ep_{2R})%
=\a_{BL}(\MU)(1+\ep_{BL})=\a_{3C}(\MU)(1+\ep_{3C})=\a_G.\nonumber\\
\label{eq13a}
\ea 
Here, $\a_G = g^2(M_U)/{4\pi}$ is the GUT fine-structure
constant. The impact of various contributions in eqns.
(\ref{eqA.4}) and (\ref{eqA.4b}) in lowering the intermediate
scale in SUSY $SO(10)$ GUTs will be discussed in detail in
subsequent sections.

Using eqns. (\ref{eqA.1}) -  (\ref{eq13a})
one obtains for the mass scales
\cite{parida1},
\ba
\ln{\MR\over \MZ}&=&{1\over(AB'-A'B)}[(AL_S-A'L_\T)+(A'J_2-AK_2) 
-{2\pi\over\l_G}(A\ep''-A'\ep')+(A'J_\D-AK_\D)],\nonumber\\
\label{eqA.5}\\
\ln{\MU\over\MZ}&=&{1\over (AB'-A'B)}[(B'L_\T-BL_S)+(BK_2-B'J_2)%
-{2\pi\over\l_G}(B'\ep'-B\ep'')+(BK_\D-B'J_\D)],\nonumber\\ 
\label{eqA.6}
\ea

 where
\ba L_S&=&{2\pi\over\l(\MZ)}\left(1-{8\over 3}{\l(\MZ)\over\l_S(\MZ)}\right),
\nonumber\\
L_\T&=&{2\pi\over \a(\MZ)}\left(1-{8\over 3}\sin^2\T_W(\MZ)\right),\nonumber\\
A&=&a'_{2R}+{2\over 3}a'_{BL}-{5\over 3}a'_{2L},\nonumber\\
B&=&{5\over 3}(a_Y-a_{2L})-A ,
\nonumber\\
A'&=&\left(a'_{2R}+{2\over 3}a'_{BL}+a'_{2L}-{8\over
3}a'_{3C}\right),\nonumber\\
B'&=&{5\over 3}a_Y+a_{2L}-{8\over 3}a_{3C} - A' .\label{eqA.7}
\ea

\ba
J_2&=&2\pi\left[\T'_{2R}+{2\over 3}\T'_{BL}-{5\over 3}\T'_{2L}
+{5 \over 3}(\T_{Y}-\T_{2L})\right],\nonumber\\
K_2&=&2\pi\left[\T'_{2R}+{2\over 3}\T'_{BL}+\T'_{2L}-{8\over 3}\T'_{3C}\right.
+\left.{5\over 3}\T_{Y}+\T_{2L}-{8\over 3}\T_{3C}\right],\nonumber\\
\ep'&=&\ep_{2R}+{2\over 3}\ep_{BL}-{5\over 3}\ep_{2L},\nonumber\\
\ep''&=&\ep_{2L}+\ep_{2R}+{2\over 3}\ep_{BL}-{8\over 3}\ep_{3C},\nonumber\\
J_\D&=&-2\pi\left[\D'_{2R}+{2\over 3}\D'_{BL}-{5\over 3}\D'_{2L}
+{5\over 3}(\D_{Y}-\D_{2L})\right],\nonumber\\
K_\D&=&-2\pi\left[\D'_{2R}+{2\over 3}\D'_{BL}+\D'_{2L}-{8\over 3}\D'_{3C}
\right.
\left.+{5\over 3}\D_{Y}+\D_{2L}-{8\over 3}\D_{3C}\right].\label{eqA.8}
\ea
 
\subsection{The minimal SUSY $SO(10)$ models}\label{minsusyso10}

In this subsection we apply the RG
evolution detailed above to the specific minimal $SO(10)$ models
keeping only the one- and two-loop contributions in eqns.
 (\ref{eqA.1}) -  (\ref{eqA.8}).

The symmetry breaking proceeds through three steps. These are 
\begin{itemize}
\item
In the 
first step, the $SO(10)$ symmetry is broken at $M_U$ by the $vev$ of a ${\bf
210}$ multiplet. As noted earlier, it is chosen to be along the
neutral component of $\{15,1,1\}$ under  ${\cal G}_{PS}$ which is
even under D-parity \cite{dpar}.  Thus, the gauge symmetry is
broken to ${\cal G}_{LR}$ and,  with unbroken D-parity,
left-right discrete symmetry survives preserving $g_{2L} =
g_{2R}$. 

\item
In the second step, the breaking is different for the two models.
\begin{enumerate}
\item
In Model I (the doublet model), the 
$vev$ of the neutral component of ${\ov {\chi_R}}
\subset {\bf \ov {16}}$ which transforms as $(1,1, 2,-1)$ under ${\cal
G}_{LR}$ breaks $SU(2)_R\times U(1)_{B-L} \to U(1)_Y$
at $M_R$.  The left-handed doublets $\chi_L(1,2,1,-1)\op
{\ov {\chi_L}}(1,2,1,1)$ and other components of $ \chi_R (1,1,2,
-1)\op {\ov  {\chi_R}}(1,1, 2,1)$ not absorbed by the RH gauge
bosons remain light with masses around the intermediate scale
$M_R$. 
\item
In Model II (the triplet model) the $vev$ is assigned to the neutral component of a
field $\ov{\Delta}_R \equiv$ (1,1,3,2) contained in a ${\bf
\ov{126}}$. In this alternative, the left-handed triplets
$\Delta_L(1,3,1,-2)\op {\ov {\Delta_L}}(1,3,1,2)$ contained in
the ${\bf 126}$ and ${\bf \ov{126}}$ as well as other components
of $ \Delta_R (1,1,3, -2)\op {\ov  {\Delta_R}}(1,1, 3,2)$ not
absorbed by the RH gauge bosons remain light and contribute to
the gauge coupling evolution from $M_R$.
\end{enumerate}

\item
Finally, the standard doublet Higgs contained in the
bi-doublet $\phi(1,2,2,0) \subset {\bf 10}$ drives the symmetry
breaking of ${\cal G}_{std} \to SU(3)_C \times U(1)_{em}$ at the
electroweak scale. For simplicity, in the remainder of this
section it is assumed that the supersymmetry scale, $M_S$, is the
same as $M_Z$. 

\end{itemize}

 One major difficulty in obtaining the parity conserving ${\cal G}_{LR}$
intermediate symmetry  originates from the mass 
spectra predictions in the triplet model
with  certain colored Higgs components of $G_{PS}$ multiplets in 
$\{15,3,1\}+\{15,1,3\} \subset {\bf 210}$ being at the $M_R$
scale~\cite{min3}. 
We note that a similar  difficulty also arises in the minimal doublet model 
unless these states are made superheavy through the presence of additional 
$SO(10)$ Higgs
representations or non-renormalizable terms in the superpotential
as discussed in sec.~\ref{remarkfermion}. Assuming that these additional scalars are made
superheavy, our RG analysis
applies   with the minimal particle content between $M_Z$ to $M_U$ 
as described above.

For Model I, the MSSM one- and  two-loop beta-function
coefficients below the scale $(M_R)$ are given by,
\ba
&&\left(\br{c} a_{Y}\\ a_{2L}\\ a_{3C}\er\right)=\left(\br{c} {33\over 5}\\
1\\ -3\er\right), ~~~~
b_{ij}=\left(\br{ccc} {199\over 25}&{27\over 5}&{88\over 5}\\ {9\over 5}&25&
24\\
{11\over 5}&9&14\er\right),i,j \subset \std. \label{eq10}  
\ea
Above $M_R$ till $M_U$ the beta-function coefficients are
\ba
&&\left(\br{c} a'_{BL}\\ a'_{2L}\\ a'_{2R}\\ a'_{3C}\er\right)
=\left(\br{c} 9\\ 2\\ 2\\ -3\er\right),~~
b'_{ij}=\left(\br{cccc} {23/2}&{27/2}&{27/2}&{8}\\{9/2}&{32}&{3}&
{24}\\{9/2}&{3}&{32}&{24}\\{1}&{9}&{9}&{14}\er\right), i, j \subset \lr. 
 \label{eq11}
\ea
Using  $\a_S(M_Z) = 0.1187$, $\a (M_Z) = 1/127.9$, and
$\sin^2\T_W = 0.2312$, the one-loop solutions
yield
\ba
M_R^0 = 1.3\times 10^{16}~ {\rm GeV~~~~~,} ~~~~~
M_U^0 = 2.9\times 10^{16} ~{\rm GeV}.~~~~~ \label{eq12}
\ea
The GUT fine structure constant is $\a_G \simeq 1/24.25$. When
two-loop contributions are included then, as noted earlier, no
intermediate symmetry breaking scale is permitted at all. 

For Model II, below $M_R$ the one- and two-loop beta function
coefficients are still given by eq.  (\ref{eq10})
while between $M_R$ and $M_U$ we have  
\ba
&&\left(\br{c} a'_{BL}\\ a'_{2L}\\ a'_{2R}\\ a'_{3C}\er\right)
=\left(\br{c} 24\\ 5\\ 5\\ -3\er\right),~~
b'_{ij}=\left(\br{cccc} {115}&{81}&{81}&{8}\\{27}&{73}&{3}&
{24}\\{27}&{3}&{73}&{24}\\{1}&{9}&{9}&{14}\er\right), i, j \subset \lr . \label{eq11a}\ea
In this case, the one-loop evolution results in \footnote{Here, $M_S = M_Z$ has been assumed. 
If $M_S$ is set at 1 TeV, then one
finds $M_R^0 = 5.0\times 10^{15} (1.6\times 10^{15})$  GeV and
$M_U^0 = 1.9\times 10^{16} (6.2\times 10^{15})$ GeV, at the
one-loop  level in Model I (Model II).}
\ba
M_R^0 = 7.9\times 10^{15}~ {\rm GeV~,} ~~~~~
M_U^0 = 1.9\times 10^{16} ~{\rm GeV}, \label{eq12a}
\ea
with the GUT fine structure constant $\a_G \simeq 1/24.00$.
As in Model I, inclusion of two-loop effects disallows any
intermediate scale.

We shall now turn to the implication of this high intermediate
left-right symmetry breaking in the context of neutrino masses
and leptogenesis. Then we will exhibit ways by which the 
difficulties can be evaded.

\section{Low scale left-right symmetry breaking}\label{lowscale}

As noted in the previous section, in the minimal supersymmetric
$SO(10)$ models the left-right symmetry breaking intermediate scale
cannot be lower than $10^{15}$ GeV. We shall briefly illustrate the
 application of Model II for successful explanation of fermion masses
 and mixings with such a high value of $M_R$.

In Model II, the left-right symmetry is broken by the
$vev$ of the right-handed triplet Higgs scalar $\bar {\Delta}_R \equiv
(1,1,3, 2) \subset {\bf \overline {126}}$. The left-handed  
triplet Higgs scalar $\bar {\Delta}_L \equiv
(1,3,1, 2)$ required by left-right symmetry is also present in
${\bf \overline {126}}$. The bi-doublet Higgs that breaks the
electroweak symmetry and the Higgs that breaks the $SO(10)$ group
are $\phi \equiv (1,2,2,0) \subset {\bf 10}$ and $\Phi \equiv (1,1,1,0)
\subset {\bf 210}$.
Since we are concerned with neutrino masses and leptogenesis,
consider the Yukawa interactions of the left-
and right-handed leptons:
\ba
\psi_L \equiv \left(\br{c} \nu \\ e \er\right)_L \equiv (1,2,1,-1) \subset {\bf 16},\nonumber\ea
\ba 
\psi_R \equiv \left(\br{c} \nu \\ e \er\right)_R \equiv (1,1,2,-1) \subset {\bf 16}. \label{eq13}\ea
The relevant Yukawa couplings are given by
\ba
{\cal L}_Y = f \ov \psi_L \psi_R \phi + \tilde{f} \ov {\psi^c_L} \psi_L \bar
{\Delta}_L
+ \tilde{f} \ov {\psi^c_R} \psi_R \bar {\Delta}_R .\label{eq14} \ea

In eqn.(\ref{eq14}), the field $\psi ^c$ is the charge conjugation of the field $\psi$ defined
as 
\be
\psi ^c = C \psi^*,
\ee

where, $C=i\g _2\g_0$ in the \emph{Dirac-Pauli} representation. Then the neutrino mass matrix 
can be written as
\ba
M_\nu =\left(\br{cc} \nu & \nu^c \er\right)_L\left(\br{cc} m_L & m_D \\ m_D & m_R \er\right)\left(\br{c} \nu \\ \nu^c \er\right)_L,
\label{eq15}\ea

where, $m_L = \tilde{f} \langle {\bar{\Delta}_L} \rangle;~~m_R =
\tilde{f} \langle {\bar {\Delta}_R} \rangle$ and $m_D = f \langle
\phi \rangle$. Generation indices have been suppressed. The
right-handed neutrinos then remain massive, while the left-handed
neutrino masses are see-saw suppressed
\ba
m_N &=& m_R, \nonumber \\
m_\nu &=& m_L - {m_D^2 \over m_R}. \label{eq16}
\ea

The first term $m_L = \tilde{f} v_L$ is also naturally small, since
$$v_L = \langle \bar {\Delta}_L \rangle = \kappa v^2 / v_R. $$

With supersymmetry in $SO(10)$, $\kappa$ is model dependent
and some fine-tuning of this parameter is needed in
the triplet model to achieve type II see-saw dominance, successful
prediction  of large neutrino mixings and parameterization of all
fermion masses and mixings including CP-violation \cite{min3,
min2, min1, goh}. With asymptotic parity invariance in the high
scale theory, the gravitino constraint is often ignored in the
triplet model \cite{ji}.  Moreover, the observed smallness of
neutrino masses may  work against bringing the left-right symmetry
breaking scale closer to $10^{9}$ -- $10^{10}$ GeV in the
triplet model.

In Model I, we  will explore an alternative approach where, without
fine-tuning of the Yukawa couplings of the see-saw formula, the
left-right symmetry breaking scale can be sufficiently lowered to
meet the requirements of resonant leptogenesis while satisfying
the gravitino constraint and maintaining consistency with
experimentally observed small values of neutrino masses.

As discussed in subsequent sections, both the $SO(10)$ representations
$\bf {210}$ and $\bf {54}$ are necessary to break $SO(10) \to
{\cal {G}}_{LR}$ in Model I as well as in Model II, to  prevent
certain undesirable  scalar components of $\bf {210}$ being
lighter than the GUT scale and upsetting successful gauge coupling
unification.  

In Model I, neutrino masses arise from the Yukawa Lagrangian:
\begin{equation}
{\cal L}_Y = f \ov \psi_L \psi_R \phi + y \left( \ov\psi_L S \chi_L
+ \ov\psi_R S \chi_R \right) +  M  S^T S + H.c. \label{eqdl}
\end{equation}
where ${\chi_L}(1,2,1,-1)$ and ${\chi_R}(1,1,2,-1)$ are in the
{\bf 16} dimensional Higgs representation, $\phi$ is in a {\bf
10}, and $S$ stands for $SO(10)$ singlets, of which there are three. 

The left-handed neutrinos $\nu_L$ and the right-handed 
neutrinos $N=\nu_R$ now mix with the new singlet fermions
$S$ through the mass matrix:
\ba
M_\nu =\left(\br{ccc} \nu & N^c & S \er\right)_L\left(\br{ccc} 0 & m_D & yv_L \\ m_D & 0 & yv_R \\ yv_L & yv_R & M \er\right)\left(\br{c} \nu \\ N^c \\ S \er\right)_L.
\label{eqdn}\ea

Here the Dirac neutrino mass, $m_D$, the Yukawa coupling,
$y$, and the singlet fermion mass, $M$, are $3\times 3$ matrices.
Light left-handed neutrino masses matching the experimental data
arise from this mass matrix through the double see-saw and type
III see-saw mechanisms, as has been widely discussed in the
literature \cite{valle, min4, albr}.  The model gives desired
values of neutrino masses even for low left-right symmetry
breaking scales without fine-tuning of the Yukawa couplings.

\section{Different techinques to achieve low intermediate scale}\label{techniq}

 We have advanced the following 
possibilities which may lead to left-right symmetry breaking at energies
much lower than in the the minimal models:
\begin{itemize}
\item {\sl Threshold Correction:} In the conventional analysis,
one assumes that different states within a GUT multiplet have the
same mass.  This is not exact and small splittings usually do
arise. The threshold effect due to a superheavy mass state
contributes to a small log at one-loop level; but in $SO(10)$
where big-sized representations like ${\bf 210}$ or ${\bf 126 +
\overline {126}}$ or both are used, the  one-loop contributions
by a large number of superheavy components lead to substantial
modification of the gauge couplings near the GUT scale. Both the
doublet and the triplet $SO(10)$ models belong to this
category. Thus threshold effects in each of them might
significantly change the allowed values of $M_R$ obtained from
the unification constraint.
\item {\sl Non-renormalizable interactions at the Planck
scale:} Since the unification scale is close to the scale of
quantum gravity, there may arise gauge invariant but
non-renormalizable interaction terms in the Lagrangian suppressed
by inverse powers of the Planck scale or a string
compactification scale. They affect the  gauge coupling values at
the GUT scale and change the predictions of the minimal models.
\item {\sl Additional light fields:} If there are any
additional light multiplets in the theory,
they can modify the evolution of the gauge
couplings and can allow a lowered $M_R$.
\end{itemize}
In the following, we have given details of these possibilities
and shown that with each of them it is possible to get
lower scale left-right symmetry breaking which in some cases 
could even be low enough to be within
striking range of the LHC/ILC.


\subsection{Threshold effects }\label{threshold}

  Conventionally, superheavy GUT multiplets are considered to be
degenerate. In general, however, the members of a representation
could possess somewhat different masses spread around the GUT
scale giving rise to sizable  modifications of the
gauge coupling constant predictions and the mass scales
{\em via} threshold effects \cite{weinberg,hall,ovrut}.  In the
absence of precise information of the actual values of these
masses, one may assume that all the components of a particular
submultiplet are degenerate, but different submultiplets have
masses that are spread closely around the scale of symmetry
breaking \cite{hall}.  In an alternate method, one introduces a
set of effective mass parameters to capture the threshold effects
\cite{carena}.  Such an approach has been used at the SUSY $SU(5)$
scale to examine uncertainties in the GUT model predictions
\cite{langacker}. This procedure is extended here to the $\lr$
symmetry breaking scale in the form of  eq. (\ref{eqA.4})
\cite{parida1}.

Below, we examine to what extent threshold corrections
could lower the scale of left-right symmetry breaking. We assume
all superheavy gauge bosons to possess degenerate masses
identical to the unification scale $M_U$.  \\

{\bf Model I:} For the particle content of Model I, from
eq. (\ref{eqA.7}) one obtains\\
\be A = B = 14/3, ~A' = 18, ~B' = 2,   ~AB' - A'B = -224/3 ,\label{eq17} \\
\ee
Using these, one has from eqns.  (\ref{eqA.5}),   (\ref{eqA.6}), and
 (\ref{eqA.8}) the following expressions for
threshold corrections on $M_R$ and $M_U$:\\
\ba
\D \ln{M_R\over M_Z} &=& {\pi\over 14}\left[ {10\over 3} \D'_{BL} -
8\D'_{2L} + {14 \over 3}\D'_{3C} + {25\over 3}
\D_Y - 13 \D_{2L} + {14 \over 3} \D_{3C} \right ], \nonumber \\
\D \ln{M_U\over M_Z} &=& {\pi\over 28}\left[ {4\over 3} \D'_{BL} +
8\D'_{2L} - {28 \over 3}\D'_{3C} +{10\over 3}\D_Y + 6\D_{2L} -
{28 \over 3} \D_{3C} \right ].  \label{eq18}
\ea
The quantities appearing on the RHS of eq. (\ref{eq18}) are
readily calculated using eq. (\ref{eqA.4}), given the superheavy
components
of $\bf {210}\op  {16} \op {\ov  {16}}\op
 {10}$. In this manner one gets  \cite{parida1},
\be  b''_{2L} = b''_{2R} = 53, ~~b''_{3C} = 56, ~~b''_{BL} = 50,
\label{eq19}\\
\ee
leading to
\ba
\D \ln{M_R\over M_Z} &=& {1\over 7}\left[ {125\over 3}
\ln{M_{1}\over M_U} - 106 \ln{M_{2}\over M_U} + {196 \over 3} \ln
{M_{3}\over M_U} \right], \nonumber \\
\D \ln{M_U\over M_Z} &=& {1\over 7}\left[ {25 \over 3} \ln{M_{1}\over M_U} +
53 \ln{M_{2}\over M_U} - {196 \over 3} \ln {M_{3}\over M_U}
\right]. \label{eq20}
\ea
\noindent

The pair of equations in  (\ref{eq20}) provide enough room 
to find solutions which will lead to a significant lowering of the scale $M_R$
while keeping $M_U$ within the Planck scale \footnote{One must also ensure 
that the ratios ${M_i\over M_U}, i=1,2,3$ lie within an appropriate
range, say 0.1 to 10, and ought not exceed the Planck mass.}.

As an illustration, one can consider a one parameter solution satisfying:
\begin{equation}
{M_U\over M_1} = {M_U\over M_3} = {M_2\over M_U} =\eta~. 
\end{equation}

One finds from  eq. (\ref{eq20}) 
\begin{equation}
\D \ln{M_R\over M_Z} = - 30.42 \ln \eta, \;\;\;
\D \ln{M_U\over M_Z} =  15.71\ln\eta~. \label{eq21}
\end{equation}

Note that, in the absence of threshold corrections, at the
two-loop level $\ln {M_U^0 \over M_Z} = 33.178$ and $\ln {M_R^0 \over
M_Z} = 32.916$. To ensure that $M_U \leq M_{Pl} =
1.2 \times 10^{19}$ GeV one must satisfy  
$\left(\D\ln{M_U\over M_Z}\right) \le 6.24 $. Thus, from 
eq. (\ref{eq21}) $ \eta
\le 1.48$ leading to $\left(\D\ln{M_R\over M_Z}\right) \ge
- 12.07$ implying
\be  M_R \ge  1.0 \times 10^{11} ~~{\rm {GeV}}, ~~ M_U \le 1.2 \times 10^{19}
~~{\rm {GeV}}. \label{eq22} \ee

\begin{table}[tbh]
\begin{center}
\begin{tabular}{|c|c|c|c|c|c|}\hline
$M_R$&$M_U$& ${M_1 \over M_U}$ &${M_2 \over M_U}$ &${M_3 \over
M_U}$ &${\a _{G}^{-1}}$  \\ (GeV)&(GeV) & &&& \\ \hline
$ 10^{11}$&$1.2 \times 10^{19}$
&${(1.48)}^{-1}$&1.48&${(1.48)}^{-1}$&23.7 \\ 
$10^{9}$&$10^{18}$&0.272&1.770&0.831 & 23.7 \\ 
$10^{7}$&$10^{18}$&0.158&1.950&0.832 & 23.7 \\ 
$10^{7}$&$5\times10^{16}$&0.151&2.750&1.524 & 27.7 \\ 
$10^{5}$&$5\times10^{18}$&0.180&3.30&1.076 & 26.7 \\ 
$10^{3}$&$10^{19}$&0.154&4.760&1.301 & 28.7 \\ \hline
\end{tabular}
\caption{\sf \small{Examples of low  intermediate scale, $M_R$, coupling
constant unification solutions
triggered by GUT-scale threshold effects in Model I (the doublet
model).}}
\label{table1}
\end{center}
\end{table}

This simple example implies that with one parameter $\eta$, $M_R$
lower than that given in eq. (\ref{eq22}) corresponds to
unification scales higher than the Planck mass.  Even this bound
on $M_R$ can be further lowered   by one order when smaller
threshold effects from lower scales \cite{baer, felipe} are
included leading to  $M_R \simeq 10^{10}$ GeV with near Planck
scale grand unification in the minimal doublet model. In
principle, there are three distinct mass scales $M_i, i =1,2,3,$
that enter in the threshold corrections, see  eq. (\ref{eq20}),
and there is much more flexibility to further lower $M_R$. We
return to such solutions later.

It is interesting to examine how gauge coupling constants are matched
by threshold corrections to reach their common unification value 
in spite of such substantial changes in  both the mass scales.
Using  eq. (\ref{eqA.4}) and eq. (\ref{eq19}), for $ \eta
= 1.48$
the GUT-threshold corrections for individual couplings are \cite{parida1}\\
\be
\D'_{BL} = -{25\over \pi} \ln\eta = -3.16,\;\;
\D'_{2L} = {53\over 2\pi} \ln\eta = 3.35, \;\;
\D'_{3C} = -{28 \over \pi} \ln\eta = -3.54.\label{eq23}
\ee

The gauge couplings extrapolated from $M_Z$
to $M_R =  10^{11}$ GeV are,\\
\be
\a_{BL}^{-1}(M_R) = 53.4,  ~~\a_{2L}^{-1}(M_R) = 26.3,
~~~\a _{3C}^{-1}(M_R) = 18.4.\label{eq24}
\ee

With GUT-threshold effects, the one loop-evolution of the coupling constants
 from $M_R$ to the new value of  $M_U$,
\ba
{1\over\a _i(M_U)}&=&{1\over\a _i(M_R)}-{a'_i\over 2\pi}\ln{M_U\over M_R}+\D'_i,
 ~~~~~~ i= 2L, BL, 3C.\label{eq25}
\ea

Then using eq. (\ref{eq20}) - eq. (\ref{eq24})  in eq. (\ref{eq25}),
\be
{1\over\a _{BL}(M_U)}= 23.1, \;\;\;
{1\over\a _{2L}(M_U)}= 23.5, \;\;\;
{1\over\a _{3C}(M_U)}= 23.7.\label{eq26}
\ee

The one parameter solution has the virtue of simplicity. However,
as noted earlier, in eq.  (\ref{eq20}) -- see also eq. (\ref{eqA.4})
--  three distinct mass scales $M_i, i =1,2,3,$ are, in general,  required to
capture the effect of the threshold corrections at the
unification scale. Table \ref{table1} depicts a whole set of such solutions.
For every solution, the effective mass splittings are within a
tolerable range and the unification scale has been increased by
the threshold corrections. The value of the unified gauge coupling
is also shown. 

{\bf Model II:} The threshold effect analysis for Model II  (the
triplet model) can be carried out along the same lines as in Model
I.  Thus, from eq. (\ref{eqA.7}) one finds:
\ba
 A = 38/3,~~~B = -10/3, \nonumber\\
~A' = 34,~~~~B' = -14,~~~~AB' - A'B &=& -64
.\label{eq17a} \ea
In place of eq. (\ref{eq18}) one now has
\ba
\D \ln{M_R\over M_Z} &=& {\pi\over 2}\left[ {8\over 9} \D'_{BL} -
3\D'_{2L} + {19 \over 9}\D'_{3C} \right ], \nonumber \\
\D \ln{M_U\over M_Z} &=& {\pi\over 2}\left[ {4\over 9} \D'_{BL} -
\D'_{2L} + {5 \over 9}\D'_{3C} \right ].  \label{eq18a}
\ea
The one-loop  beta-function coefficients from Model II required
for an evaluation of the RHS are:
\be  b''_{2L} = b''_{2R} = 116, ~~b''_{3C} = 122, ~~b''_{BL} = 101 .
\label{eq19a}\\
\ee
Thus, from the superheavy components of $\bf 
{210}\op  {126} \op {\ov {126}}\op  {10}$ one gets \cite{parida1}:
\ba
\D \ln{M_R\over M_Z} &=& \left[ {202 \over 9} \ln{M_{1}\over M_U} -
{87} \ln{M_{2}\over M_U} + {1159 \over 18} \ln {M_{3}\over
 M_U} \right], \nonumber \\
\D \ln{M_U\over M_Z} &=& \left[ {101 \over 9} \ln{M_{1}\over M_U}
-{29} \ln{M_{2}\over M_U} + {305 \over 18} \ln {M_{3}\over
 M_U} \right].~~~~~~\label{eq20a}
\ea
\noindent

Eqns. (\ref{eq20a}) depend, as in the case of Model I, on the three
mass scales $M_i, i=1,2,3$ which can be chosen appropriately to
ensure a solution with a low intermediate scale $M_R$. A few
typical examples are presented in Table \ref{table2}. It is noteworthy that
the gauge coupling at unification is larger for these solutions
than for the ones in Table \ref{table1}. 
\begin{table}[tbh]
\begin{center}
\begin{tabular}{|c|c|c|c|c|c|}\hline
$M_R$&$M_U$& ${M_1 \over M_U}$ &${M_2 \over M_U}$ &${M_3 \over
M_U}$ &${\a _{G}^{-1}}$  \\ (GeV)&(GeV) & &&& \\ \hline
$5\times 10^{9}$&$1.58\times10^{16}$ &2.204 &1.200 &0.659 &15.0 \\ 
$10^{10}$&$1.58\times10^{16}$ &2.065&1.160 &0.659 & 15.0 \\ 
$10^{11}$&$1.58\times10^{16}$ &1.661&1.050 &0.656 & 15.0 \\ \hline
\end{tabular}
\caption{\sf \small {Examples of low  intermediate scale, $M_R$, coupling
constant unification solutions
triggered by GUT-scale threshold effects in Model II (the triplet
model).}}
\label{table2}
\end{center}
\end{table}

  Before moving on, let us remark that in many of the threshold
effect driven solutions in Model I the unification scale is
pushed to higher values. It is well known that suppression of
Higgsino mediated supersymmetric proton decay modes like $p \to
K^+ {\ov {\nu}}$, $p \to K^0 {\mu}^+$ etc. is a generic problem
in minimal SUSY GUTs and the amplitudes are proportional to
$M_U^{-2}$.  The higher unification scales help to evade this
problem in a natural and effective fashion with a suppression
factor $({M_U^0\over M_U})^2 = 10^{-2} - 10^{-4}$.

\subsection{ Planck scale effects}\label{planck}

Since the GUT scale is close to the Planck mass, it is possible
that gravity induced non-renormalizable terms could change the
usual field theoretic predictions of gauge coupling
unification. These interactions are suppressed by inverse powers
of the Planck mass.  For example, consider the gauge invariant
 Lagrangian consisting of the $\rm {dim}.5$
non-renormalizable operators (NRO),
\ba
{\cal L}_{NRO}&=& -{\eta_1\over
2M_G}Tr\left(F_{\mu\nu}\Phi_{210}F^{\mu\nu}\right) 
-{\eta_2\over
2M_G} Tr\left(F_{\mu\nu}\Phi_{54}F^{\mu\nu}\right).
\label{eq27}
\ea
The effective gauge coupling constants at the unification point
get changed due to these non-renormalizable terms.
In particular, these interactions determine the
parameters in eq.   (\ref{eq13a}) and one finds  \cite{shafi,
parpat},
\ba\ep_{2L}&=&\ep_{2R}=-{3\over 2}\ep_2,
~\ep_{3C}= \ep_2-\ep_1,~~\ep_{BL}=2\ep_1+\ep_2,\nonumber\\
\ep'&=& {4\over 3}\ep_1 + {5 \over 3}\ep_2, ~~ \ep'' = 4\ep_1 - 5\ep_2,
\nonumber
\ea
where
\be 
\ep_1={{3\eta_1}\over 4}{M_U\over M_{G}}\left[{1\over
{4\pi\a _G}}\right]^{1\over 2},
\;\;\;
 \ep_2={{3\eta_2}\over 4}{M_U\over M_{G}}\left[{1\over
{15\pi\a _G}}\right]^{1\over 2},\label{eq28}
\ee
leading to the following analytic expressions for the corrections
on the mass scales,
\ba\left(\D\ln{M_R\over M_Z}\right)_{gr}&=&{2\pi(A'\ep'-A\ep'')\over\a _G
(AB'-A'B)}, 
= -{\pi\over {7\a _G}}\left[\ep_1 + 10\ep_2\right],\nonumber \\
\left(\D\ln{M_U\over M_Z}\right)_{gr}&=&{2\pi(B\ep''-B'\ep')\over\a _G(AB'-A'B)}
= {\pi\over{7\a _G}}\left[5\ep_2-3\ep_1\right].~~~~~~~~\label{eq29}
\ea
While the change in the mass scales are governed by the
above relations the individual coupling constants near the
GUT scale change as,\\
\be
 \D_{2L}^{\rm {(gr)}} = {3\ep_2\over 2\a _G}, \;\;\;
 \D_{BL}^{\rm {(gr)}} = -{\left(2\ep_1+\ep_2\right)\over \a _G}, \;\;\;
 \D_{3C}^{\rm {(gr)}} =  {\left(\ep_1-\ep_2\right)\over \a _G}. \label{eq30}
\ee
Using the most natural scale for the  two NRO's as the Planck
mass, $M_G = 1.2\times 10^{19}$  GeV,  and eq. (\ref{eq28}) -
eq. (\ref{eq30}) we searched for gravity corrected solutions for
low intermediate mass scale and high GUT scale with the
constraint $|\eta_{1,2}| \simeq O(1)$.

\begin{table}[tbh]
\begin{center}
\begin{tabular}{|c|c|c|c|c|}\hline
$M_R$&$M_U$&  $\eta_1$ &$\eta_2$&${\a _{G}^{-1}}$ \\
(GeV)&(GeV) & && \\ \hline
$10^{9}$&$3.16\times 10^{18}$&0.305&0.96&25.00 \\
$10^{7}$&$3.16\times 10^{18}$&0.494&1.16&25.64 \\
$10^{6}$&$8\times 10^{17}$&2.728&4.77&25.32 \\ 
$10^{5}$&$3.16\times 10^{18}$&0.671&1.34&25.32 \\ \hline
\end{tabular}
\caption{\sf \small {Sample coupling constant unification solutions with low 
intermediate scales, $M_R$,
obtained for Model I (the doublet model) through Planck scale
induced interactions parameterized by $\eta_1$ and $\eta_2$ (see
text).}}
\label{table3}
\end{center}
\end{table}

For example with $\epsilon_1 = 0.15$, $\epsilon_2 = 0.174$, $M_G
= M_{Pl.}$  we have   $M_R = 10^7$ GeV and $M_U = 10^{18.4}$ GeV,
corresponding to $\eta_1 = 0.494$ and $\eta_2 = 1.160$.  The
corrections to the coupling constants are obtained through
$\D_{BL}^{\rm {(gr)}} = - 11.47$, $\D_{2L}^{\rm {(gr)}} =  6.52$,
and $\D_{3C}^{\rm {(gr)}} = 0.6$ .  When these are added to
one-loop extrapolated values from $M_Z$ to $M_U ~(\equiv
10^{18.4}$ GeV), the three coupling constants match consistently
with their common value  $\a _G^{-1} \simeq 25$.  All solutions
with high unification scales require $|\eta_{1,2}| \simeq O(1)$
as shown in Table \ref{table3}. Thus, $\rm {dim.}5$ operators are capable of
lowering the left-right symmetry breaking scale to  $M_R = 10^5 -
10^9$ GeV, making Model I consistent with large neutrino mixing
and leptogenesis when the minimal doublet model is supplemented
by the addition of a ${\bf 54}$.

We find that  
high values of $M_U \simeq 10^{18}$ GeV  require smaller 
$\eta_{1,2} \simeq O(1)$ while a lower  $M_U \simeq 10^{16}$ GeV  
requires unnaturally  larger values of the parameters. The preferred 
solutions with naturally large values of $M_U$ exhibit the virtue of 
suppression of Higgsino mediated proton decay by factors $({M_U^0\over M_U})^2 =
10^{-3} - 10^{-4}$.

We now extend the triplet model by the addition of a Higgs representation 
{\bf 54} and including the effects of the two non-renormalizable operators
of eq. (\ref{eq27}). The changes in the mass scales are given by
\ba\left(\D\ln{M_R\over M_Z}\right)_{gr}&=& 
 -{\pi\over {12\a _G}}\left[-2\ep_1 + 45\ep_2\right],\nonumber \\
\left(\D\ln{M_U\over M_Z}\right)_{gr}&=&
 -{\pi\over{12\a _G}}\left[2\ep_1+15\ep_2\right].\label{eq31}
\ea

Unlike for  the doublet model, we find that gravitational corrections alone 
do not succeed in substantially reducing the $M_R$ scale. This
behaviour of the triplet model can be understood in terms of the
larger Higgs representations -- {\bf 126} and ${\bf \overline{126}}$
--  involved and the consequent tension with perturbativity (see
Sec.\ref{sec:lb}). 

\subsection{Doublet model with additional light multiplets}\label{extrascalar}

The third and final alternative that we discuss for obtaining a
low intermediate scale in Model I is through additional light
chiral submultiplets. We find that if there are appropriate light
states in the particle spectrum then the unification of gauge
couplings is consistent with a significant lowering of $M_R$.

In earlier work  attempts have been made to obtain
intermediate scales much lower than the GUT scale by
spontaneous breaking of SUSY $SO(10)$ in the first step and the
gauge group $\lr$ in the second step with or without \cite{lee}
left-right discrete symmetry.  The crucial point of this
chapter is that we require the  left-right symmetric gauge group
with $ g_{2L} = g_{2R}$ to survive to low intermediate scales in
order to evade the gravitino problem and at the same time obtain
low mass $W_R^{\pm}$ gauge bosons to possibly even provide
testable signals at collider energies in the near future.

We present below two models which meet these requirements. The
models are identical up till the scale $M_R$ and consist of the
MSSM particles.  They differ in the number and type of
additional chiral multiplets which contribute in the range  $
M_R$ to $M_U $. In this subsection, we choose to distinguish
between the SUSY scale, $M_S$ (which is chosen at 1 TeV),  and
$M_Z$. The RG evolution of the couplings from $M_Z$ to $M_S$ is
governed by the  one- and two-loop coefficients:
\ba
&&\left(\br{c} a_{Y}\\ a_{2L}\\ a_{3C}\er\right)=\left(\br{c} {21\over 5}\\
-3\\ -7\er\right), ~~
b_{ij}=\left(\br{ccc} {104\over 25}&{18\over 5}&{44\over 5}\\ {6\over 5}&8&
12\\
{11\over 10}&{9 \over 2}&-26\er\right),\, i,j \subset \std, \label{eq10b}
\ea
while from $M_S$ to the scale $M_R$ eq. (\ref{eq10}) 
is applicable. In eq.(\ref{eq10b}) the beta-function coefficients have been 
derived assuming two light doublets in the nonSUSY model below $M_S$ which 
 emerges naturally from the MSSM existing above $M_S$. \\

{\bf Model A:}
  In addition to the  MSSM particles, we assume that
supermultiplets with  
the following gauge quantum numbers are light
with masses at the $M_R$ scale:
\ba
\sigma (3, 1, 1, 4/3)\op {\ov {\sigma}}(\ov 3, 1, 1, -4/3)
& \subset & {\bf 45, 210}, \nonumber\\
\eta(1, 1, 1, 2) \op {\ov {\eta}}(1, 1, 1, -2) & \subset & {\bf 120}.\label{eq33}
\ea

The one- and two-loop
coefficients including these fields are,
\ba
&&\left(\br{c} a'_{BL}\\ a'_{2L}\\ a'_{2R}\\ a'_{3C}\er\right)
=\left(\br{c} 16\\ 2\\ 2\\ -2\er\right),\label{eq34}
\ea
\ba
&& b'_{ij}=\left(\br{cccc} {241/6}&{27/2}&{27/2}&{88/3}\\{9/2}&{32}&{3}&
{24}\\{9/2}&{3}&{32}&{24}\\{11/3}&{9}&{9}&{76/3}\er\right),\,
 i, j=BL, 2L, 2R, 3C.\label{eq35}\ea

At two-loop level the evolution of gauge couplings and their
unification have been shown in Fig. \ref{f:coupl} for $M_R = 10^4$ GeV.
Some sample solutions to the RGEs for gauge couplings with allowed values of
$M_R$, $M_U$ and the GUT fine structure  constant ($\a _G$) are
presented in Table \ref{table4}.  We find that with the grand unification
scale $M_U = 2\times 10^{16}$ GeV, an intermediate scale in the
range of $M_R = 5$ TeV - $10^{10}$ GeV is possible in this model
with excellent unification of the gauge couplings. In spite of
the presence of additional fields, the gauge couplings at the GUT
scale remain perturbative in a manner similar to the minimal GUT
with $\alpha_G^{-1} = 22.22 - 20.40 $.
\vskip0.1cm

{\bf Model B:}
In addition to the MSSM particles we assume
that there are additional superfields  with their masses at the
$M_R$ scale which transform as:
\ba
\xi (6,1, 1, 4/3)\op {\ov {\xi}}(\ov 6,1, 1, -4/3, )  & \subset &
{\bf 54}, \nonumber\\
\eta(1, 1, 1, 2) \op {\ov {\eta}}(1,1, 1,-2)  & \subset & {\bf 120},\nonumber\\
C(1,2,2,0)  & \subset & {\bf 10, 120, 126}, \nonumber \\ 
 D_L(1,3,1,0)\op D_R(1,1,3,0)  & \subset & {\bf 45, 210},\label{eq36}
\ea
where we have used a pair of $C(1,2, 2, 0)$.

The one- and two-loop coefficients in this scenario are
\ba
&&\left(\br{c} a'_{BL}\\ a'_{2L}\\ a'_{2R}\\ a'_{3C}\er\right)
=\left(\br{c} 20\\ 6\\ 6\\ 2\er\right),\label{eq37}
\ea
\ba
b'_{ij}=\left(\br{cccc} {305/6}&{27/2}&{27/2}&{344/3}\\{9/2}&{70}&{9}&
{24}\\{9/2}&{9}&{70}&{24}\\{43/3}&{9}&{9}&{332/3}\er\right), i, j=BL, 2L, 2R, 3C. \label{eq38}\ea

\begin{table}[tbh]
\begin{center}
\begin{tabular}{|l|c|c|r|}\hline
Model&${\rm M_R}$&${\rm M_U}$&${\a _{G}^{-1}}$  \\
&(GeV)&(GeV)& \\ \hline
&$10^{9}$&$1.15\times 10^{16}$&22.22\\
A&$10^{5}$&$1.10\times 10^{16}$&20.83\\
&$10^{4}$&$ 10^{16}$&20.40\\\hline
&$10^{9}$&$1.82\times 10^{16}$&7.58\\
B&$10^{8}$&$2.00\times 10^{16}$&10.13\\ \hline
\end{tabular}
\caption{\sf \small {Sample coupling constant unification solutions for low intermediate scales, $M_R$, in two models with additional light multiplets at the intermediate
scale (see text).}}
\label{table4}
\end{center}
\end{table}

Gauge coupling evolution and unification in this case is shown in
Fig. \ref{f:coupl} for an example with $M_R = 10^{8}$ GeV. A
couple of sample solutions with $M_R$ which satisfy the gravitino
constraint are presented in Table \ref{table4}. For this alternative, the
intermediate scales are typically in the range of $M_R = 10^7$
GeV - $10^{10}$ GeV. 
\begin{figure}[htbp]
  \centering
\psfig{figure=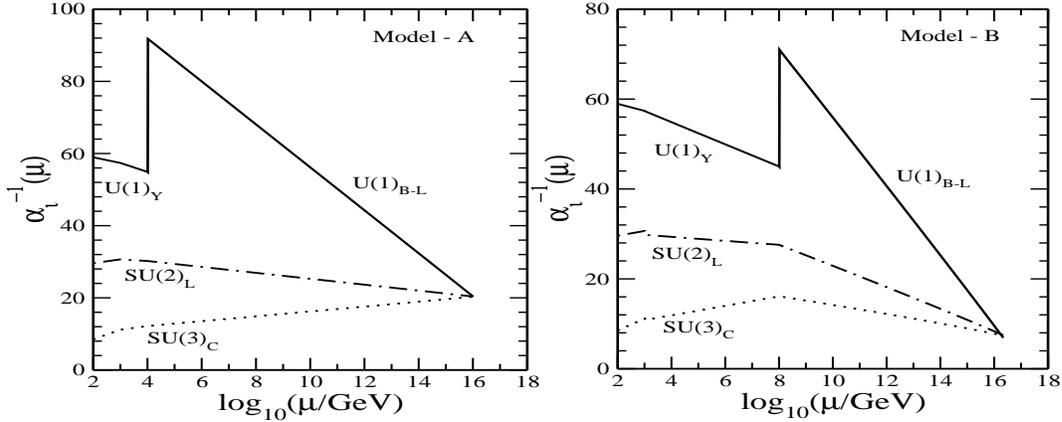,width=15cm,height=8cm,angle=270}
   \caption{\sf \small{Evolution of the gauge couplings in
models with additional light multiplets. The left- (right-) panel
corresponds to Model A (Model B).}}
  \label{f:coupl}
\end{figure}
A very precise unification of the gauge
couplings has been found  when further small SUSY threshold
effects at the TeV scale are taken into account
\cite{baer}. Because of these  effects, the resulting gauge couplings show 
small discontinuities
at $M_S = 10^3$ GeV as shown in the Fig. \ref{f:coupl} for Model B. The gauge
couplings near the GUT scale approach strong coupling 
($ \alpha_G \simeq 0.1 $) as shown in Table \ref{table4} and Fig.
\ref{f:coupl}.

We show in the next section that the intermediate scale in the
triplet model has a lower bound at $10^9$ GeV which is expected
to be increased by  additional Higgs scalars at the $M_R$ scale.

From the above two examples and earlier investigations it is
clear that right-handed mass scales as low as $M_R= 5$ TeV $-
10^{10}$ GeV are viable when additional light chiral multiplets at
the $M_R$ scale are admitted. As already noted, such low scales
are necessary for the successful implementation of leptogenesis
in the doublet model (Model I). Obviously, these models may have
interesting new signatures at LHC and future collider
experiments.  It is noteworthy that all the light multiplets
exploited in the two models are contained in $SO(10)$
representations which have been invoked in the literature anyway
for various purposes.


\section{Lower bound on intermediate scale in the triplet model}
\label{sec:lb}

As pointed out earlier, the  higher dimensional Higgs
representations like ${\bf {210}}$ and/or ${\bf {126} + \ov
{126}}$ result in large threshold corrections at the GUT scale
even if their superheavy components are only few times heavier or
lighter than $M_U$. In this respect, threshold corrections in the
triplet model with ${\bf {126} + \ov {126}}$ are more significant
compared to those in the doublet model which uses ${\bf {16}+ \ov
{16}}$. Normally, one would therefore expect to obtain lower
$M_R$ in the former model.

In this section we show that this is not true and, in fact,
establish that $M_R$ cannot be lower than $10^9$ GeV in the
triplet model. This lower bound is set by the perturbative
renormalization group constraint when parity survives in the
left-right gauge group as happens in the case of $\cal {G}_{LR}$.
As the GUT threshold effects contribute only at the unification
scale, we use the two-loop equation for $\a _{BL}$ between $M_R$
and $ M_U$ with the corresponding coefficients given in eq.
(\ref{eq10}) and eq. (\ref{eq11a}).  It is seen that if $M_R \le
10^{9}$ GeV, $\a _{BL}$ exceeds the perturbative limit ($\simeq
1$) before the GUT scale is reached.

Analytically, this  behavior of the gauge coupling becomes transparent 
by noting that 
the position of the Landau pole ($\mu_0$), where $g_{BL}(\mu_0) = \infty$,
is given by,
  
\ba
\mu_0 &=& M_R ~\exp\left[{2\pi\over a'_{BL}}{1\over \a _{BL}(M_R)}\right] .
\label{eq38a}
\ea

Here
\ba
{1\over \a _{BL}(M_R)}= {5\over 2}\left({1\over
\a _Y(M_Z)}-\Theta_Y + \D_Y\right) 
-{3\over2}\left({1\over\a _{2L}(M_Z)}-\Theta_{2L}+\D_{2L}\right)\nonumber \\
- {1\over 4\pi}\left(5a_Y - 3a_{2L}\right)\ln {M_R\over M_Z}.
\label{eq39} \ea

Using  eq. (\ref{eq39}) we calculate $\a _{BL}^{-1}(M_R)$ for
$M_R = 10^3$ GeV to $10^{11}$ GeV from low energy data  ignoring
the small threshold effect due to superpartners and use them in
eq. (\ref{eq38a}) to estimate the value of $\mu_0$. Our two-loop
estimations of the pole position are shown in Table \ref{table5} for the
triplet model with $a'_{BL}=24$.
\begin{table}[tbh]
\begin{center}
\begin{tabular}{|c|c|c|}\hline
$M_R$&$\a^{-1}_{BL}(M_R)$&$\mu_0$\\
(GeV)&&(GeV) \\\hline
$10^{3}$&$97.429$&$7.76\times 10^{13}$\\
$10^{5}$&$86.407$&$4.56\times 10^{14}$\\
$10^{7}$&$75.406$&$2.56\times 10^{15}$\\
$10^{8}$&$69.907$&$6.16\times 10^{15}$\\
$10^{9}$&$64.409$&$1.44\times 10^{16}$\\
$10^{10}$&$58.912$&$3.46\times10^{16}$\\
$10^{11}$&$53.415$&$8.31\times10^{16}$\\ \hline
\end{tabular}
\caption{\sf \small Location of Landau poles, $\mu_0$, signifying violation of
perturbativity, for different choices of the 
intermediate scale $M_R$ in the triplet model.}
\label{table5}
\end{center}
\end{table}
  The two-loop corrections
predict slightly lower values of $\mu_0$ than eq. (\ref{eq38a}).
For intermediate scales $M_R = 1$ TeV  to  $10^9$  GeV, the pole
positions are found  in the range $7.76\times 10^{13}$ GeV to
$1.44\times 10^{16}$ GeV indicating that for the $U(1)_{BL}$
gauge coupling perturbation theory breaks down below the GUT
scale for these values of $M_R$.  When $M_R \gsim 10^{10}$ GeV,
the pole positions occur clearly above the GUT scale with $\mu_0
\gsim  3.46\times 10^{16}$ GeV.   In other words, with only the
minimal particle content  needed to maintain supersymmetry and
left-right symmetry below the GUT scale, from the requirement of
perturbativity the triplet model leads
to the conservative lower bond on the intermediate scale,
\be
M_R > 10^9 ~ {\rm GeV}. 
\ee

 Inclusion of additional new scalar degrees of freedom anywhere
between $M_R$ to $M_U$ would increase the one-loop
beta-function coefficient of the $U(1)_{B-L}$ gauge coupling and
bring down the pole position further. This, in turn, would
further tighten the lower bound on $M_R$ beyond $10^{9}$ GeV.
This is why, unlike in the doublet model, the presence of additional
light scalars near $M_R$ cannot reduce the value of the
intermediate scale in the triplet model.

In contrast to the triplet model for which $a'_{BL}=24$, the
doublet model has $a'_{BL}=9$  which enhances the argument of
the exponential on the RHS of eq. (\ref{eq38a}) by a factor
$\simeq 24/9 = 2.66$ compared to the triplet model for the same
value of $M_R$. Such a factor in the argument pushes the Landau
pole to a position much above the GUT scale. Thus, even with $M_R
= 1$ TeV, whereas the triplet model pole position is at
$\mu_0\simeq 1.18\times 10^{14}$ GeV which is approximately two
orders below the GUT scale, in the the doublet model the pole
occurs at $\mu_0 \simeq  3.3\times 10^{32}$ GeV.  Although this
latter scale for the doublet model is expected to be
substantially lower because of the contribution of superheavy
particles near the GUT scale, it is clear that the coupling
constant never hits a Landau pole below the
GUT-Planck  scales $\simeq 10^{18}$ GeV. This tallies with the
results in sec.~\ref{techniq}  where solutions have
been obtained using threshold and gravitational corrections.

With such a lower bound on $M_R$ in the triplet model, this
version of  SUSY $SO(10)$ rightly deserves its description as a high
scale theory. The SUSY $SO(10)$ triplet model appears to fit
ideally for description of quark-lepton masses and mixings
through high-scale $b-\tau$ unification and type II see-saw
dominance or even through  type I see-saw mechanism~\cite{min2, goh, babu} .

Since $M_R \gsim 10^{9}$ GeV in the triplet model, the lightest
right-handed neutrino mass could satisfy the gravitino
constraint, but in this case generating the quark and lepton
masses and mixings  has to be re-examined.  While a detailed
analysis of neutrino data is yet to emerge in the doublet model,
it is well known that reproducing small neutrino masses is no
problem even if the right-handed neutrinos are near the TeV
scale.  With such a low value of $M_R$  the desired criteria of TeV
scale resonant leptogenesis is fulfilled and through the
$W_R^{\pm}$ and  $Z_R$ bosons and the light Higgs scalars,
$\chi_L^{\pm}, \chi_L^{0}$, $\chi_R^{\pm}$, and $ \chi_R^{0}$,
the model can be tested  at the LHC and ILC.
The superpartner of the lightest right-handed neutrino in the
doublet model may also be a good candidate for dark matter.

\section{Remarks on light scalars and fermion masses in minimal $SO(10)$}\label{remarkfermion}

One of the most appealing features of the minimal supersymmetric
$SO(10)$ model is that one can calculate the pattern of symmetry
breaking and predict fermion mass relations at the GUT scale
\cite{fmsen}. Concomitant with these, in the minimal model, is
an intermediate left-right breaking scale, $M_R$, constrained to
be rather close to the GUT scale $M_U$.  Can the virtues of the
model be made to survive when $M_R$ is lowered?

Let us briefly summarize the salient features with reference to
Model II. The Higgs fields are: $$ \Phi \equiv {\bf 210}, ~~~~~
\Sigma \equiv {\bf 126}, ~~~~~ \bar \Sigma \equiv {\bf
\overline{126}}, ~~~~~ H \equiv {\bf 10}, $$ where $\Delta_{L,R}
\subset \Sigma$ and $\phi \subset H$. The fermions belong to the
representation $\Psi \equiv$ {\bf 16}.  The complete
superpotential of the model can then be written as
\begin{equation}
W = W_Y + W_H,
\end{equation}
where the Yukawa couplings are in $W_Y$ and the scalar potential
can be derived from $W_H$. They can be written as (we follow the
notations of ref. \cite{min3})
\ba
W_Y &=& Y_{10} \Psi \Psi H + Y_{126} \Psi \Psi \bar \Sigma  ,
\nonumber \\
W_H &=& {m_{\Phi} \over 4!} \Phi \Phi + {\lambda \over 4!} \Phi \Phi \Phi
+ {M \over 5!} \Sigma \bar \Sigma + {\eta \over 4!} \Phi \Sigma
\bar \Sigma \nonumber \\
&& + m_H H H + { 1 \over 4!} \Phi H ~(\alpha \Sigma + \beta \bar \Sigma).
\label{eq47}
\ea
As usual, minimization of the scalar potential  gives the
allowed values of the $vev$ of the different fields. In addition,
fermion mass relations are also determined in terms of the
parameters of the model.

 It may appear 
that the solutions presented earlier with lowered left-right symmetry 
breaking scales are in conflict with results on fermion masses.
However, this need not be the case. 
For example, when gravitational corrections are
included, there may well be non-renormalizable terms in the
superpotential, suppressed by the Planck scale, which can
contribute to the Yukawa couplings after the GUT symmetry 
breaking by the field $\Phi$. 
Thus, in the presence of such corrections, the
superpotential will have to be supplemented by

\begin{equation}
W_Y^{G} = {1 \over M_{Pl}} (Y^G_{10} \Psi \Psi H \Phi + 
Y^G_{126} \Psi \Psi \bar{\Sigma} \Phi) + \cdots .
\end{equation}

These new interactions will be suppressed by ${\langle \Phi
\rangle / M_{Pl}}$. But  $\langle \Phi \rangle \sim M_U$ is close to
the Planck scale, as we have illustrated, and hence the
suppression need not be too much. In addition, the non-renormalizable
couplings $Y^G$ could also be large. Then the fermion mass
relations obtained for the minimal supersymmetric $SO(10)$ models
could be radically affected.  Fermion mass relations can also get
changed in the presence of new Higgs scalars. Thus the
low intermediate mass scales, $M_R$, obtained in the present
analysis need not be inconsistent with the fermion mass
relations.

At the tree level, the minimal triplet
model predicts ~\cite{min3}  masses near
$M_R$ for  additional states belonging to ${\bf 210}$ with the
quantum numbers
\ba
E_L(3, 3, 1, 4/3)\op {\ov E}_L(\ov 3, 3, 1, -4/3), \nonumber\\
E_R(3, 1, 3, 4/3) \op {\ov E}_R(\ov 3, 1, 3, -4/3).\label{eq48}
\ea

We have checked that with the minimal Higgs content, the
renormalizable doublet model also leads to similar light Higgs
scalars. It has been further  noted in ref. \cite{min3} that
these states prevent having  parity conserving ${\cal G}_{LR}$ at
any value of the  intermediate scale below $M_U$.  We remark that
their presence at $M_R$ sufficiently lower than $M_U$, apart from
being in conflict  with  $\sin^2\theta_W(M_Z)$ and $\a _S(M_Z)$,
spoils perturbative gauge coupling evolutions by developing
Landau poles in the coupling constants in the region $M_R < \mu <
M_U$.  This difficulty could be  avoided \footnote{These states represent pseudo-Goldstone
bosons and may also acquire masses near the $M_U$ scale through
loops.} by extensions of the
minimal doublet or the triplet model through the inclusion of
non-renormalizable operators and/or additional $SO(10)$ Higgs
representations,  like $\bf {54}$. For example, the presence of the
non-renormalizable term in the superpotential
\ba
W_{gr} &=& {\lambda^G \over {4!M_G}} \Phi^4, \nonumber  
\ea 
with $M_G = M_{Pl}$, or  (string) compactification scale, can
lift the masses of these light scalars close to the GUT scale
when the ${\bf 210}$ gets $vev$ along the direction
$\langle\Phi^0\{15,1,1\}\rangle~\sim~M_U$, leading to $M_E
={{2\lambda^G m_{\Phi}^2}/{\lambda^2 {M_G}}}$. Then their
contributions are added to GUT-threshold effects,  as discussed
earlier.

\section{Summary and conclusion}\label{summprs}

In this chapter we have discussed the question of low intermediate
left-right symmetry breaking scales, as preferred by
leptogenesis, in the minimal supersymmetric $SO(10)$ GUTs with
only doublet Higgs scalars as well as with triplet scalars.
In view of the presence of additional scalar components predicted
from mass spectra analysis \cite{min3} which disrupt perturbativity and 
gauge coupling
unification, the minimal renormalizable triplet model with Higgs 
representations
${\bf 210 \oplus 126 \oplus {\ov {126}} \oplus 10}$ is ruled out as a 
candidate 
for any value of left-right symmetry breaking intermediate scale. 
With the added presence of a Higgs representation ${\bf 54}$ and/or 
non-renormalizable interactions, these unwanted scalar components are
made superheavy and 
we find, in agreement with previous work, that in the minimal
models, at the one-loop level gauge coupling unification requires
the scale of left-right symmetry breaking to be close to the GUT
scale \footnote{ Here it has been assumed that the light scalar
components in $ \{15,3,1\} \oplus\{15,1,3\} \subset {\bf 210}$,
emerging from mass spectra predictions, are made superheavy. 
 This is possible if, for example, the minimal models are extended 
by the addition of a Higgs representation $\bf {54}$ in each case. But the
situation would be worse still in both the models if the scalar components  
remain light in the absence of  ${\bf 54}$ or suitable nonrenormalizable terms
in the superpotential.}. Inclusion
of the two-loop contributions  eliminates
even this possibility as no solution can be found at all with an
intermediate scale.  On the other hand, evading the gravitino
problem, which would otherwise plague successful big bang
nucleosynthesis, would require $M_R \leq 10^9~ {\rm GeV}$.  We
have pointed  out that this impasse can be circumvented in the
case of the doublet model by including threshold corrections near
the GUT scale, including non-renormalizable interactions due to
gravity induced Planck scale effects, or by adding new light
scalar multiplets.  In the last alternative, the additional light
submultiplets used are present in representations commonly used in
$SO(10)$ non-minimal models, but they are different from those which emerge
from mass spectra analysis~\cite{min3}.  These considerations
allow the left-right symmetry breaking scale to be low, as low as
even a few TeV, making it phenomenologically interesting.  The
unification scale obtained in the doublet model using the first two methods 
turns out to be large,
making it safe for Higgsino mediated proton decay as well as with
fermion mass relations.  
In the triplet model, although threshold
effects can easily decrease the intermediate scale, we find a
perturbative lower bound, $M_R > 10^9$ GeV, below which the
intermediate scale cannot be lowered. With this bound, the
triplet model with an added ${\bf 54}$ and/or nonrenormalizable interactions
 emerges as a high scale theory of SUSY $SO(10)$
description of fermion masses and mixings. In this model the
possibility of meeting the gravitino constraint can be fulfilled
provided neutrino masses and mixings  are successfully reproduced
with $M_R \gsim 10^{9}$ GeV.  With $M_R$ in the TeV region in the
doublet model, apart from successful resonant leptogenesis with
full compliance of the gravitino constraint, the model
predictions  can be tested through their various manifestations
\vspace{0.3cm}
at the LHC and ILC.\\


\newpage
\mbox{}
\chapter{ $SU(6)$, Triquark states, and the pentaquark }



\section{Introduction}

Since long, baryon spectroscopy has been an arena to learn about
low-energy quantum chromodynamics. The purported observation of a
narrow baryon state of strangeness +1 at a mass around 1540 MeV,
$\Theta^+$, by several experiments \cite{theta} brought renewed
attention to this theatre. The evidence in support of this new
state is now of conflicting nature, loaded more in the direction
of non-observation \cite{nonob, rev}. Within the quark picture, the
positive strangeness ($\equiv \bar{s}$) of the $\Theta^+$ baryon
puts it in an exotic category and entails an interpretation in
terms of a minimum of four quarks and an antiquark -- a
pentaquark state ($udud\bar{s}$).

Soon after, three other states which also demand a pentaquark
classification were also observed.  These are the $\Xi^{--}
(dsds\bar{u})$ and $\Xi^{0} (dsus\bar{d})$ both at 1862 MeV
\cite{na49} and the $\Sigma^c (udud\bar{c})$ \cite{hera} with
mass 3099 MeV.

Though exotic states such as the pentaquark have a long history,
particular attention was drawn to a possible $\Theta^+$-like
state in the $SU(3)$ version of the chiral soliton model
\cite{soliton}. Subsequently, the experimental results have
stimulated the exploration of many ideas, e.g., quark clusters,
colour hyperfine interactions,   Goldstone boson exchange, QCD
sum rules, lattice methods, etc., which have been reviewed in the
literature \cite{jm}.

For the $\Theta^+$, within the quark model framework, two models
\cite{kl,jw} have achieved special prominence. It is convenient to
discuss these using the language of $SU(6)$ of colour-spin, $SU(3)$
of colour, and $SU(2)$ of spin. Thus, for example, a quark
transforms as (6,3,2), where the three integers within the
parentheses identify the representations of the above $SU(6)$,
$SU(3)$, and $SU(2)$, respectively. To avoid cluttering, the flavour
$SU(3)$ structure is not explicitly shown. Our interest will
be on the triquark state which is an ingredient of the
Karliner-Lipkin model \cite{kl}.

An alternative possibility is the Jaffe--Wilczek (JW) model
\cite{jw}. Here the four quarks are assumed to form
two diquark clusters, each in the $(21,\bar{3},1)$
representation.  Of the four possible combinations
for a two-quark cluster -- (21,6,3), (15,6,1), $(15,\bar{3},3)$,
$(21,\bar{3},1)$ -- this is the one of the lowest energy. The two
diquark clusters and the remaining antiquark -- each one of which is
in colour $\bar{3}$ -- combine to form the colour singlet
pentaquark state $(qq)(qq)(\bar{q})$, {\em e.g.,} $\Theta^+
\equiv (ud)(ud)(\bar{s})$. A relative orbital angular momentum,
L=1, is assumed between the diquarks; this is in tune with the
observed narrow width of the state.  Another consequence is that
the pentaquark parity  
is predicted to be positive.  Note that the colour-spin symmetric
nature of the $(21,\bar{3},1)$ diquark requires it to be
antisymmetric, $\bar{3}$, in flavour to satisfy the generalized
Pauli principle.  The two diquarks (colour $\bar{3}$ bosons)
combine to form colour 3 to match up with the antiquark. This,
and L=1, requires the combination to be in a flavour symmetric
$\bar{6}$ state. The overall pentaquark flavour must be in
$\bar{6} \otimes \bar{3} = 8 + \overline{10}$.  The quantum
numbers of $\Theta^+$ can be accommodated only in the
$\overline{10}$.

In the Karliner-Lipkin (KL) model the quark clustering is
different. Here, it is postulated that there is one diquark cluster
with the same quantum numbers as in the JW model. The difference
is that the remaining two quarks and the antiquark are assumed to
form a triquark cluster $(qq\bar{q})$ with the quantum numbers
(6,3,2) which is in a flavour $\bar6$. The pentaquark state is
the colour singlet  $(qq)(qq\bar{q})$ combination. To explain the
narrowness of the observed states, a relative orbital angular
momentum, L=1, is postulated between the clusters so that the
parity of the state is predicted to be positive in this model as
well. The flavour structure of the states is the same as in the
JW model.

In this work, we set two goals. First, we take a detailed look at
the group theoretic properties of the triquark state.  We derive
expressions for the $SU(6)$ unitary scalar factors and Racah
coefficients related to the Clebsch-Gordan coefficients relevant
for this state \cite{Majee:2007gi}. Second, we use these results to estimate masses
for pentaquark states. We indicate how flavour symmetry breaking
may be incorporated in the analysis.

In the next section we present the $SU(6)$
unitary scalar factors and Racah coefficients, which have been
derived {\em ab initio}. In
section \ref{s:hypint} we recall the nature of the colour-spin hyperfine
interaction while in the following section we use it to estimate
the hyperfine energies for baryons, mesons, diquarks, and
triquarks. In section \ref{s:pqmass} the different threads are brought together for
estimating pentaquark masses. In section \ref{s:result} we discuss the results.
We end in section \ref{s:concl} with our conclusions.

\section{Some group-theoretic results}\label{s:grp} 
In this section, we collect some results about $SU(6)$ unitary scalar
factors and Racah-like coefficients which will be useful for the
subsequent discussion. Though our motivation in obtaining these
results is the triquark state, they may find some use in
other applications of the $SU(6)$ group.

\subsection{$SU(6)$ unitary scalar factors}
To minimise the complexities, we first
summarize the notations.  A member of an $SU(2)$ multiplet is
denoted by $\{(2I+1),I_3\}$; {\em e.g.,} the  $s_z = +{1 \over
2}$ state of a spin-half particle is $\{{2}, +{1 \over 2}\}$.

For $SU(3)$, the sub-representations are designated by the
$SU(2)^c$ representation\footnote{The superscript `$c$' has been
added to indicate the subgroups of $SU(3)$.} and the `hypercharge',
$Y^c$. Thus, one uses the combination $\{R_3,\alpha, I_3^c\}$
where $R_3$ is the $SU(3)$ representation and $\alpha \equiv [(2I^c
+ 1),Y^c]$. For illustration, a quark state with $I_3^c = +{1
\over 2}$ and $Y^c = {1 \over 3}$ will be denoted as  
$\{3,[{2}, {1 \over 3}],+{1 \over 2}\}$.  

Putting the above together,
an $SU(6)$ state is denoted by
$(R_6,\{R_3,\alpha, I_3^c\},\{(2I+1),I_3\})$ where $R_6$ is the
$SU(6)$ representation while $\{R_3,\alpha,I_3^c\}$ and
$\{(2I+1),I_3\}$ characterize the corresponding $SU(3)$ and $SU(2)$
sub-representations. The quark state mentioned above, will be
$(6,\{3,[{2}, {1 \over 3}], +{1 \over 2}\}, \{{2},\pm{1 \over
2}\})$, where the $SU(3)$ ($SU(2)$) quantum numbers are enclosed in
the first (second) braces. In most of the following, it will be
possible to suppress $\alpha, I_3^c$ and $I_3$ -- {\em e.g.,} the
quark state $\equiv$ (6,3,2). This is because the unitary scalar
factors and the Racah coefficients are independent of $\alpha,
I_3^c$ and $I_3$.

The $SU(6)$ unitary scalar factors are generalisations of the
$SU(3)$ isoscalar factors.  The Clebsch-Gordan (CG)  coefficients
of $SU(2)$ are well known. If $ i
\otimes  j =  k \oplus
\ldots $, where  $i,j,k$ are $SU(2)$ representations,  we use
$CG(SU(2)_{i,j,k})$ as an abbreviation for the usual
$C^{i,j,k}_{i_3,j_3,k_3}$ \cite{co}. 
 
Using the $SU(2)$ submultiplets within an $SU(3)$ representation, the
CG coefficients for $SU(3)$ can be expressed in
terms of products of isoscalar factors and $SU(2)$ CG coefficients.
Schematically, for the case $ P
\otimes  Q =  R \oplus
\ldots $:
\begin{equation}
CG(SU(3)_{P,Q,R}) = \left[\matrix
{P & Q & R  \cr
\alpha_P & \alpha_Q & \alpha_R}
\right] \times CG(SU(2)_{I_P, I_Q, I_R }),
\end{equation}

where the $\alpha_i, ~i=P,Q,R$ indicate the sub-representations of
the $SU(3)$ representations $P,Q,R$. The first factor on the
right-hand-side is the $SU(3)$ isoscalar factor. It is independent
of $I_{P3},I_{Q3},I_{R3}$. Tables of $SU(3)$ isoscalar factors 
have been available for long \cite{dS}.

Similarly, in $SU(6)$, if $ X \otimes  Y =  Z \oplus
\ldots $ then 
\begin{eqnarray}
CG(SU(6)_{X,Y,Z}) &=& \left[\matrix
{X & Y & Z\cr
(P_X,I_X) & (P_Y,I_Y) & (P_Z,I_Z)}
\right] \nonumber \\ 
& & \nonumber \\
& & \times CG(SU(3)_{P_X,P_Y,P_Z}) \times
CG(SU(2)_{I_X,I_Y,I_Z}). 
\end{eqnarray}
Here, the first factor on the right-hand-side is an $SU(6)$ unitary
scalar factor -- the generalization of the $SU(3)$ isoscalar
factor.  $P_X (I_X)$ indicates the $SU(3)$ ($SU(2)$) sub-representation
within the $SU(6)$ multiplet $X$.

Since the triquark state is made out of two quarks ($q_1, q_2$)
and an antiquark ($\bar{q}_3$), the following $SU(6)$ combinations
arise:
\begin{equation}
qq {\rm ~ state:~~} 6 \otimes  6 =  21 \oplus  15 
\label{qq}
\end{equation}
\begin{equation}
qq\bar{q} {\rm ~ state:~~} 21 \otimes  \bar{6} =  120 \oplus  6_1^\phi, \;\; 
15 \otimes  \bar{6} =  84 \oplus  6_2^\phi.
\label{qqbarq}
\end{equation}
or, alternatively,
\begin{equation}
q\bar{q} {\rm ~ state:~~} 6 \otimes  \bar{6} =  35 \oplus  1 
\label{qbarq}
\end{equation}
\begin{equation}
q\bar{q}q {\rm ~ state:~~} 35 \otimes  6 =  120 \oplus  84 \oplus
6_1^\psi, \;\; 1 \otimes  6 =   6_2^\psi.
\label{qbarqq}
\end{equation}
The superscripts $\phi$ and $\psi$ will be clarified in the next
subsection where we identify the Racah coefficients which relate
$(6_1^\phi, 6_2^\phi)$ to $(6_1^\psi, 6_2^\psi)$.
 
For the purpose of the triquark, the $SU(6)$ CG coefficients 
for the product  21 $\otimes$  $\bar{6}$
=  120 $\oplus$  6 are necessary. We have not been able to find the
$SU(6)$ unitary scalar factors for this product in the published
literature \cite{cm}.  Here, therefore, their  {\em ab initio}
calculated values are presented. We follow the generalized
Condon-Shortley phase convention \cite{bb} and obtain:
\begin{equation}
\left[\matrix
{21 & \overline{6} & 6 \cr
(6,3) & (\overline{3},2) & (3,2)}
\right] = \sqrt{6 \over 7},\;\;\;
\left[\matrix
{21 & \overline{6} & 6\cr
(\overline{3},1) & (\overline{3},2) & (3,2)}
\right] = \sqrt{1 \over 7}.
\label{usf1}
\end{equation}

Also,
\begin{equation}
\left[\matrix
{21 & \overline{6} & 120 \cr
(6,3) & (\overline{3},2) & (3,2)}
\right] = \sqrt{1 \over 7},\;\;\;
\left[\matrix
{21 & \overline{6} & 120\cr
(\overline{3},1) & (\overline{3},2) & (3,2)}
\right] = -\sqrt{6 \over 7}.
\label{usf2}
\end{equation}

For the sake of completeness, the $SU(6)$ unitary scalar
factors for the case  15 $\otimes$  $\bar{6}$
=  84 $\oplus$  6 are:
\begin{equation}
\left[\matrix
{15 & \overline{6} & 6\cr
(6,1) & (\overline{3},2) & (3,2)} 
\right] = \sqrt{2 \over 5},\;\;\;
\left[\matrix
{15 & \overline{6} & 6\cr
(\overline{3},3) & (\overline{3},2) & (3,2)}
\right] = \sqrt{3 \over 5}.
\label{usf3}
\end{equation}
and
\begin{equation}
\left[\matrix
{15 & \overline{6} & 84\cr
(6,1) & (\overline{3},2) & (3,2)} 
\right] = \sqrt{3 \over 5},\;\;\;
\left[\matrix
{15 & \overline{6} & 84\cr
(\overline{3},3) & (\overline{3},2) & (3,2)}
\right] = -\sqrt{2 \over 5}.
\label{usf4}
\end{equation}
\vskip 30pt

\subsection{Racah coefficients for the triquark cluster}
\subsubsection{$SU(2)$ and $SU(3)$} 

In this subsection, after recapitulating the concept of Racah
coefficients, 
using angular momentum as an illustration, the necessary results
useful for the triquark case are presented.

When three angular momenta $j_1,j_2,j_3$ are added, one can
obtain the same final angular momentum $j$ by, for example, (a)
combining $j_1$ and $j_2$ first to get $j_{12}$ and adding $j_3$
to it, or by (b) first adding $j_1$ and $j_3$ to obtain $j_{13}$
and then combining it with $j_2$, or by (c) adding $j_2$ and
$j_3$ to obtain $j_{23}$ and then adding $j_1$ to it.  The states
of the representation $j$ obtained by these three different
routes, may be denoted by $|j_1,j_2,j_3;j_{12},j,m\rangle$,
$|j_1,j_2,j_3;j_{13},j,m\rangle$, and
$|j_1,j_2,j_3;j_{23},j,m\rangle$, respectively. These three
sets of states are related to each other by unitary
transformations whose coefficients, $U$, are called the {\em
normalized} Racah coefficients. For example,

\begin{equation}
U(j_1,j_2,j_3,j;j_{12},j_{13}) =
\langle j_1,j_2,j_3;j_{12},j,m|j_1,j_2,j_3;j_{13},j,m\rangle  .
\label{eq:rdef}
\end{equation}

The triquark state is of the structure ($q_1 q_2 \bar{q}_3$). 
Since the quarks (antiquarks) transform as 6 ($\bar{6}$) of
colour-spin $SU(6)$, for the analysis of these states one requires
the Racah coefficients for $SU(6)$ for the product $6
\times 6 \times  \bar{6}$. 

For most purposes,
it actually suffices if one has the colour $SU(3)$ and spin $SU(2)$
Racah coefficients. 

The same final triquark state may be reached
by first combining $q_1$ and $q_2$ (colour: $3 \times 3 = \bar3 +
6$ and spin: $2 \times 2 = 3 + 1$) and then combining with each
of these possibilities the antiquark state $\bar{q}_3$. An
alternate way of obtaining the same state is to first pair $q_1$
with $\bar{q}_3$ (colour: $3 \times \bar3 = 8 + 1$ and spin: $2
\times 2 = 3 + 1$) and then adjoining $q_2$ to the result. A
third possibility is obtained by interchanging $q_1
\leftrightarrow q_2$ in the previous alternative. 

We concentrate, in the interest of the pentaquark application, on
the triqark state which transforms like a colour $SU(3)$ triplet
and an $SU(2)$ doublet. The basis states in this sector may be
denoted as:
\begin{equation}
\left( \matrix{|\phi_1\rangle\cr |\phi_2\rangle\cr
|\phi_3\rangle\cr |\phi_4\rangle\cr }
\right) \equiv 
\left( \matrix{|(q_{1}q_{2})^{\bar{3}}_{1}(\bar q_{3})^{\bar 3}_{2}
\rangle_{(3,2)} 
\cr |(q_{1}q_{2})^{6}_{1}(\bar q_{3})^{\bar 3}_{2} \rangle_{(3,2)} \cr
|(q_{1}q_{2})^{\bar{3}}_{3}(\bar q_{3})^{\bar 3}_{2} \rangle_{(3,2)}
\cr |(q_{1}q_{2})^{6}_{3}(\bar q_{3})^{\bar 3}_{2} \rangle_{(3,2)} \cr}
\right)
\label{base1}
\end{equation}
and
\begin{equation}
\left( \matrix{|\psi_1\rangle\cr |\psi_2\rangle\cr
|\psi_3\rangle\cr |\psi_4\rangle\cr }
\right) \equiv
\left( \matrix{|(q_{1}\bar{q}_{3})^{1}_{1}(q_{2})^{3}_{2} \rangle_{(3,2)} 
\cr |(q_{1}\bar{q}_{3})^{8}_{1}(q_{2})^{3}_{2} \rangle_{(3,2)} \cr
|(q_{1}\bar{q}_{3})^{1}_{3}(q_{2})^{3}_{2} \rangle_{(3,2)} 
\cr |(q_{1}\bar{q}_{3})^{8}_{3}( q_{2})^{3}_{2} \rangle_{(3,2)} \cr}
\right),\;\;\; 
\left( \matrix{|\chi_1\rangle\cr |\chi_2\rangle\cr
|\chi_3\rangle\cr |\chi_4\rangle\cr }
\right) \equiv
\left( \matrix{|(q_{2}\bar{q}_{3})^{1}_{1}(q_{1})^{3}_{2} \rangle_{(3,2)} 
\cr |(q_{2}\bar{q}_{3})^{8}_{1}(q_{1})^{3}_{2} \rangle_{(3,2)} \cr
|(q_{2}\bar{q}_{3})^{1}_{3}(q_{1})^{3}_{2} \rangle_{(3,2)} 
\cr |(q_{2}\bar{q}_{3})^{8}_{3}( q_{1})^{3}_{2} \rangle_{(3,2)} \cr}
\right).
\label{base2}
\end{equation}
The notation used here, for example, is that the triquark state with
$SU(3)$ ($SU(2)$) multiplicity $c'$ ($s'$) obtained through the diquark
combination $(q_{1}q_{2})$ with $SU(3)$ and $SU(2)$ multiplicity $c$ and
$s$, respectively, is represented as $|(q_{1}q_{2})^{c}_{s} (\bar
q_{3})^{\bar 3}_{2}\rangle_{(c',s')}$.

These
possibilities are related by Racah-like coefficients which are
found by explicit calculation to be:
\begin{equation}
\left( \matrix{|\phi_1\rangle\cr |\phi_2\rangle\cr
|\phi_3\rangle\cr |\phi_4\rangle\cr }
\right) 
=\left( \matrix{-\frac{1}{2\sqrt3} &  \frac{1}{\sqrt6} & \frac{1}{2}
& -\frac{1}{\sqrt2} \cr
\frac{1}{\sqrt6} &  \frac{1}{2\sqrt3} & -\frac{1}{\sqrt2}
& -\frac{1}{2} \cr
\frac{1}{2} &  -\frac{1}{\sqrt2} & \frac{1}{2\sqrt3}
& -\frac{1}{\sqrt6} \cr
-\frac{1}{\sqrt2} &  -\frac{1}{2} & -\frac{1}{\sqrt6}
& -\frac{1}{2\sqrt3} \cr}
\right)
\left( \matrix{|\psi_1\rangle\cr |\psi_2\rangle\cr
|\psi_3\rangle\cr |\psi_4\rangle\cr }
\right) 
\label{conv1}
\end{equation}
and

\vskip 10pt

\begin{equation}
\left( \matrix{|\phi_1\rangle\cr |\phi_2\rangle\cr
|\phi_3\rangle\cr |\phi_4\rangle\cr }
\right)
=\left( \matrix{-\frac{1}{2\sqrt3} &  \frac{1}{\sqrt6} & \frac{1}{2}
& -\frac{1}{\sqrt2} \cr
-\frac{1}{\sqrt6} &  -\frac{1}{2\sqrt3} & \frac{1}{\sqrt2}
& \frac{1}{2} \cr
-\frac{1}{2} &  \frac{1}{\sqrt2} & -\frac{1}{2\sqrt3}
& \frac{1}{\sqrt6} \cr
-\frac{1}{\sqrt2} &  -\frac{1}{2} & -\frac{1}{\sqrt6}
& -\frac{1}{2\sqrt3} \cr}
\right)
\left( \matrix{|\chi_1\rangle\cr |\chi_2\rangle\cr
|\chi_3\rangle\cr |\chi_4\rangle\cr }
\right).
\label{conv2}
\end{equation}

\subsubsection{$SU(6)$ Racah coefficients}

One can use the 
unitary scalar factors in eqs. (\ref{usf1}) - (\ref{usf2}) to write:
\begin{equation}
|q_1q_2\bar{q}_3\rangle_{(6_1^\phi,3,2)} = \sqrt{6 \over 7}
~|\phi_4\rangle +
\sqrt{1 \over 7} ~|\phi_1\rangle,
\;\;|q_1q_2\bar{q}_3\rangle_{(120,3,2)} = \sqrt{1 \over 7}
~|\phi_4\rangle -
\sqrt{6 \over 7} ~|\phi_1\rangle.
\label{tri1}
\end{equation}

From eqs. (\ref{usf3}) - (\ref{usf4}) the states obtained if the
diquarks are in the 15 of $SU(6)$ are:
\begin{equation}
|q_1q_2\bar{q}_3\rangle_{(6_2^\phi,3,2)} = \sqrt{2 \over 5}
~|\phi_2\rangle +
\sqrt{3 \over 5} ~|\phi_3\rangle,
\;\;|q_1q_2\bar{q}_3\rangle_{(84,3,2)} = \sqrt{3 \over 5}
~|\phi_2\rangle - \sqrt{2 \over 5} ~|\phi_3\rangle.
\label{tri2}
\end{equation}

Using eq. (\ref{conv1}) one then has:
\begin{equation}
|q_1q_2\bar{q}_3\rangle_{(6_1^\phi,3,2)} = -\sqrt{7 \over 12}|\psi_1 \rangle
-\sqrt{ 2\over 21}|\psi_2 \rangle -\sqrt{ 1\over 28}|\psi_3 \rangle
-\sqrt{ 2\over 7}|\psi_4 \rangle
\label{eq:six1}
\end{equation}
and
\begin{equation}
|q_1q_2\bar{q}_3\rangle_{(6_2^\phi,3,2)} = \sqrt{5 \over 12}|\psi_1 \rangle
-\sqrt{ 2\over 15}|\psi_2 \rangle -\sqrt{ 1\over 20}|\psi_3 \rangle
-\sqrt{ 2\over 5}|\psi_4 \rangle
\label{eq:six2}
\end{equation}
Thus, one arrives at the Racah coefficients:
\begin{equation}
\left( \matrix{|(6_1^\phi,3,2)\rangle \cr |(6_2^\phi,3,2)\rangle}
\right)
=\left( \matrix{\sqrt{\frac{5}{12}} & -\sqrt{\frac{7}{12}}
\cr
\sqrt{\frac{7}{12}} & \sqrt{\frac{5}{12}} }
\right)
\left( \matrix{|(6_1^\psi,3,2)\rangle \cr |(6_2^\psi,3,2) \rangle}
\right)
\label{racah6}
\end{equation}

The non-trivial unitary scalar factors corresponding to eq.
(\ref{qbarqq}) can be written as:
\begin{equation}
\left[\matrix
{35 & 6 & \alpha \cr
i & ({3},2) & (3,2)}
\right] = U_{i,\alpha},
\end{equation}
with $i = 1,2,3$ corresponding to (8,1), (1,3), and (8,3) while
$\alpha = 1,2,3$ to 120, 84, and $6_1^\psi$. Then,
\begin{equation}
U
=\left( \matrix{
-\sqrt{\frac{9}{28}} &  -\sqrt{\frac{8}{21}}  & \sqrt{\frac{25}{84}}  \cr
\sqrt{\frac{9}{20}} &  -\sqrt{\frac{8}{15}}  & -\sqrt{\frac{1}{60}}  \cr
-\sqrt{\frac{8}{35}} &  -\sqrt{\frac{3}{35}}  & -\sqrt{\frac{24}{35}}  \cr}
\right),
\label{usfpsi}
\end{equation}

Now we turn to the application of these results to the pentaquark.

\section{Colour-spin hyperfine interaction}\label{s:hypint}

Besides colour electric forces between all quarks and antiquarks,
there exists a colour-spin hyperfine (colour magnetic)
interaction \cite{dgg}.  In the KL model, it is assumed
that this interaction is operative inside the clusters but, due
to the larger separation, the hyperfine interaction between
clusters is negligible\footnote{Inclusion of the inter-cluster
hyperfine interaction has also been considered \cite{cheung}.}.
The colour-spin $SU(6)$ hyperfine interaction energy is:
\begin{equation}
V = -\sum_{i > j} v_{ij}
(\vec{\sigma_i}.\vec{\sigma_j})(\vec{\lambda_i} . \vec{\lambda_j}).
\label{eq:hyperf}
\end{equation}
Here, $\vec{\sigma}$ and $\vec{\lambda}$ are the Pauli and Gell-Mann
matrices, and $i$ and $j$ run over the constituent quarks and
antiquarks. The common practice is to take $v_{ij} \equiv v$
(flavour symmetry).  $v$ captures information about the radial
dependence of the bound state wave-function. For a composite
system of $n_q$ quarks and $n_{\bar{q}}$ antiquarks, the
hyperfine energy contribution is given by:
\begin{equation}
E_{hyp} = \left[D(q+\bar{q}) - 2D(q) - 2D(\bar{q}) + 16(n_q +
n_{\bar{q}}) \right] v/2,\label{eq:ehyp}
\end{equation}
where
\begin{equation}
D(R_6,R_3,s) = C_6(R_6) - C_3(R_3) - {8 \over 3}s(s+1).
\end{equation}
$C_6$ and $C_3$ are the quadratic Casimir
operators of $SU(6)$ and $SU(3)$ respectively, and $s$, is the spin
of the state.  The effect of this hyperfine interaction on
multiquark exotic states has been a topic of research over
several decades \cite{jaffe,hs}.

The mass estimate for the pentaquark proceeds along the following
pattern. There are three contributions: (a) the masses of the
constituent quarks, (b) the colour-spin hyperfine energy, and (c)
the energy due to the P-wave excitation. The practice has been to
estimate (a) from the masses of the decay products, (baryon +
meson), since their quark content is the same as that of the
parent; but here the hyperfine interaction contribution to the
baryon and meson mass must be first subtracted out, as detailed
in section V. Thus, the
hyperfine interaction enters directly in (b) and also indirectly
in (a) through the way it is extracted.

\section{Hyperfine energies}
\subsection{Mesons and Baryons}
As noted, the hyperfine interaction contributions to the
meson $(q\bar{q})$ and baryon $(qqq)$ masses are required for the
estimation of the pentaquark mass. These can be readily
calculated using eq.  (\ref{eq:ehyp}). For example, in the
flavour symmetry limit, one finds:
\begin{equation}
E_{N(70,1,2)} = -8v,\;\;E_{\Delta(20,1,4)} = 8v,\;\;E_{\pi(1,1,1)} = 
-16v,\;\;E_{\rho(35,1,3)} =\frac{16}{3}v, 
\label{eq:had}
\end{equation}
where in the parentheses the $SU(6)$, $SU(3)$, and $SU(2)$ properties
of the particle have been indicated. 

\subsection{The diquark cluster}
As already mentioned, the  diquark $(qq)$ is usually chosen to be
in the (21,$\bar3$,1) representation which is symmetric in $SU(6)$.
In addition, a diquark can be in the (21,6,3), (15,6,1), and
(15,$\bar3$,3) but these have higher energy. One finds from eq.
(\ref{eq:ehyp}) that the hyperfine energies for these four states
are:
\begin{equation}
E_{(21,\bar{3},1)} = -8v,\;\; E_{(21,6,3)}= -\frac{4}{3}v, \;\;
E_{(15,6,1)}=4v, \;\; E_{(15,\bar{3},3)}= \frac{8}{3}v.
\label{eq:diq}
\end{equation}

\subsection{The triquark cluster}
The triquark cluster in the Karliner-Lipkin model is a member of
the (6,3,2) multiplet and contains two quarks and an antiquark.
The two quarks are assumed to combine to a symmetric 21 of
colour-spin $SU(6)$. For $SU(6)$ 21 $\otimes \bar6$ = 6 $\oplus$ 120,
and the triquark (120,3,2) carries higher hyperfine energy. If
the two quarks are combined in an antisymmetric fashion,
producing a 15 of $SU(6)$, then\footnote{In $SU(6)$, $15 \otimes
\bar6 = 6 \oplus 84$.  In the absence of flavour symmetry, the
triquark is a superposition of these and the 6 and 120 (see
later).} the triquark can be in (6,3,2) or (84,3,2).

More important is the fact that in the existing literature, the
triquark in the (6,3,2) is {\em assumed} to be made with the two
quarks within the cluster forming a (21,6,3). In actuality, so
long as flavour symmetry of the hyperfine interaction holds, the
lowest energy eigenstate of $SU(6)$ receives contributions from
both the $(21,6,3)$ and the $(21,\bar3,1)$ combinations -- see
eq.  (\ref{usf1}) -- and this triquark has the form given in
the first expression in eq. (\ref{tri1}). The other possible
triquark states are the second expression in eq. (\ref{tri1}) and
the  ones in eq. (\ref{tri2}).

\subsubsection{The triquark hyperfine energy}

The calculation of the triquark hyperfine energy using
eq. (\ref{eq:ehyp}) is complicated by the fact that the operator
$D(q+\bar{q})$ and $D(q)$ do not commute; e.g., in eq. (\ref{tri1})
an eigenstate of $D(q+\bar{q})$ is expressed as a linear combination
of those of $D(q)$. 

To circumvent this difficulty, we use the following procedure. We
consider the contribution of eq. (\ref{eq:hyperf}) for the triquark
state term by term as:
\begin{equation}
V =  V_{12} (\vec{\sigma_1}.\vec{\sigma_2})(\vec{\lambda_1} .
 \vec{\lambda_2}) +
V_{13} (\vec{\sigma_1}.\vec{\sigma_3})(\vec{\lambda_1} .
\vec{\lambda_3}) + V_{23}
(\vec{\sigma_2}.\vec{\sigma_3})(\vec{\lambda_2} .
\vec{\lambda_3}).  
\label{eq:thf}
\end{equation}

The hyperfine energy from each term is most readily
calculated in the basis where the two contributing
quarks/antiquarks are first combined \cite{hs2}; i.e., corresponding to the
three terms in the r.h.s. of eq. (\ref{eq:thf}) these are the
$|\phi \rangle$, $|\psi \rangle$, and $|\chi \rangle$ bases
of Sec. \ref{s:grp}, respectively. They are related to each other
through eqs. (\ref{conv1}) and (\ref{conv2}).
In terms of these basis states, one can immediately write down
the expectation value of the Hamiltonian in eq. (\ref{eq:thf}).
Thus\footnote{This form was noted in  \cite{hs2}}, one has:
\begin{equation}
\langle\phi | V|\phi\rangle
=\left( \matrix{\frac{4}{3}V_{12}+ \frac{20}{3}V^{\phi}_+& 4\sqrt{2}V^{\phi}_- & 
\frac{10}{\sqrt{3}}V^{\phi}_- & 2\sqrt{6}V^{\phi}_+ \cr
4\sqrt{2}V^{\phi}_- & -\frac{8}{3}V_{12}+ \frac{8}{3}V^{\phi}_+ & 2\sqrt{6}V^{\phi}_+ 
&\frac{4}{\sqrt{3}}V^{\phi}_-  \cr
\frac{10}{\sqrt{3}}V^{\phi}_- & 2\sqrt{6}V^{\phi}_+  & -4V_{12} 
& 0 \cr
2\sqrt{6}V^{\phi}_+ &  \frac{4}{\sqrt{3}}V^{\phi}_-  & 0
& 8V_{12} \cr}
\right),
\label{expH1}
\end{equation}
where $V^{\phi}_{\pm} = V_{13} \pm V_{23}$. Analogously,
\begin{equation}
\langle\psi | V|\psi\rangle
=\left( \matrix{\frac{8}{3}V_{12}+ \frac{2}{3}V_{13}+ \frac{28}{3}V_{23}& 
\frac{16}{3\sqrt{2}}V^{\psi}_- & 
\frac{4}{\sqrt{3}}V_{12} - \frac{14}{\sqrt{3}}V_{23} &
\frac{8}{\sqrt{6}}V^{\psi}_+
\cr
\frac{16}{3\sqrt{2}}V^{\psi}_- & -\frac{16}{3}V_{13} &
\frac{8}{\sqrt{6}}V^{\psi}_+ & 0 \cr
\frac{4}{\sqrt{3}}V_{12} - \frac{14}{\sqrt{3}}V_{23} &
\frac{8}{\sqrt{6}}V^{\psi}_+ & -2V_{13} & 0 \cr
\frac{8}{\sqrt{6}}V^{\psi}_+ &  0  & 0 & 16V_{13} \cr}
\right),
\label{expH2}
\end{equation}
where $V^{\psi}_{\pm} = V_{12} \pm V_{23}$. $\langle\chi |
V|\chi\rangle$ is similar and is not presented here.
 
The eigenvalues and eigenvectors of this matrix
give the triquark energy and its corresponding group theoretic
configuration, respectively.

The method which we follow can be smoothly adopted to the case of
flavour symmetry violation by appropriately changing the individual coupling
strengths in the three terms of eq. (\ref{eq:thf}). In the
flavour symmetry limit, $V_{12} = V_{23} = V_{13} = v$, whence
$V_-^{\phi} = V_-^{\psi} = 0$. It is seen from 
eq. (\ref{expH1}) that $(\phi_1, \phi_4)$ decouple from
$(\phi_2, \phi_3)$ in this limit.

\section{Pentaquark masses}\label{s:pqmass}
\subsection{Hyperfine interaction couplings}
Needless to say, the strength of the colour-spin hyperfine
interaction, $v$, is an important ingredient of the pentaquark mass
estimation. The procedure has generally been to assume that it
takes a universal value which is estimated by ascribing the
$\Delta - N$ mass splitting to this interaction. Using eq.
(\ref{eq:had}),
\begin{equation}
v_3 = \frac{m_\Delta - m_N}{16} \simeq 18.3\; {\rm MeV}.
\label{eq:vbar}
\end{equation}
While this can be a first approximation, it should be borne in
mind that $v$ is determined by the radial dependence of the bound state
wave-function and thus is  most likely different 
for two-body and three-body bound states. Indeed, using eq.
(\ref{eq:had}) for the meson sector one has, 
\begin{equation}
v_2 = \frac{m_\rho - m_\pi}{64/3} \simeq 29.6\; {\rm MeV}.
\label{eq:vmes}
\end{equation}
This is actually an overestimate of $v_2$ since it is well known
that  the pion mass is too small for a simple quark model
interpretation. Eq.  (\ref{eq:vmes}) is only for the purpose of
illustration\footnote{We extract $v_2$ from heavier mesons in the
next subsection.}.  However, it does indicate that it may not be
unreasonable to expect that  $v_2 \neq v_3$ would give a better
approximation to reality.  In the following, in addition to
discussing the results for the choice $v_2 = v_3$,  for the
sake of comparison, we also use a $v_2$ for the diquarks
different from the $v_3$ for the triquarks.

\subsection{Flavour symmetry breaking}
In the limit of exact flavour symmetry, the splitting between the
lowest lying pseudoscalar mesons and the corresponding vector
mesons with the same quark content would be flavour independent.
A measure of flavour symmetry breaking can be obtained from
\begin{equation}
x_f = \frac{m_{K^*} - m_K}{m_\rho - m_\pi} \simeq 0.63.
\label{eq:flav}
\end{equation}
This suggests that the hyperfine interaction involving an
$s$-quark or antiquark carries a suppression by the
factor $x_f$. 
In eqs. (\ref{eq:vmes}) and (\ref{eq:flav}) the use of $m_\pi$
makes the precise values inaccurate. To improve upon this, we use
the masses of the heavier mesons $\rho$, $\phi$, $K^*$, and $K$.
Using eq. (\ref{eq:had}), the hyperfine contributions for these
states are, respectively,  
\begin{equation}
E_{\rho} =\frac{16}{3}v_2, \;\;E_{\phi}=\frac{16}{3}x_{f2}^2v_2,
\;\; E_{K^*} =\frac{16}{3}x_{f2}v_2, \;\;E_{K} =-16x_{f2}v_2,
\label{eq:mesf}
\end{equation}
Here we have added a subscript to $v$ and $x_f$ to indicate that
these values of the hyperfine parameters apply for
two-quark and/or antiquark systems. Using the masses of the mesons, one
can solve for the hyperfine interaction parameters $(v_2,
x_{f2})$ as well as the quark masses. In this manner, one gets:
\begin{equation}
v_2 = 23.62 ~{\rm MeV}, \;\; x_{f2} = 0.782, \;\; 
m_{u,d} = 322 ~{\rm MeV}, \;\; m_s = 471 ~{\rm MeV}.
\label{eq:mespar}
\end{equation}
These values are used in our subsequent calculations.

There are two three-body systems which enter in this analysis.
One is the triquark state and the other the baryon to which the
pentaquark decays. Just as for mesons, one can estimate the
values of $v_3$ and $x_{f3}$ from the $N - \Delta$ and $\Sigma -
\Sigma^*$ mass splittings which are given by:
\begin{equation}
E_\Delta - E_N = 16v_3,\;\; E_{{\Sigma}^*} - E_{\Sigma} =
\frac{16}{3}v_3(2x_{f3} + 1),\;\; E_{{\Xi}^*} - E_{\Xi} =
\frac{16}{3}v_3x_{f3}(x_{f3} + 2).
\label{eq:barf}
\end{equation}
As a consistency check, we use the
values so obtained to calculate the $\Xi - \Xi^*$ splitting and
find that the agreement is not satisfactory. Therefore, we use all
of the three above splittings to arrive at the best-fit values:  
\begin{equation}
v_3 = 17.89 ~{\rm MeV}, \;\; x_{f3} = 0.708.
\label{eq:barpar}
\end{equation}
In the following, these have been used for the triquark and
baryons.

\subsection{P-wave excitation}
The energy due to the P-wave excitation can be estimated from the
recently observed $D_s^*$ state at 2317 MeV, which is believed to
be an orbital excitation of the state at 2112 MeV. This
gives\footnote{Alternatively, one might use $ E_{P} =
m_{\Lambda({\frac{1}{2}})^-} - m_{\Lambda({\frac{1}{2}})^+}
\simeq (1406 - 1116)\; {\rm MeV} = 290\; {\rm MeV}$. This will
increase all pentaquark mass estimates below by $\sim $ 85 MeV.}
\begin{equation}
E_{P}= m_{D_s^*}(P) -
m_{D_s^*}(S) \simeq (2317 - 2112)\; {\rm MeV} = 205\; {\rm MeV}.
\label{eq:pw}
\end{equation}
\section{Results}\label{s:result}
\subsection{The flavour antidecuplet and the octet}
Putting together the inputs from the previous sections, one can
readily obtain the masses of the pentaquark states in the
Karliner-Lipkin model. For example, for $\Theta^+$, using eqs.
(\ref{eq:had}) and (\ref{eq:diq}):
\begin{equation}
m_{\Theta^+} = \{(m_N + 8v_3) + (m_s + m_q)\} + E_P  - 8v_2 + 
E_{tri}(v_3,x_{f3}),
\label{klmass}
\end{equation}
where the expression in the curly brackets is the contribution
from the quark masses.  The last (penultimate) term is the hyperfine
energy of the triquark (diquark). For other
pentaquarks, the r.h.s. in eq.  (\ref{klmass}) has to be
appropriately modified to reflect the quark content of the state
and, when necessary, deviations from flavour symmetry have to be
incorporated in eq.  (\ref{eq:thf}) to obtain the correct
$E_{tri}(v_3,x_{f3})$.

\begin{table}[hbt]
\begin{center}
\begin{tabular}{|c|c|c|c|c|c|c|c|}
\hline
Pentaquark &\multicolumn{7}{|c|}{Mass (in MeV)} \\ \cline{2-8} 
states & $\Theta^+$ &  $N_{10}$ & $\Sigma_{10}$ & $\Xi_{10}$ &
 $N_{8}$ & $\Sigma_{8}$ &  $\Xi_{8}$  \\ \hline
Lowest    &1601&1358&1626 & 1783&2057& 2217 & 2326\\ \hline 
$SU(6)$ Excited   &1789&1573&1840 & 1966&2321& 2439 & 2512\\ \hline 
\end{tabular}
\caption{\sf \small Pentaquark lowest lying state and first colour-spin
excited state masses for the reference values of the parameters
in eqs. (\ref{eq:mespar}) and (\ref{eq:barpar}).}
\label{t1}
\end{center}
\end{table}

As noted earlier, the pentaquark states fill an octet and an
antidecuplet of flavour. Excepting for the three states, $\Theta^+
\equiv udud\bar{s}$, $\Xi^{--} \equiv dsds\bar{u}$, and $\Xi^{+}
\equiv usus\bar{d}$, all other states in the antidecuplet have
partners in the octet with identical isospin and hypercharge. In
estimating the masses,  we have assumed {\em ideal} mixing between
the partners and ascribed the {\em lighter} member to the
antidecuplet. Note that isospin symmetry is assumed unbroken, so it is
enough to present the mass of one member of an isomultiplet. The
masses of the pentaquark states at the reference values of the
parameters -- see eqs. (\ref{eq:mespar}) and (\ref{eq:barpar}) --
are given in Table \ref{t1}.

In Fig. \ref{f1}, in the left panel the antidecuplet pentaquark
masses are shown as a function of the flavour symmetry violation
parameter $x_f$, which assumes the value unity in the symmetry
limit. In view of the closeness of the estimates of $x_f$ in eqs.
(\ref{eq:mespar}) and (\ref{eq:barpar}), for this figure we have
taken $x_{f3} = x_{f2} = x_{f}$. The triquark interaction
strength has been kept fixed at  $v_3 = 17.89$ MeV. The bands
arise from a variation of the strength of the diquark hyperfine
interaction, $v_2$, with the lower edge corresponding to $v_2 =
v_3$ and the upper to  $v_2 = 23.62$ MeV (see eq.
(\ref{eq:mespar})).  For this figure, $E_P$ has been chosen as 209
MeV, following eq. (\ref{eq:pw}). It is observed that the
triquark corresponding to the lowest eigenvalue of the hyperfine energy
Hamiltonian -- eq. (\ref{expH1}) --  is predominantly a combination
of the states $\phi_1$ and $\phi_4$ (see eq. (\ref{base1})) which
are antisymmetric in the quark flavours.

\begin{center}
\begin{figure}[thb]
\includegraphics[width=0.5\textwidth,height=0.4\textheight,angle=270]
{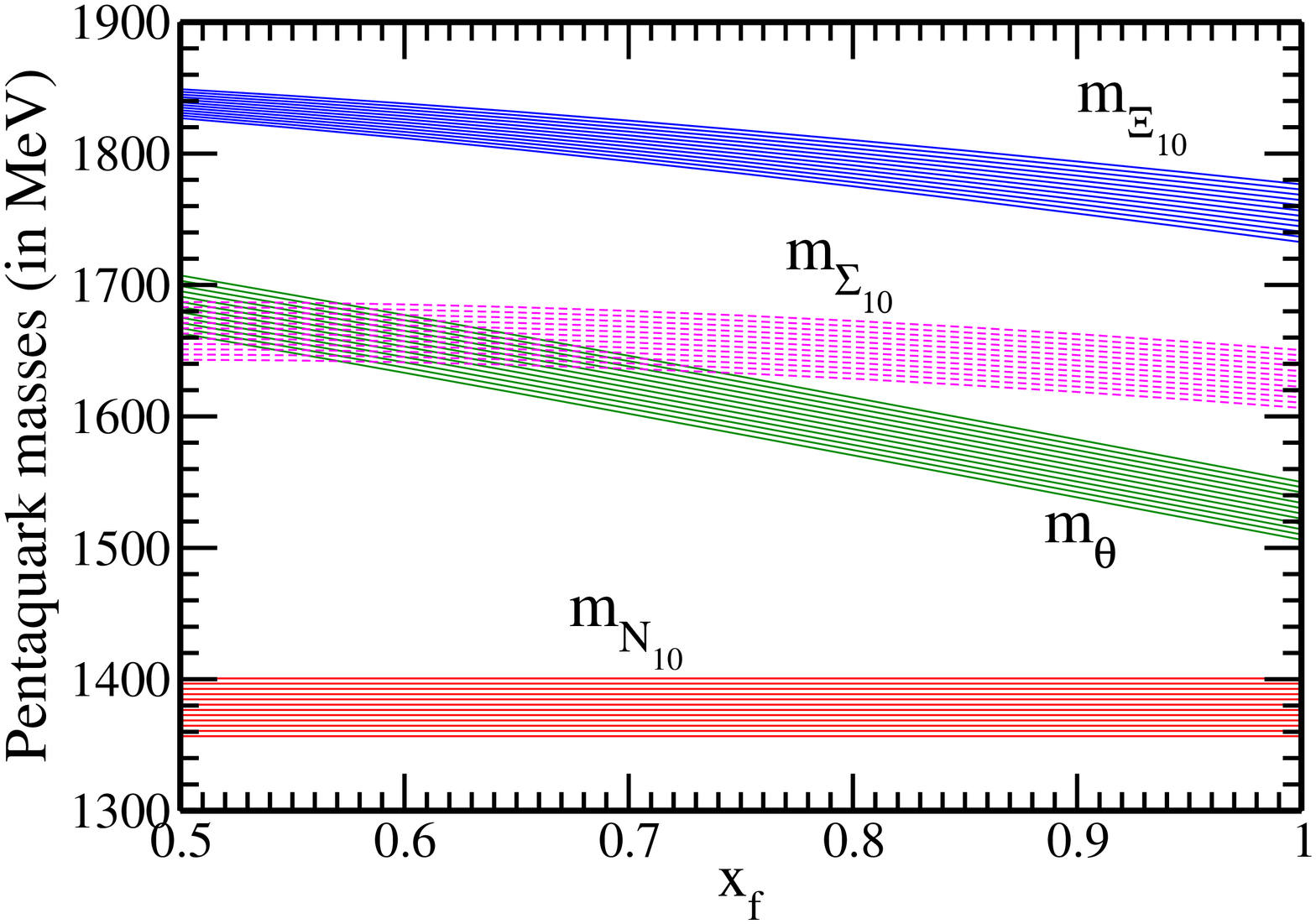}
\vskip -8.5cm
\hskip 8.5cm
\includegraphics[width=0.5\textwidth,height=0.4\textheight,angle=270]
{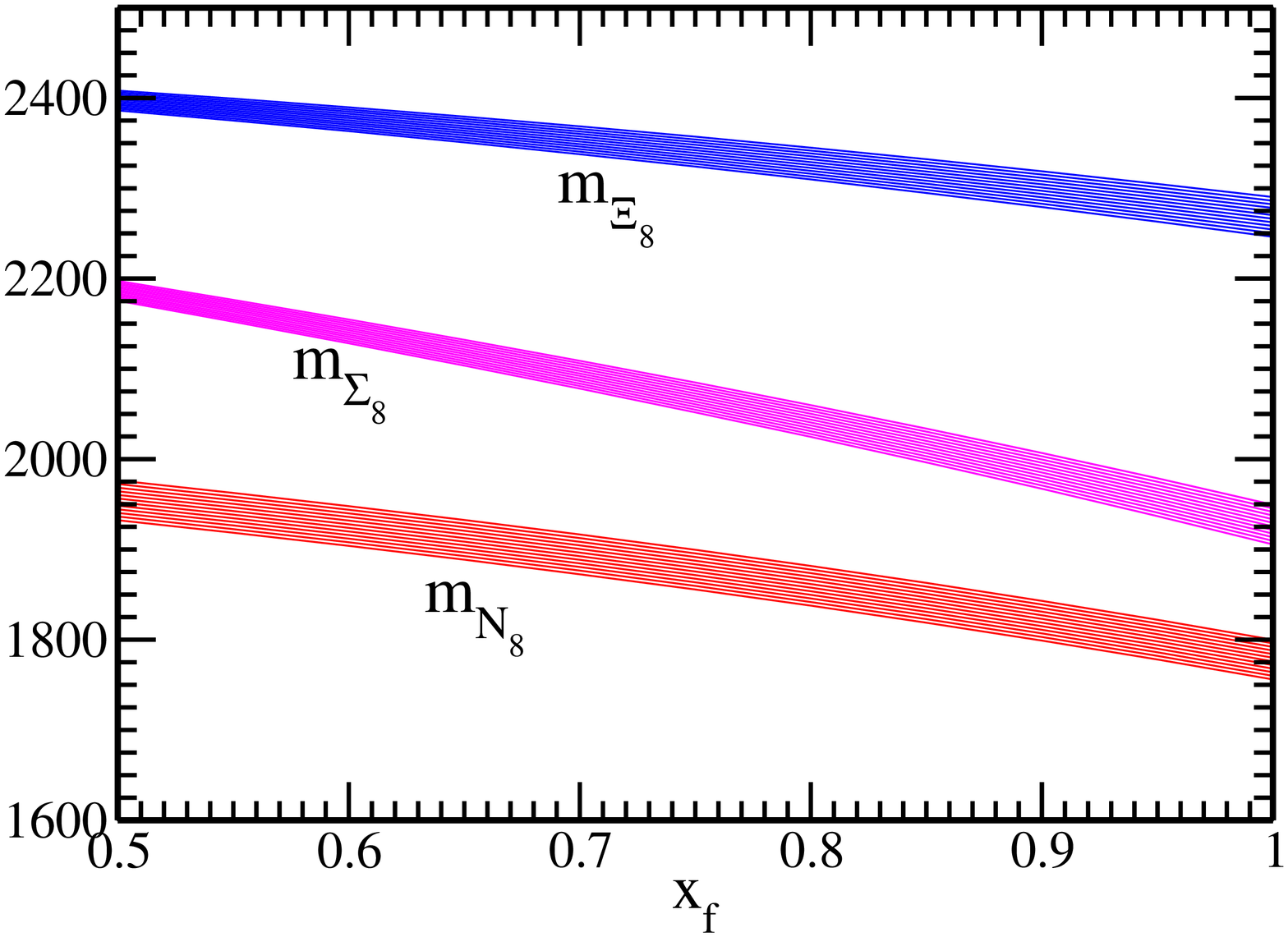}
\caption{\sf \small{Dependence of pentaquark masses on the deviation
from flavour symmetry ($x_f = 1$). The left (right) panel
corresponds to flavour antidecuplet (octet) pentaquarks. The
bands are obtained when the diquark hyperfine
interaction strength is varied over the range  17.89 MeV
$\leq v_2 \leq 23.62$ MeV (see text).}}
\label{f1}
\end{figure}
\end{center}

\vskip -40pt
Note that, $N_{10}$, the non-strange member of the
antidecuplet\footnote{This state could have been proposed as a
possible interpretation of the Roper resonance at 1440 MeV.} is
predicted to be at a mass of 1355 MeV for $v_2 = v_3$ which is
enhanced to $\sim$ 1400 MeV when $v_2 = 23.62$ MeV is used. This
prediction is independent of the choice of $x_f$ since the state
does not have strange quarks.  For the exotic $\Xi^{--}_{10}$
state the mass prediction is in the range 1795 -- 1825 MeV for
$x_f$ = 0.7 to be compared with that of the experimentally
observed state at 1862 MeV \cite{na49}.

In the right panel of  Fig. \ref{f1} are shown the octet
pentaquark masses.  The splitting between the masses of the octet
states and the corresponding antidecuplet states is seen to be
typically around 500-600 MeV. As noted earlier, at the level of
these calculations, the masses of the I=1 and I=0 members of the
octet with S = -1 are the same.   The non-strange neutral state
in the octet, $N_8^0$, has the quark structure $(ud\bar{s})(ds)$
and its mass is consequently dependent on $x_f$.

A remark needs to be made about the symmetry property of the
triquark state for the octet pentaquarks. This feature is most
easily brought out from a consideration of the S = -2 member of
the octet, $\Xi_8$, which has the quark structure
$(us)(ss\bar{s})$. The diquark is antisymmetric in flavour so its
choice is fixed. Unlike all the other states, here the triquark
is compelled to have two identical ($s$) quarks, besides the
antiquark.  Consequently, in the notation of section II, it can
arise only from a combination of the states $\phi_2$ and $\phi_3$
(see eq. (\ref{base1})) which are symmetric in flavour.
Obviously, all states in the pentaquark octet will share this
feature in the exact flavour $SU(3)$ limit.

The H1 experiment at HERA found evidence of a
possible charmed pentaquark at mass 3099 MeV \cite{hera}. This
state has the quantum numbers of a pentaquark with the structure
$udud\bar{c}$.  Including flavour violation 
($x_f = 0.23$ for the
$\bar{c}$ quark) and taking 
$v_2$ = 23.62 MeV, $v_3$ = 17.89 MeV, we find the
predicted mass for such a state is 2757 MeV.

\subsection{Triquark $SU(6)$ excitations}

Colour triplet, spin $\frac{1}{2}$ triquarks come in four
varieties. These are the four eigenstates of the hyperfine energy
matrix in eq. (\ref{expH1}). The results presented so far are
obtained using the eigenstate with the minimum energy consistent
with symmetry requirements -- a
certain choice of colour-spin assignments for the quark clusters
-- and leads to the lowest lying pentaquarks. It is evident that
the other triquark eigenstate clusters also lead to colour
singlet spin $1 \over 2$ pentaquark states, albeit heavier. How
different are the masses in these other cases?

For illustration, we show in Table \ref{t1} the masses of the
first excited partners of the antidecuplet and octet pentaquarks
for the reference values of the hyperfine interaction parameters.
In the flavour symmetry limit ($x_{f3} = x_{f2}$ = 1), the
spacing between the excited states is independent of the flavour
and the lowest and first excited states are separated by 215 MeV
(370 MeV) for every member of the antidecuplet (octet).
 
There is no obvious argument to suppress the production of these
additional states.  It will be of interest to extend the ongoing
searches to look for such $SU(6)$ colour-spin excited partners, a
novelty of QCD and the pentaquark system.

\section{Conclusions}\label{s:concl}

A pentaquark interpretation of the $\Theta^+$ leads to
predictions of several other colour singlet states in a similar
mass range which populate an antidecuplet and an octet of flavour
$SU(3)$. In this work, the masses of these pentaquark states have
been calculated in a triquark-diquark (Karliner-Lipkin) model with 
refined estimates, up to first order, of the color-spin $SU(6)$ 
hyperfine interaction contributions.

Motivated by the structure of these states,  the $SU(6)$ unitary
scalar factors relevant for the $qq\bar{q}$ triquark structure
and the Racah coefficients, not available in the literature, have
been calculated {\em ab initio}. Using these results, the
colour-spin $SU(6)$ hyperfine contributions have been obtained
taking two variations from the simplest picture. One of these
concerns the deviation from flavour symmetry. The other
originates from a possible difference in the strength of the
hyperfine interaction for two- and three-quark bound states which
can be related to  the known splittings in baryonic and mesonic
systems. Both of these variations do affect the pentaquark mass
predictions.  An element of uncertainty is introduced in these
mass estimates by the P-wave excitation energy for which we have
used the information from the $D$-meson system.

The triquark states within the antidecuplet and the octet are
chosen, for good reason,  to be the lowest eigenstate of the
hyperfine energy Hamiltonian satisfying  symmetry
requirements.  The other eigenstates are possible triquark states
of $SU(6)$ colour-spin excitations. The masses of colour singlet,
spin $1\over 2$ pentaquarks resulting from these triquark
excitations have also been estimated.

Irrespective of whether the claimed observation of the $\Theta^+$
baryon is vindicated or not, pentaquarks can prove to be the
tip of a revealing iceberg of new hadronic states illuminating
novel facets of QCD. 


\newpage
\chapter{Summary and Conclusions}
The authenticity of the existence of a theory beyond the standard model is now
beyond doubt. We have already had experimental evidence, \emph{like} 
neutrino mass, in its favour from the electro-weak sector. In addition, we 
have different theoretical as well as conceptual problems \emph{like} the hierarchy problem etc.
 discussed in the Introduction. Beyond the standard model, thus, is an obvious 
area one should look into in order to explain the present and future 
experimental data as well as to have a clear picture about the physics.
As the LHC is getting all set to roar in 2008, expectations are
mounting as we prepare ourselves to get a glimpse of new and
unexplored territory. New physics of different incarnations,
especially supersymmetry and/or extra dimensions, are crying out for
verification.
So based on the current 
experimental data we put some constraints on different parameters of  
new physics possibilities and also discuss how the characteristics of different 
standard model phenomena change in the presence of such new physics.

The first two works of my thesis, discussed in chapters 2 and 3, are 
devoted to the physics of extra dimensions.
In chapter 2 we have discussed how the evolutions of different 
gauge, Yukawa and quartic coupling constants are affected in the 
presence of extra dimensions. How they differ from the conventional
SM behaviour has been the subject of our investigation in that work. 
In that chapter we have performed a diagram by diagram book-keeping 
leading to the evolution equations. We have observed that low gauge 
coupling unification scales can be achieved due to the power law evolution of the
 coupling constants. The unification scale depends on
$R$, and is approximately given by $\Lambda \sim (25-30)/R$. The 
$`$triviality' and $`$vacuum stability' bounds on the Higgs mass have
been studied in the context of power law evolution. This limits the
Higgs mass in the range $148 ~ \ltap m_H ~\ltap 186$ GeV at the 
one-loop level. We had also pointed out for the first time that if low 
energy SUSY is realised in Nature, then the requirement of perturbative 
gauge coupling unification pushes the inverse radius of compactification 
all the way up to $\sim 10^{10}$ GeV. Thus if superpartners of the SM 
particles are observed at the LHC, the nearest KK states within the UED 
framework are predicted to lie beyond the boundary of any observational 
relevance.

In chapter 3 we have probed how much the upper limit on the lightest neutral 
Higgs mass, ${\rm m}_h$, could be relaxed, should the MSSM be 
embedded in one ($S^1/Z_2$) or two ($T^2/Z_4$) extra dimensions. As the large 
contribution will come from the top 
family, we had in our model only allowed third generation of 
fermions with their superpartners to access the extra dimension. This
also helps to keep the theory perturbative in the intended zone of $R$.
The KK towers of the  top quark and stop squarks provide a 
positive contribution to ${\rm m}_h^2$ raising it by several tens of GeV.
Ignoring the left-right scalar mixing and assuming moderate 
$\tan\beta \sim (5-10)$, we obtain $\Delta {\rm m}_h^2 ({\rm KK})
\sim (60~{\rm GeV})^2 \times (M_S R)^2$ in the 5d scenario.
Including the left-right scalar mixings, i.e., non-zero $\mu$ and
trilinear parameters, somewhat enhances the magnitude of the
correction. In the 6d theory with two extra dimensions 
compatcified on a chiral square the correction gets sizably enhanced 
due to a denser packing of KK states. At the same time the low tan~$\b$ 
region can be revived in this scenario.  

The problem of low intermediate left-right symmetry breaking scales, 
as preferred by leptogenesis, in the minimal supersymmetric $SO(10)$ GUTs with
only doublet Higgs scalars as well as with triplet scalars is discussed
in chapter 4. The minimal renormalizable triplet model with Higgs 
representations ${\bf 210 \oplus 126 \oplus {\ov {126}} \oplus 10}$ is 
excluded as a candidate for any low value of left-right symmetry breaking 
intermediate scale. 
We find in agreement with previous work, that in the minimal models,
 at the one-loop level gauge coupling 
unification requires the scale of left-right symmetry breaking to be 
close to the GUT scale. Inclusion of the two-loop contributions  eliminates
even this possibility as no solution can be found at all with an
intermediate scale.  On the other hand, evading the gravitino
problem, which would otherwise plague successful big bang
nucleosynthesis, would require $M_R \leq 10^9~ {\rm GeV}$.  We
have pointed  out that this impasse can be circumvented in the
case of the doublet model by including threshold corrections near
the GUT scale, including non-renormalizable interactions due to
gravity induced Planck scale effects, or by adding new light
scalar multiplets.  In the last alternative, the additional light
submultiplets used are present in representations commonly used in
$SO(10)$ non-minimal models, but they are different from those which emerge
from mass spectra analysis~\cite{min3}.  These considerations
allow the left-right symmetry breaking scale to be low, as low as
even a few TeV, making it phenomenologically interesting.  The
unification scale obtained in the doublet model using the first two methods 
turns out to be large,
making it safe for Higgsino mediated proton decay as well as 
fermion mass relations.  
In the triplet model, although threshold
effects can easily decrease the intermediate scale, we find a
perturbative lower bound, $M_R > 10^9$ GeV, below which the
intermediate scale cannot be lowered. 

Besides the discussion of different new physics beyond the standard 
model we have also tried to explore some features of quantum 
chromodynamics in chapter 5. 
In this work, the masses of the pentaquark states have 
been calculated in a triquark-diquark (Karliner-Lipkin) model with
refined estimates of the colour-spin $SU(6)$ hyperfine interaction
contributions. Motivated by the structure of these states, the $SU(6)$ 
unitary scalar factors relevant for the $qq\bar{q}$ triquark structure
and the Racah coefficients, not available in the literature, have
been calculated {\em ab initio}. The result is used to determine
the tree level pentaquark masses properly. Irrespective of whether the 
claimed observation of the $\Theta^+$ baryon is vindicated or not, pentaquarks 
can prove to be the tip of a revealing iceberg of new hadronic states 
illuminating novel facets of QCD. 

\newpage
\bc
{\Large \underline {List of Publications}}
\ec

\begin{itemize}
\item
{\bf Power law blitzkrieg in universal extra dimension scenarios.} \\
Authors: {\sf Gautam Bhattacharyya, Anindya Datta, Swarup Kumar Majee, Amitava Raychaudhuri}\\
Published in {\it Nucl.Phys.{\bf B760}:117-127,2007}.\\
e-Print: hep-ph/0608208

\item
{\bf Low intermediate scales for leptogenesis in SUSY SO(10) GUTs.}\\
Authors: {\sf Swarup Kumar Majee, Mina K. Parida, Amitava Raychaudhuri, Utpal Sarkar}\\
Published in {\it Phys.Rev.{\bf D75}:075003,2007.}\\
e-Print: hep-ph/0701109

\item
{\bf Extra-dimensional relaxation of the upper limit of the lightest supersymmetric neutral Higgs mass.}\\
Authors: {\sf Gautam Bhattacharyya, Swarup Kumar Majee, Amitava Raychaudhuri}\\
Published in {\it Nucl.Phys.{\bf B793}:114-130,2008.}\\
e-Print: arXiv:0705.3103 [hep-ph]

\item
{\bf SU(6), Triquark states, and the pentaquark.}\\
Authors: {\sf Swarup Kumar Majee, Amitava Raychaudhuri}\\
Published in {\it Phys.Rev.{\bf D77}:074016,2008.}\\
e-Print: arXiv:0711.3910 [hep-ph]

\end{itemize}

\end{document}